\documentclass[a4paper, 12pt, sort&compress]{elsarticle}

\usepackage{graphicx}
\usepackage{subfig}
\usepackage{amsmath}
\usepackage{amssymb}
\usepackage{bbm, bbold}
\usepackage{float}
\usepackage{algorithm}
\usepackage{setspace}
\usepackage{bm}
\usepackage{xcolor}
\usepackage{multirow}
\usepackage[shortcuts]{extdash}
\usepackage{url}
\usepackage{hyperref}

\newcommand{\pderiv}[2]{\frac{\partial #1}{\partial #2}}
\newcommand{\dev}[1]{\mathbf{#1}_{\mathrm{dev}}}

\newcommand{\IBi}{\int_{\mathcal{B}_{0}}}
\newcommand{\IBt}{\int_{\mathcal{B}_{t}}}

\newcommand{\T}{\text{T}}
\newcommand{\dV}{\text{d}V}
\newcommand{\dv}{\text{d}v}
\newcommand{\dA}{\text{d}A}

\newcommand{\PsiVol}{\Psi^\text{vol}}
\newcommand{\PsiDev}{\Psi^\text{dev}}

\newcommand{\mycolor}[1]{\textcolor{black}{#1}}

\newcommand{\myred}[1]{{\leavevmode\color{black}#1}}

\RequirePackage[margin=20mm]{geometry}

\graphicspath{{./figures-pdf/}}

\begin{document}

\fontfamily{ptm}\selectfont

\begin{frontmatter}

\title{A \mycolor{linearised} consistent mixed displacement-pressure formulation for hyperelasticity}

\author[add1]{Chennakesava Kadapa\corref{cor1}}
\ead{c.kadapa@swansea.ac.uk}

\author[add2]{Mokarram Hossain}

\cortext[cor1]{Corresponding authors}

\address[add1]{Swansea Academy of Advanced Computing, Swansea University, SA1 8EN, United Kingdom}
\address[add2]{Zienkiewicz Centre for Computational Engineering (ZCCE), Swansea University, SA1 8EN, United Kingdom}
\begin{abstract}
We propose a novel mixed displacement-pressure formulation based on an energy functional that takes into account the relation between the pressure and the volumetric energy function. We demonstrate that the proposed two-field mixed displacement-pressure formulation is not only applicable for nearly and truly incompressible cases but also is consistent in the compressible regime. Furthermore, we prove with analytical derivation and numerical results that the proposed two-field formulation is a simplified and efficient alternative for the three-field displacement-pressure-Jacobian formulation for hyperelastic materials whose strain energy density functions are decomposed into deviatoric and volumetric parts.
\end{abstract}


\end{frontmatter}

\section{Introduction}
Hyperelastic models are the widely used constitutive equations for rubber-like polymers and soft tissues, and these models are the building blocks for the constitutive relations of fast-growing soft and smart materials such as electroactive and magnetoactive polymers. The deformation behaviour of materials modelled with hyperelastic constitutive equations ranges from the compressible regime in which materials undergo significant volumetric changes to the incompressible regime where the volumetric change is zero.

Hyperelastic models in the incompressible regime require sophisticated finite element formulations for computing accurate numerical results. In the literature, there exist various techniques for addressing the issue of incompressibility. For the truly incompressible case, the mixed displacement-pressure formulation is the only computationally-efficient option, which, however, results in saddle-point problems \cite{book-fem-BrezziFortin,book-fem-ZienkiewiczVol2,KadapaPhDThesis,KadapaCMAME2016elast,KadapaIJNME2019mixed}. To overcome the numerical issues in solving saddle-point problems, the widely followed approach in computational solid mechanics is to impose the incompressibility constraint weakly using quasi-incompressible approximations. Some popular approaches for simulating  quasi-incompressible hyperelastic materials are $\bm{F}$-bar formulation \cite{NetoIJSS1996,NetoIJNME2005Fbarpatch}, enhanced-strain method \cite{SimoJAM1986}, average nodal strain formulation \cite{BonetCNME1998,PiresCNME2004}, reduced integration method with hourglass control \cite{FlanaganIJNME1981,BelytschkoCMAME1984hglass}, enhanced assumed strain (EAS) methods \cite{WriggersCM1996, KorelcCM2010}, the two-field displacement-pressure formulation \cite{ChenCMAME1997,KadapaPhDThesis,KadapaCMAME2016elast}, the three-field formulation \cite{SimoCMAME1985}, mixed stabilised formulations \cite{ChiumentiCMAME2002,CerveraCMAME2010a,CerveraCMAME2010b,ScovazziIJNME2016,ScovazziCMAME2017velocity,AbboudIJNME2018,FrancaNM1988,KlassCMAME1999,MasudJAM2005}, energy-sampling stabilisation \cite{PakravanIJNME2017b,BijalwanIJMSI2019}, least-squares formulations \cite{KadapaIJNME2015,KadapaPhDThesis,ManteuffelSIAMJNA2006,MajidiSIAMJNA2001,MajidiSIAMJNA2002}, $\bm{F}$-bar projection methods \cite{ElguedjCMAME2008}. Some noteworthy recent contributions to incompressible and quasi-incompressible computational solid mechanics are in \cite{LeiEC2016, MehnertMMS2017, WulfinghoffCMAME2017, WriggersCM2017, JiangIJCM2018, BayatAMSES2018, BayatCM2018, CoombsCMAME2018, SevillaIJNME2018, ToghipourCMA2018, BijalwanIJMSI2019, MoutsanidisCPM2019, DalIJNME2019, OnishiIJNME2017, OnishiIJCM2019, SevillaCS2019, connolly2018, connolly2019, ViebahnAMSES2019, RajagopalMAMS2018}.

While the two-field mixed displacement-pressure formulation based on the perturbed Lagrangian approach, cf. \cite{ChenCMAME1997,KadapaPhDThesis,KadapaCMAME2016elast,book-fem-BonetWood,SchroderCM2017}, yields accurate results for the nearly and truly incompressible models, the accuracy of results deteriorates for problems that undergo significant stretches in the compressible regime, i.e., for Poisson's ratio, $\nu \leq 0.4$. This deterioration of accuracy is due to the absence of the relation between the pressure and the volumetric energy function in the perturbed Lagrangian approach. Although such an issue can be overcome by employing the weak statement for the two-field mixed formulation \cite{KadapaIJNME2019mixed,AbboudIJNME2018,ScovazziIJNME2016}, this approach results in unsymmetric matrix systems for the majority of volumetric energy functions. Note that any solver for unsymmetric matrices is expensive when compared to the one for symmetric matrices. \mycolor{Although a mixed displacement-pressure formulation using complementary energy functions yields symmetric matrix systems, the difficulties associated with computing the complementary energy functions limit the applicability of such a formulation to a few simple volumetric energy functions \cite{OrtigosaCMAME2016}.}

The three-field mixed displacement-pressure-Jacobian formulation is one of the widely-used numerical approaches in recently emerged computational electromechanics and magnetomechanics for rubber-like materials \cite{BisharaMMS2018,ParkIJSS2012,ParkSS2013,SeifiCMAME2018,AskIJNLM2012,PelteretIJNME2016,JabareenPIUTAM2015,mehnert2018,mehnert2019}. The formulation takes into account the relation between the pressure and the volumetric energy functional and also results in symmetric matrix systems. However, since the mechanical part of the constitutive models for rubber-like polymers is decomposed into deviatoric and volumetric parts, the three-field formulation gives no particular computational advantage for such materials modelling because of the fact that the coupling term between the displacement and Jacobian variables vanishes. Besides, the three-field formulation is not applicable for simulating the truly incompressible materials.

Recently, Schr\"oder et al. \cite{SchroderCM2017} proposed a novel consistent two-field mixed displacement-pressure formulation that is independent of the penalty function. This approach, however, is limited to incompressible and quasi-incompressible cases as it generalises the classical perturbed Lagrangian approach for different penalty functions in the incompressibility limit. Motivated by the work of Schr\"oder et al. \cite{SchroderCM2017}, in this paper, we present a new generalised mixed displacement-pressure formulation that (i) is applicable for compressible, quasi-incompressible as well as truly incompressible cases, (ii) takes into account the relation between the pressure and the volumetric energy function, (iii) yields symmetric matrix systems irrespective of the volumetric energy function, and (iv) proves to be an efficient alternative for the three-field formulation. Such a consistent formulation is necessary for simplified implementations and for viscoelastic and elastoplastic material models in which the deformation behaviour can vary between compressible and incompressible regimes during the course of a simulation.

This paper is organised as follows. The basics of finite strain elasticity are introduced in Section \ref{section-governing-equations}. The mixed displacement-pressure formulation based on the perturbed Lagrangian approach and the issued associated with it are discussed in Section \ref{section-pertLagr}, followed by the displacement-pressure formulation based on the weak Galerkin statement in Section \ref{section-Galerkin}. \myred{ The proposed mixed displacement-pressure formulation is derived in Section \ref{section-proposedform}, and is compared against the well-established three-field mixed displacement-pressure-Jacobian formulation in Section \ref{section-3field}. The accuracy of the proposed formulation is demonstrated with two numerical examples in Section \ref{section-examples}. The paper is concluded with Section. \ref{section-summary} with a summary of observations made and conclusions drawn.
}

\section{Governing equations for finite strain elasticity} \label{section-governing-equations}
\subsection{Deformation, strain and stress}
With $\mathcal{B}_{0}$ as the reference configuration of the solid body under consideration, and $\mathcal{B}_{t}$ as its new configuration under the action of external forces, the nonlinear deformation map that takes a point $\bm{X} \in \mathcal{B}_{0}$ to a point $\bm{x} \in \mathcal{B}_{t}$ is denoted as $\mathcal{X} : \mathcal{B}_{0} \rightarrow \mathcal{B}_{t}$. Now, the displacement field ($\bm{u}$) and the deformation gradient ($\bm{F}$) are defined as
\begin{align} \label{eqn-displ-definition}
\bm{u} &:= \mathcal{X}(\bm{X}) - \bm{X} = \bm{x} - \bm{X}, \\
\bm{F} &:= \pderiv{\mathcal{X}}{\bm{X}} = \pderiv{\bm{x}}{\bm{X}} = \bm{I} + \pderiv{\bm{u}}{\bm{X}},
\end{align}
where, $\bm{I}$ is the second-order identity tensor. Two important strain measures that are used in this work are the right Cauchy-Green deformation tensor ($\bm{C}$) and the determinant of deformation gradient ($J$), and they are defined as,
\begin{align}
\bm{C} := \bm{F}^{\T} \, \bm{F}; \quad \mathrm{and} \quad J := \det(\bm{F}).
\end{align}

For truly incompressible materials, the deformation of the solid is such that the total volume change at any point in the domain is zero. This can be represented mathematically as the \textit{incompressibility constraint} in the finite strain regime as,
\begin{align} \label{eqn-constraint-fic}
J = 1.
\end{align}

For modelling hyperelastic materials in the incompressible finite strain regime, the deformation gradient, $\bm{F}$, is decomposed into deviatoric and volumetric components as
\begin{equation}
\bm{F} = \bm{F}_{\mathrm{vol}} \, \bm{F}_{\mathrm{dev}},
\end{equation}
with
\begin{equation}
\bm{F}_{\mathrm{vol}} := J^{1/3} \, \bm{I}, \quad \mathrm{and} \quad \bm{F}_{\mathrm{dev}} := J^{-1/3} \bm{F}.
\end{equation}

Using the above definitions, the modified versions of the deformation gradient and the right Cauchy-Green tensor are related as
\begin{align}
\overline{\bm{F}} &:= J^{ -1/3} \, \bm{F} \\
\overline{\bm{C}} &:= \overline{\bm{F}}^{\T} \, \overline{\bm{F}}.
\end{align}

In this work, the strain energy function for the hyperelastic materials is assumed to be decomposed into a deviatoric part, $\Psi^{\text{dev}}$, and a volumetric part, $\Psi^{\text{vol}}$, as
\begin{align} \label{eqn-energy-total}
\Psi(\overline{\bm{C}},J) = \Psi^{\text{dev}}(\overline{\bm{C}}) + \Psi^{\text{vol}}(J).
\end{align}
Here, the deviatoric part of the function is often represented as a function of the invariants of $\overline{\bm{C}}$ and the volumetric part is only a function of the Jacobian $J$. For comprehensive details on the widely-used strain energy functions for hyperelastic materials, we refer the reader to the works of Steinmann et al. \cite{SteinmannAAM2012}, Hossain and Steinmann \cite{HossainJMBM2013}, Hossain et al. \cite{HossainJMBM2015}, Marckmann and Verron \cite{MarckmannRCT2006}, Doll and Schweizorhof \cite{DollJAM2000} and Moerman et al. \cite{Moerman2019}, and references cited therein.

Now, two important stress measures of interest in the present work are defined as
\begin{align}
\text{First Piola-Kirchhoff stress tensor}, \quad  &  \bm{P} := \pderiv{\Psi}{\bm{F}}, \label{eqn-PK1} \\
\text{Cauchy stress tensor}, \quad  &  \bm{\sigma} := \frac{1}{J} \, \bm{P} \, \bm{F}^{\T}. \label{eqn-Cauchy}
\end{align}

\subsection{Equilibrium equations}
In the absence of dynamic effects, the balance of linear momentum and the corresponding boundary conditions in the reference configuration are given as
\begin{align}
- \nabla_{\bm{X}} \cdot \widehat{\bm{P}} &= \bm{f}_{0}, && \text{in } \quad \mathcal{B}_{0}, \label{eqn-momentum-ref} \\
\bm{u}              &= \bm{u}_{\text{D}}, && \text{on } \quad \partial \mathcal{B}_{0}^{\text{D}}, \\
\widehat{\bm{P}} \cdot \bm{n}_{0} &= \bm{t}_{0}, && \text{on } \quad \partial \mathcal{B}_{0}^{\text{N}} \label{eqn-neumannbc}
\end{align}
where, $\bm{f}_{0}$ is the body force per unit undeformed volume, $\bm{n}_{0}$ is the unit outward normal on the boundary $\partial \mathcal{B}_{0}$, $\bm{u}_{\text{D}}$ is the prescribed value of displacement on the Dirichlet boundary $\partial \mathcal{B}_{0}^{\text{D}}$ and $\bm{t}_{0}$ is the specified traction force per unit undeformed area on the Neumann boundary $\partial \mathcal{B}_{0}^{\text{N}}$. Here, $\partial \mathcal{B}_{0}^{\text{D}}$ and $\partial \mathcal{B}_{0}^{\text{N}}$ are such that $\partial \mathcal{B}_{0}^{\text{D}} \bigcup \partial \mathcal{B}_{0}^{\text{N}} = \partial \mathcal{B}_{0}$ and $\partial \mathcal{B}_{0}^{\text{D}} \bigcap \partial \mathcal{B}_{0}^{\text{N}}=\emptyset$.

In Eqns (\ref{eqn-momentum-ref}) and (\ref{eqn-neumannbc}), $\widehat{\bm{P}}$ is the effective first Piola-Kirchhoff stress for the mixed formulation, and is defined as
\begin{align}
\widehat{\bm{P}} := \pderiv{\Psi^{\text{dev}}(\overline{\bm{C}})}{\bm{F}} + p \, J \, \bm{F}^{-\T} = \overline{\bm{P}} + p \, J \, \bm{F}^{-\T},
\end{align}
where, $\overline{\bm{P}}$ is deviatoric component of the first Piola Kirchhoff stress which is computed solely from the deviatoric part of the strain energy function, and $p$ is an independent approximation for the hydrostatic pressure. For the compressible and the nearly incompressible cases, the pressure is related to the volumetric energy function via the relation
\begin{align} \label{eqn-constraint-nic}
p = \pderiv{\Psi^{\text{vol}} (J) }{J}.
\end{align}


\section{Perturbed Lagrangian formulation} \label{section-pertLagr}
For the truly incompressible materials, the volumetric energy function vanishes. Therefore, the incompressibility constraint given by equation (\ref{eqn-constraint-fic}) needs to be imposed either strongly by using the Lagrange multiplier approach or weakly by using the penalty approach. The penalty approach results in a formulation that consists of displacement degrees of freedom (DOFs) only; therefore, such an approach suffers from the well-known issues of volumetric locking and spurious pressure fields observed with the pure displacement formulation. Hence, for the truly incompressible case, the Lagrange multiplier approach is the only viable choice. In this approach, the Lagrange multiplier becomes an independent variable for the pressure field.

With $p$ as the Lagrangian multiplier, the potential energy functional  is given as
\begin{align}  \label{eqn-mixedform-func-lagmul}
\Pi_{\text{LagMul}}(\bm{u},p) &= \IBi \, \left[ \Psi^{\text{dev}}(\overline{\bm{C}}) + p \, [J-1] \right] \, \dV - \Pi_{\mathrm{ext}},
\end{align}
where, $\Pi_{\mathrm{ext}}$ is the potential energy corresponding to the applied body and surface loads. It is given as
\begin{align}
\Pi_{\mathrm{ext}} = \IBi \bm{u}^{\T} \, \bm{f}_{0} \, \dV +  \int_{\partial \mathcal{B}^{\text{N}}_{0}} \bm{u}^{\T} \, \bm{t}_{0} \, \dA.
\end{align}

\begin{figure*}
\centering
\includegraphics[clip, scale=0.8]{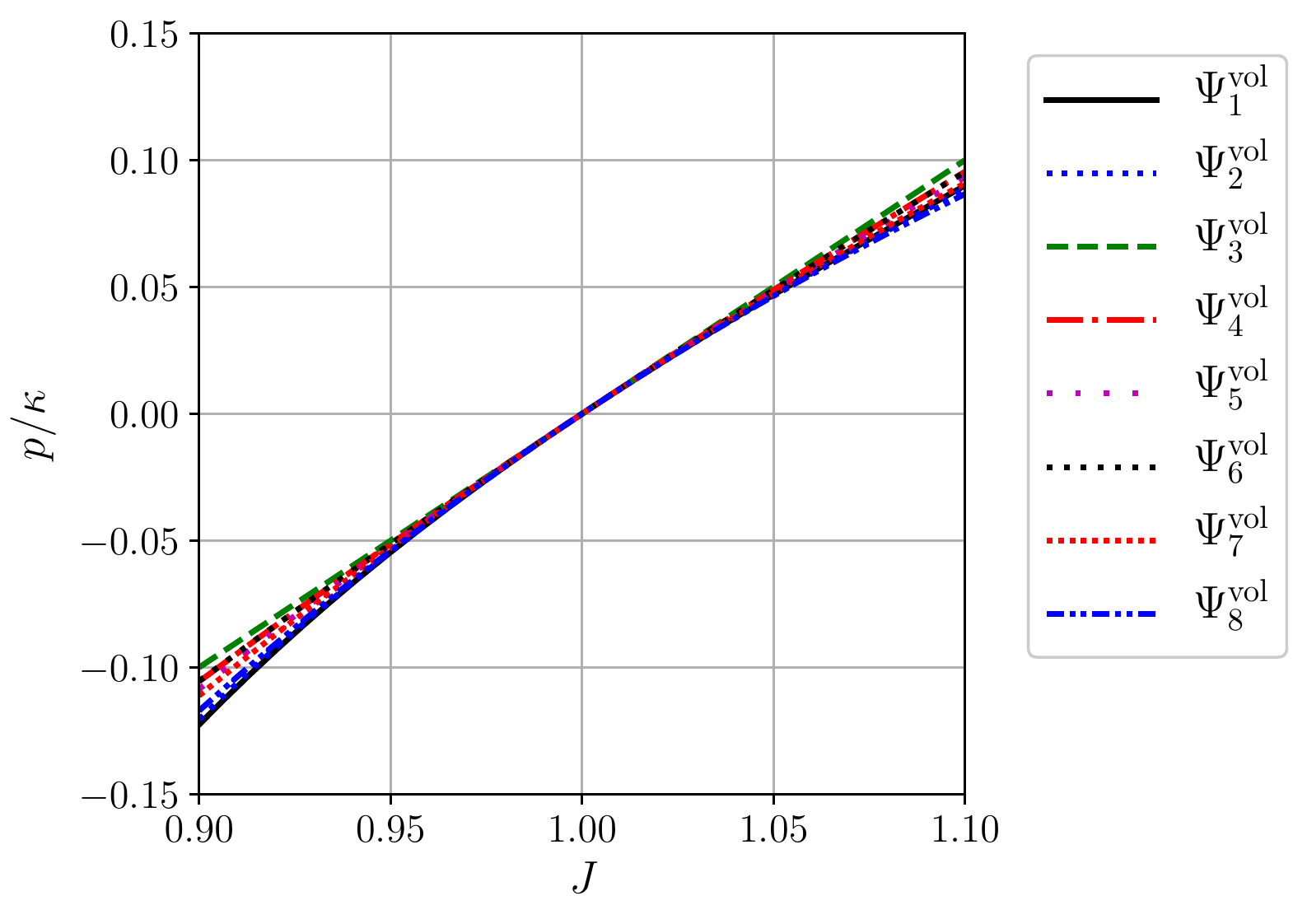}
\caption{Variation of $p$ against $J$ for different volumetric energy functions (see Table. \ref{table-volmfunctions}) in the incompressible limit, i.e., $J\approx1.0.$}
\label{fig-volfuncs-plot1}
\end{figure*}

To overcome the numerical difficulties in solving the saddle-point problems resulting from the Lagrangian multiplier approach, the term corresponding to the constraint equation is modified slightly, and such an approach is known as the perturbed Lagrangian approach \cite{ChenCMAME1997, book-fem-BonetWood, KadapaCMAME2016elast, KadapaIJNME2019mixed, KadapaPhDThesis, Bercovier, Wriggers, Zienkiewicz, Simo, Tur}. The energy functional for the perturbed Lagrangian approach is given by
\begin{align}  \label{eqn-mixedform-func-perlag}
\Pi_{\text{PerLag}}(\bm{u},p) &= \IBi \, \left[ \Psi^{\text{dev}}(\overline{\bm{C}}) + p \left [J-1-\frac{p}{2\kappa}\right] \right] \dV - \Pi_{\mathrm{ext}}
\end{align}
where, $\kappa$ is the bulk modulus. Note that such a modification introduces slight compressibility into the model, thus, facilitating a numerical framework for modelling nearly incompressible cases, i.e., for $\nu \approx 0.5$ or equivalently, $J \approx 1.0$. For such cases, the difference in the values of $p$ computed from different volumetric energy functions, as shown in Fig. \ref{fig-volfuncs-plot1}, can be considered to be negligible. Note also that the energy functional for the perturbed Lagrangian approach (\ref{eqn-mixedform-func-perlag}) reduces to that of the Lagrange multiplier approach (\ref{eqn-mixedform-func-lagmul}) for the truly incompressible case, i.e., for $\nu=0.5$ for which $\kappa=\infty$.

\begin{figure*}
\centering
\includegraphics[clip, scale=0.8]{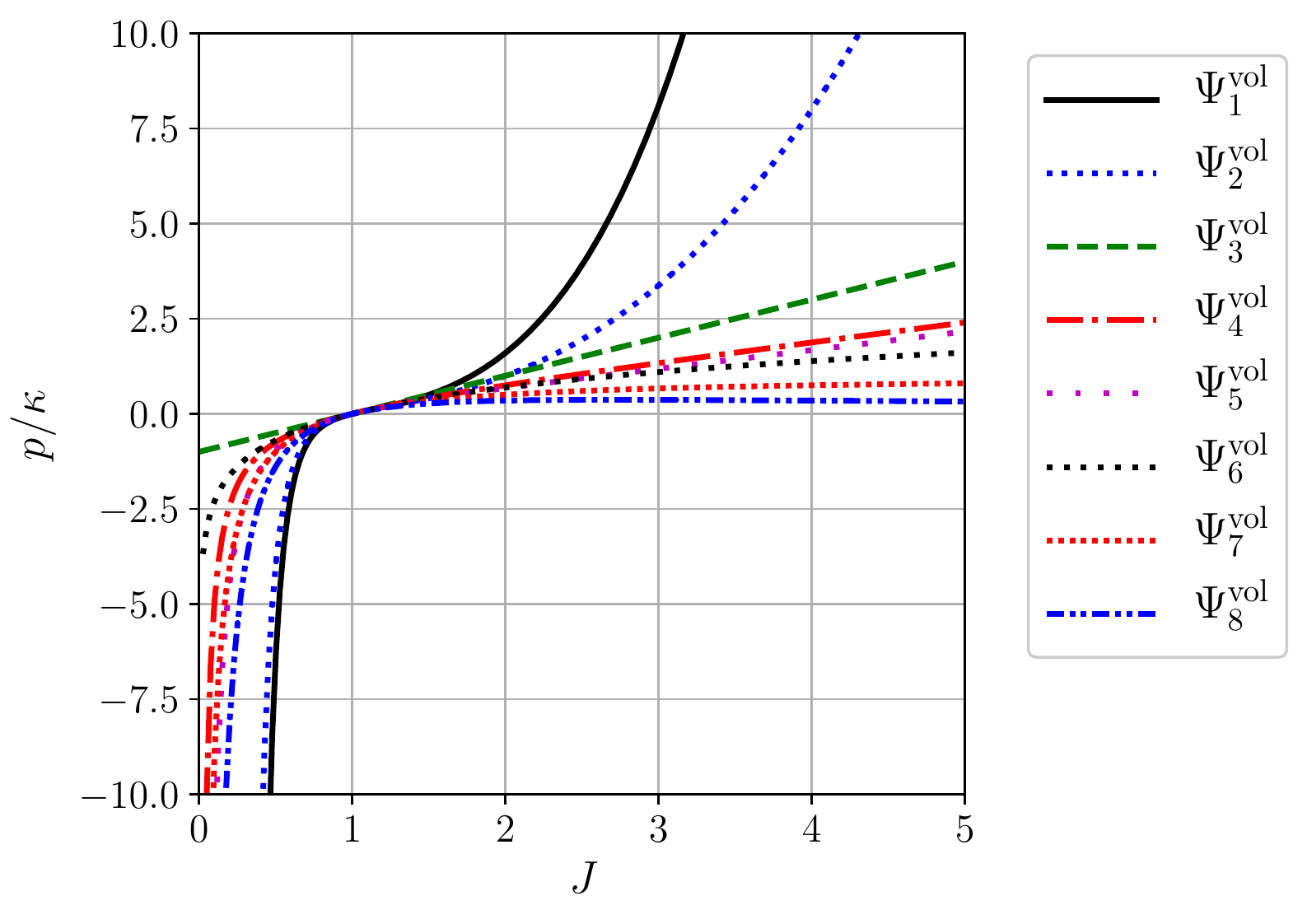}
\caption{Variation of $p$ against $J$ for different volumetric energy functions (see Table. \ref{table-volmfunctions}) for $0 < J \leq 5.0.$}
\label{fig-volfuncs-plot2}
\end{figure*}

However, despite its suitability for both nearly and truly incompressible materials modelling, the perturbed Lagrangian approach fails to represent the physical behaviour of the material in the compressible regime. This is due to the lack of accountability for the relation between the pressure and the volumetric energy function, i.e., $p = \pderiv{\PsiVol}{J}$, in the perturbed Lagrangian approach (\ref{eqn-mixedform-func-perlag}). Because of the significant differences in the pressure values computed using different volumetric energy functions in the compressible regime, as illustrated in Fig. \ref{fig-volfuncs-plot2} for $0.0 < J \leq 5.0$, the perturbed Lagrangian approach produces incorrect results for problems simulated with compressible material models that experience severe stretches. \mycolor{To overcome this issue}, we propose a new generalised mixed displacement-pressure formulation that is applicable for compressible, nearly incompressible as well as truly incompressible cases.
\renewcommand{\arraystretch}{1.5}
\begin{table*}
\centering
\begin{tabular}{|p{5cm}|p{3cm}|p{4cm}|p{2cm}|}
\hline
$\PsiVol$    &   $p=\pderiv{\PsiVol}{J}$    &   $\pderiv{^2\PsiVol}{J^2}$  & Reference \\
\hline
$\PsiVol_1 = \frac{\kappa}{50} \, [ J^5 + \frac{1}{J^5} - 2]$  & $\frac{\kappa}{10} \, [ J^4 - \frac{1}{J^6}]$    &  $\frac{\kappa}{5} \, [2 \, J^3+\frac{3}{J^7}]$   & \cite{HartmannIJSS2003} \\
$\PsiVol_2 = \frac{\kappa}{32} \, [ J^2 - \frac{1}{J^2}]^2$  & $\frac{\kappa}{8} \, [ J^2 - \frac{1}{J^2}] \, [J+\frac{1}{J^3}]$    &  $\frac{\kappa}{8} \, [3 \, J^2+\frac{5}{J^6}]$  & \cite{Ansys2000} \\
$\PsiVol_3 = \frac{\kappa}{2} \, [J-1]^2$                    &  $\kappa \, [J-1]$   &  $\kappa$  & \\
$\PsiVol_4 = \frac{\kappa}{4} \, [ J^2 - 1 - 2 \, \ln(J)]$   &  $\frac{\kappa}{2} \, [J-\frac{1}{J}]$   &  $\frac{\kappa}{2} \, [1+\frac{1}{J^2}]$    & \cite{SimoCMAME1991} \\
$\PsiVol_5 = \frac{\kappa}{4} \, [ [J-1]^2 + \ln(J)^2]$      &  $\frac{\kappa}{2} \, [J-1+\frac{\ln(J)}{J}]$   &  $\frac{\kappa}{2 \, J^2} \, [J^2+1-\ln(J)]$    &  \cite{SimoCMAME1982}  \\
$\PsiVol_6 = \kappa \, [ J \, \ln(J) - J + 1]$               & $\kappa \, \ln(J)$    &  $\frac{\kappa}{J}$   & \cite{LiuEC1994}  \\
$\PsiVol_7 = \kappa \, [ J - \ln(J) - 1]$                    & $\kappa \, [1-\frac{1}{J}]$    &  $\frac{\kappa}{J^2}$   & \cite{MeiheIJNME1994} \\
$\PsiVol_8 = \frac{\kappa}{2} \, [\ln(J)]^2$                 &  $\kappa \, \frac{\ln(J)}{J}$   &  $\frac{\kappa}{J^2} \, [1-\ln(J)]$   &  \cite{SimoCMAME1985}  \\
\hline
\end{tabular}
\caption{Volumetric energy functions, and their first and second derivatives with respect to $J$.}
\label{table-volmfunctions}
\end{table*}
\renewcommand{\arraystretch}{1.0}

\section{Weak Galerkin formulation} \label{section-Galerkin}
The mixed displacement-pressure Galerkin formulation for the finite strain hyperelasticity in the original configuration can be stated as:
find the displacement $\bm{u} \in \mathcal{S}_{\bm{u}}$ and pressure $p \in \mathcal{S}_{p}$ such that for all the test functions $\bm{w} \in \mathcal{T}_{\bm{u}}$ and $q \in \mathcal{T}_{p}$,
\begin{align} \label{eqn-galerkinform}
\IBi \mathop{}\nabla_{\!\bm{X}} \bm{w} \colon \widehat{\bm{P}} \, \dV 
&+ \IBi q \, C \, \dV = \IBi \bm{w} \cdot \bm{f}_{0} \, \dV
+ \int_{\partial \mathcal{B}^{\text{N}}_{0}} \bm{w} \cdot \bm{t}_{0} \, \dA
\end{align}
where, $\mathcal{S}_{\bm{u}}$ and $\mathcal{S}_{p}$ are the function spaces for the displacement and pressure, and $\mathcal{T}_{\bm{u}}$ and $\mathcal{T}_{p}$ are the corresponding test function spaces. $C$ is the generic constraint that is chosen depending upon whether the material is truly or nearly incompressible. $C$ is defined as
\begin{align} \label{eqn-C-defs}
C &:= J - 1,   && \text{for  } \nu=0.5 \\
C &:= \frac{1}{\kappa} \left[ \pderiv{\Psi^{\text{vol}}}{J} - p \right],   && \text{for  } \nu<0.5.
\end{align}

For the finite element analysis, the displacement and pressure fields, $\bm{u}$ and $p$, and their test functions, $\bm{w}$ and $q$, are approximated as
\begin{align}  \label{approximations-mixedform}
 \bm{w} = \mathbf{N}_{\bm{u}} \, \mathbf{w}, \quad
      q = \mathbf{N}_p \, \mathbf{q}, \nonumber \\
 \bm{u} = \mathbf{N}_{\bm{u}} \, \mathbf{u}, \quad
      p = \mathbf{N}_p \, \mathbf{p},
\end{align}
where, $\mathbf{u}$ and $\mathbf{w}$ are the vectors of displacement degrees of freedom (DOFs), and $\mathbf{p}$ and $\mathbf{q}$ are the vectors of the pressure degrees of freedom; $\mathbf{N}_u$ and $\mathbf{N}_p$ are the matrices of basis functions, respectively, for the displacement and pressure fields. Using the above definitions, the gradient and the divergence of displacement ($\bm{u}$) and its test function ($\bm{w}$) can be written as,
\begin{align}  \label{eqn-approx-grad-u}
\nabla_{\bm{x}} \bm{u}
&= u_{i,j} \nonumber \\
&=
\begin{bmatrix}
u_{x,x}  &  u_{y,x}  &  u_{z,x}  &
u_{x,y}  &  u_{y,y}  &  u_{z,y}  &
u_{x,z}  &  u_{y,z}  &  u_{z,z}
\end{bmatrix}^{\T} \nonumber \\
&=
\begin{bmatrix}
\pderiv{u_x}{x}  &  \pderiv{u_y}{x}  &  \pderiv{u_z}{x}  &
\pderiv{u_x}{y}  &  \pderiv{u_y}{y}  &  \pderiv{u_z}{y}  &
\pderiv{u_x}{z}  &  \pderiv{u_y}{z}  &  \pderiv{u_z}{z}
\end{bmatrix}^{\T} \nonumber \\
&=
\mathbf{G}_{\bm{u}} \, \mathbf{u}
\end{align}
\begin{equation}  \label{eqn-approx-div-u}
\nabla \cdot \bm{u} = u_{i,i}
=
\pderiv{u_x}{x} + \pderiv{u_y}{y} + \pderiv{u_z}{z}
= 
\mathbf{D}_{\bm{u}} \, \mathbf{u}
\end{equation}
\begin{align}  \label{eqn-approx-grad-w}
\nabla_{\bm{x}} \bm{w} = w_{i,j} = \mathbf{G}_{\bm{u}} \, \mathbf{w}
\end{align}
\begin{align}  \label{eqn-approx-div-w}
\nabla_{\bm{x}} \cdot \bm{w} = w_{i,i} = \mathbf{D}_{\bm{u}} \, \mathbf{w},
\end{align}
where, $u_x$, $u_y$ and $u_z$ are the components of displacement, respectively, in X, Y, and Z directions, and $\mathbf{G}_{\bm{u}}$ and $\mathbf{D}_{\bm{u}}$ are the gradient-displacement and the divergence-displacement matrices, respectively. $\mathbf{G}_{\bm{u}}$ and $\mathbf{D}_{\bm{u}}$ for a single basis function ($N_{\bm{u}}$), are given as
\renewcommand{\arraystretch}{1.2}
\begin{equation}
\mathbf{G}_{\bm{u}} =
\begin{bmatrix}
\pderiv{N_{\bm{u}}}{x}  &          0        &       0        &
\pderiv{N_{\bm{u}}}{y}  &          0        &       0        &
\pderiv{N_{\bm{u}}}{z}  &          0        &       0        \\
       0       &   \pderiv{N_{\bm{u}}}{x}   &       0        &
       0       &   \pderiv{N_{\bm{u}}}{y}   &       0        &
       0       &   \pderiv{N_{\bm{u}}}{z}   &       0        \\
       0       &          0        &  \pderiv{N_{\bm{u}}}{x} &
       0       &          0        &  \pderiv{N_{\bm{u}}}{y} &
       0       &          0        &  \pderiv{N_{\bm{u}}}{z} 
\end{bmatrix}^{\T}
\end{equation}
\renewcommand{\arraystretch}{1.0}
\begin{equation}
\mathbf{D}_{\bm{u}}
=
\begin{bmatrix}
\pderiv{N_{\bm{u}}}{x}  &  \pderiv{N_{\bm{u}}}{y}  &  \pderiv{N_{\bm{u}}}{z}
\end{bmatrix}.
\end{equation}
\renewcommand{\arraystretch}{1.0}

Analogous to the gradient of displacement in Eq. (\ref{eqn-approx-grad-u}), the Cauchy stress tensor is represented in the vector form as
\begin{align}
\bm{\sigma} = \sigma_{ij} &=
\begin{bmatrix}
\sigma_{xx}  &  \sigma_{yx}  &  \sigma_{zx}  &
\sigma_{xy}  &  \sigma_{yy}  &  \sigma_{zy}  &
\sigma_{xz}  &  \sigma_{yz}  &  \sigma_{zz}
\end{bmatrix}^{\T} \nonumber \\
&=
\begin{bmatrix}
\sigma_{11}  &  \sigma_{21}  &  \sigma_{31}  &
\sigma_{12}  &  \sigma_{22}  &  \sigma_{32}  &
\sigma_{13}  &  \sigma_{23}  &  \sigma_{33}
\end{bmatrix}^{\T}.
\end{align}

\noindent
\mycolor{\textbf{Remark I:}} The formulation can also be written in terms of the symmetric part of the strain and stress measures by following the Voigt notation, see \cite{KadapaPhDThesis,KadapaIJNME2019bbar,KadapaIJNME2019mixed,KadapaCMAME2016elast,book-fem-ZienkiewiczVol2}. Such an approach, however, leads to complications in transforming the fourth- and third-order tensors resulting from the electrical, magnetic and coupled energy functions in electro- and magnetomechanics problems. For the ease of computer implementation of the finite element discretization, we present the two-field formulations in terms of full strain and stress tensors as vectors. The proposed approach avoids ambiguities in dealing with the derivatives of symmetric and unsymmetric second-order tensors.

Substituting Eqns (\ref{approximations-mixedform})-(\ref{eqn-approx-div-w}) in Eqn (\ref{eqn-galerkinform}), and invoking the vanishing condition for all the test functions, we obtain the following nonlinear equations
\begin{align} 
\IBt \, \mathbf{G}_{\bm{u}}^{\T} \, \widehat{\bm{\sigma}} \, \dv 
&= \IBi \mathbf{N}_{\bm{u}}^{\T} \, \bm{f}_{0} \, \dV + \int_{\partial \mathcal{B}^{\text{N}}_{0}} \mathbf{N}_{\bm{u}}^{\T} \, \bm{t}_{0} \, \dA,  \label{eqn-dispform-residual-Ru} \\
\IBi \mathbf{N}_{p}^{\T} \, C &= \bm{0}, \label{eqn-dispform-residual-Rp}
\end{align}
where, $\widehat{\bm{\sigma}}$ is the effective Cauchy stress, and it is defined as
\begin{align} \label{eqn-cauchy-conventionl}
\widehat{\bm{\sigma}} := \frac{1}{J} \, \widehat{\bm{P}} \, \bm{F}^{\T}
= \frac{1}{J} \, \overline{\bm{P}} \, \bm{F}^{\T} + p \, \bm{I}.
\end{align}

Assuming the displacement and pressure DOFs at the $k$th iteration of the Newton-Raphson scheme are $\mathbf{u}^{k}$ and $\mathbf{p}^{k}$, respectively, the corresponding DOFs at the $(k+1)$th iteration become
\begin{align}
\mathbf{u}^{k+1} &= \mathbf{u}^{k} + \Delta \mathbf{u}, \\
\mathbf{p}^{k+1} &= \mathbf{p}^{k} + \Delta \mathbf{p},
\end{align}
where $\Delta \mathbf{u}$ and $\Delta \mathbf{p}$ are the incremental displacement and pressure DOFs, respectively. The corresponding increments for the continuum variables are denoted as $\Delta \bm{u}$ and $\Delta p$.

For the application of Newton-Raphson scheme, the linearisation of the constraint variable $C$ yields,
\begin{align} \label{eqn-C-linear1}
C(\bm{u}^{k+1}, p^{k+1})
&\approx
C(\bm{u}^{k}, p^{k}) + \left.\pderiv{C}{J} \right\vert_{(\bm{u}^k,p^k)} \, J \, [\nabla_{\bm{x}} \cdot \Delta \bm{u}] + \left.\pderiv{C}{p} \right\vert_{(\bm{u}^k,p^k)} \, \Delta p = 0,
\end{align}
which, after making use of the finite element approximations (\ref{approximations-mixedform}) and the divergence-displacement matrix in (\ref{eqn-approx-div-u}), can be rewritten as
\begin{align} \label{eqn-C-linear2}
\left.\pderiv{C}{J} \right\vert_{(\bm{u}^k,p^k)} \, J \, \mathbf{D}_{\bm{u}} \, \Delta \mathbf{u} - \vartheta \, \mathbf{N}_{p} \, \Delta \mathbf{p} = - \, C(\bm{u}^{k}, p^{k}),
\end{align}
where, the variable $\vartheta$ is defined as
\begin{align}
\vartheta := - \left.\pderiv{C}{p} \right\vert_{(\bm{u}^k,p^k)}.
\end{align}

Following equation (\ref{eqn-C-defs}), we obtain, \\
\noindent
for $\nu=0.5$:
\begin{align} \label{eqn-C-derivatives1}
\left.\pderiv{C}{J} \right\vert_{(\bm{u}^k,p^k)} &= 1, \quad  \vartheta = - \, \left.\pderiv{C}{p} \right\vert_{(\bm{u}^k,p^k)} = 0.
\end{align}
and for $\nu<0.5$:
\begin{align} \label{eqn-C-derivatives2}
\left.\pderiv{C}{J} \right\vert_{(\bm{u}^k,p^k)} &= \frac{1}{\kappa} \, \left.\pderiv{^2\Psi^{\text{vol}}}{J^2}\right\vert_{\bm{u}^k}, \quad  \vartheta = - \, \left.\pderiv{C}{p} \right\vert_{(\bm{u}^k,p^k)} = \frac{1}{\kappa}.
\end{align}

Now, the resulting coupled matrix system for the mixed formulation can be written as
\begin{equation}  \label{matrixsystem-galerkin}
 \begin{bmatrix}
  \mathbf{K}_{\bm{u}\bm{u}}  &  \mathbf{K}_{\bm{u}p} \\
  \mathbf{K}_{p\bm{u}}  &  \mathbf{K}_{pp}
 \end{bmatrix}
 \begin{Bmatrix}  \Delta \mathbf{u} \\ \Delta \mathbf{p} \end{Bmatrix}
=
- \begin{Bmatrix}  \mathbf{R}_{\bm{u}} \\ \mathbf{R}_{p} \end{Bmatrix}
\end{equation}
where
\begin{align}
\mathbf{K}_{\bm{u}\bm{u}} &= \IBt \mathbf{G}^{\T}_{\bm{u}} \, \bm{\mathsf{e}} \, \mathbf{G}_{\bm{u}} \, \dv,  \label{matrices-kuu}  \\
\mathbf{K}_{\bm{u}p}  &= \IBt \mathbf{D}^{\T}_{\bm{u}} \, \mathbf{N}_{p} \, \dv, \\
\mathbf{K}_{p\bm{u}}  &= \IBt \, \left.\pderiv{C}{J} \right\vert_{\bm{u}^k} \, \mathbf{N}_{p}^\mathrm{T} \, \mathbf{D}_{\bm{u}} \, \dv,  \label{matrices-kpu} \\
\mathbf{K}_{pp}  &= - \, \IBi \vartheta \, \mathbf{N}_{p}^{\T} \mathbf{N}_{p} \, \dV, \\
\mathbf{R}_{\bm{u}} &= \IBt \mathbf{B}^{\mathrm{T}} \, \widehat{\bm{\sigma}}(\bm{u}^k, p^k) \, \dv - \mathbf{F}^{\text{ext}}, \\
\mathbf{R}_p     &= \IBi \mathbf{N}_p^\mathrm{T} \, C(\bm{u}^{k}, p^{k}) \, \dV,
\end{align}
and 
\begin{align} 
\mathbf{F}^{\text{ext}} = \IBi \mathbf{N}_{\bm{u}}^{\T} \, \bm{f}_{0} \, \dV + \int_{\partial \mathcal{B}^{\text{N}}_{0}} \mathbf{N}_{\bm{u}}^{\T} \, \bm{t}_{0} \, \dA.
\end{align}
In Eqn (\ref{matrices-kuu}), $\bm{\mathsf{e}}$ is the elasticity tensor of order four, and in the index notation, it is given as,
\begin{align}
\mathsf{e}_{ijkl} = \frac{1}{J} F_{jJ} \, \pderiv{\overline{P}_{iJ}}{F_{kL}} \, F_{lL} + p \, \left[ \delta_{ij} \, \delta_{kl} - \delta_{il} \, \delta_{jk} \right].
\end{align}

For the truly incompressible case and for $\PsiVol=\PsiVol_3$, the coupled matrix system (\ref{matrixsystem-galerkin}) is symmetric; for the other cases, it is unsymmetric. Thus, requiring a solver for unsymmetric sparse matrix systems which is expensive when compared to that of symmetric sparse matrix systems, see Duff et al. \cite{book-Matrices-Duff}, Gould et al. \cite{GouldACMTOMS2007}, Dongarra et al. \cite{book-linalgebra-Dongarra}, Pissanetzky \cite{book-Matrices-Pissanetzky}, Saad \cite{book-Matrices-Saad}, Barrett et al. \cite{book-Matrices-Barrett}, Nachtigal et al. \cite{NachtigalSIAMJMA1992} and Watkins \cite{book-Matrices-Watkins}, and references therein. This is another motivation behind the novel mixed formulation discussed in the following section.

\section{Proposed mixed formulation} \label{section-proposedform}
The definition of energy functional similar to the ones for the Lagrangian multiplier (\ref{eqn-mixedform-func-lagmul}) and the perturbed Lagrangian (\ref{eqn-mixedform-func-perlag}) approaches is not straightforward because of the nonlinear nature of the relation (\ref{eqn-constraint-nic}) between the pressure and the volumetric energy function. Although we can devise the following energy functional for the mixed formulation,
\begin{align} \label{eqn-mixedform-func1}
\Pi_{\text{Option1}}(\bm{u},p) &= \IBi \, \left[ \Psi^{\text{dev}}(\overline{\bm{C}}) + p \, \frac{1}{\kappa} \, \left[ \pderiv{\PsiVol}{J} - \frac{1}{2} \, p \right] \right] \, \dV  - \Pi_{\mathrm{ext}},
\end{align}
such a functional results in an unconventional form of the effective Cauchy stress as 
\begin{align}
\widehat{\bm{\sigma}} = \bm{\sigma} + \frac{1}{\kappa} \, \pderiv{^2\PsiVol}{J^2} \, p \bm{I}.
\end{align}

Therefore, for the variational approach, the energy functional that yields the conventional form of the Cauchy stress (\ref{eqn-cauchy-conventionl}) is sought. The energy functional for the proposed mixed formulation should be of the form
\begin{align} \label{eqn-mixedform-func-variational-1}
\Pi_{\text{Proposed}}(\bm{u},p) &= \IBi \, \left[ \Psi^{\text{dev}}(\overline{\bm{C}}) + \Psi_p \right] \, \dV - \Pi_{\mathrm{ext}},
\end{align}
with $\Psi_{p}$ being the corresponding energy function that takes the relation (\ref{eqn-constraint-nic}) into account. We obtain $\Psi_{p}$ by linearising $\pderiv{\PsiVol}{J}$. With the linearisation of $\pderiv{\PsiVol}{J}$ at the ($i+1$)th load step as
\begin{align} \label{eqn-Psi-linear1}
\pderiv{\PsiVol}{J}(J^{i+1}) \approx \pderiv{\PsiVol}{J}(J^{i}) + \pderiv{^2\PsiVol}{J^2} \biggr\rvert_{J^{i}} \, [J^{i+1} - J^{i}],
\end{align}
where $J^{\bullet}=\det{[\bm{F}(\bm{u}^{\bullet})]}$. The linearised form of $\left[ \pderiv{\PsiVol}{J}(J^{i+1})-p \right]$ can be written as
\begin{align} \label{eqn-Psi-linear2}
J^{i+1} - J^{i} + \frac{\pderiv{\PsiVol}{J}\biggr\rvert_{J^{i}}}{\pderiv{^2\PsiVol}{J^2} \biggr\rvert_{J^{i}}}  - \frac{1}{\pderiv{^2\PsiVol}{J^2} \biggr\rvert_{J^{i}}} \, p
=
J^{i+1} - \widehat{J} - \widehat{\vartheta} \, p,
\end{align}
where
\begin{align} \label{eqn-Psi-linear3}
\widehat{J} = J^{i} - \frac{\pderiv{\PsiVol}{J}\biggr\rvert_{J^{i}}}{\pderiv{^2\PsiVol}{J^2} \biggr\rvert_{J^{i}}}; \qquad
\widehat{\vartheta} = \frac{1}{\pderiv{^2\PsiVol}{J^2} \biggr\rvert_{J^{i}}}.
\end{align}
Since $J^{i}$ and the corresponding quantities are evaluated from the displacement from the previously converged load step, $\widehat{J}$ and $\widehat{\vartheta}$ are constants for the current load step. Therefore, there is no need for their linearisations when applying the Newton-Raphson scheme.

Now, the first variation and the associated energy functional for the relation (\ref{eqn-constraint-nic}) can be written as
\begin{align}
\delta \Psi_{p} &= \int_{\Omega} \delta p \, \left[ J - \widehat{J} - \widehat{\vartheta} \, p \right] \, d\Omega, \\
\Psi_{p} &= \int_{\Omega} p \, \left[ J - \widehat{J} - \frac{\widehat{\vartheta} \, p}{2} \right] \, d\Omega.
\end{align}
Note that the mixed formulation for the truly incompressible case can be recovered from the above equations by simply setting $\widehat{J}=1$ and $\widehat{\vartheta}=0$. Moreover, by setting $\widehat{J}=1$ and $\widehat{\vartheta}=\frac{1}{\kappa}$, the energy functional for the perturbed Lagrangian formulation (\ref{eqn-mixedform-func-perlag}) can also be recovered. Thus, the proposed formulation not only takes into account the relation between the pressure and the volumetric energy function for the compressible materials but also is applicable for the truly incompressible case.

The total energy functional for the proposed formulation can now be written as
\begin{align} \label{eqn-mixedform-func-variational-2}
\Pi_{\text{Proposed}}(\bm{u},p) &= \IBi \, \left[ \PsiDev(\overline{\bm{C}}) + p \, \left[ J - \widehat{J} - \frac{\widehat{\vartheta} \, p}{2} \right] \right] \, \dV - \Pi_{\mathrm{ext}}.
\end{align}

The first ($\delta$) and the second ($d$) variations for the  energy functional become
\begin{align}
\delta \Pi_{\text{Proposed}}
&= \IBi \, \left[ \delta F_{iJ} \, \overline{P}_{iJ} + \delta J \, p \right] \dV + \IBi \, \delta p \, \left[ J - \widehat{J} - \widehat{\vartheta} \, p \right] \dV - \delta \, \Pi_{\mathrm{ext}}
\end{align}
and
\begin{align}
d(\delta \Pi_{\text{Proposed}})
= \IBi \, \left[ \delta F_{iJ} \, \pderiv{\overline{P}_{iJ}}{F_{kL}} \, dF_{kL} + p \, d(\delta J)  \right] \dV + \IBi \, \left[ \delta p \, dJ + \delta J \, dp - \widehat{\vartheta} \, \delta p \, dp \right] \dV - d(\delta \, \Pi_{\mathrm{ext}}).
\end{align}

Using following identities,
\begin{align}
\delta J &= J \, \delta u_{i,j} \, \delta_{ij}, \\
\delta F_{iJ} &= \delta u_{i,J} = \delta u_{i,j} \, F_{jJ}, \\
\widehat{\sigma}_{ij} &= \frac{1}{J} \, \overline{P}_{iJ} \, F_{jJ} + p \, \delta_{ij},
\end{align}
and considering the approximations for the solution variables and their variations as
\begin{align}  \label{approximations-proposed}
 \bm{u} &= \mathbf{N}_{\bm{u}} \, \mathbf{u}, \qquad
       \hspace{3.9mm} p = \mathbf{N}_p \, \mathbf{p}, \\
 \delta \bm{u} &= \mathbf{N}_{\bm{u}} \, \delta \mathbf{u}, \qquad
 \delta    p = \mathbf{N}_p \, \delta \mathbf{p}, \\
 d \bm{u} &= \mathbf{N}_{\bm{u}} \, d \mathbf{u}, \qquad
 d    p = \mathbf{N}_p \, d \mathbf{p},
\end{align}
the coupled matrix system can be written as
\begin{equation}  \label{matrixsystem-proposed}
 \begin{bmatrix}
  \mathbf{K}_{\bm{u}\bm{u}}  &  \mathbf{K}_{\bm{u}p} \\
  \mathbf{K}_{p\bm{u}}  &  \mathbf{K}_{pp}
 \end{bmatrix}
 \begin{Bmatrix}  \Delta \mathbf{u} \\ \Delta \mathbf{p} \end{Bmatrix}
=
- \begin{Bmatrix}  \mathbf{R}_{\bm{u}} \\ \mathbf{R}_{p} \end{Bmatrix}
\end{equation}
where,
\begin{align}
\mathbf{K}_{\bm{u}\bm{u}} &= \IBt \mathbf{G}^{\T}_{\bm{u}} \, \bm{\mathsf{e}} \, \mathbf{G}_{\bm{u}} \, \dv, \\
\mathbf{K}_{\bm{u}p}  &= \IBt \mathbf{D}^{\T}_{\bm{u}} \, \mathbf{N}_{p} \, \dv = \mathbf{K}_{p\bm{u}}^{\T} \\
\mathbf{K}_{pp}  &= - \, \IBi \widehat{\vartheta} \, \mathbf{N}_{p}^{\T} \mathbf{N}_{p} \, \dV, \\
\mathbf{R}_{\bm{u}} &= \IBt \mathbf{B}^{\mathrm{T}} \, \widehat{\bm{\sigma}}(\bm{u}^k, p^k) \, \dv - \mathbf{F}^{\text{ext}}, \\
\mathbf{R}_p     &= \IBi \mathbf{N}_p^\mathrm{T} \, \left[ J - \widehat{J} - \widehat{\vartheta} \, p \right] \, \dV. \label{eqn-Rp-proposed}
\end{align}

It is evident from the above equations that the coupled matrix system (\ref{matrixsystem-proposed}) of the proposed two-field formulation is always symmetric irrespective of the volumetric energy function, unlike the coupled system (\ref{matrixsystem-galerkin}) of the mixed Galerkin approach which is symmetric only for $\PsiVol=\PsiVol_{3}$.\\

\noindent
\mycolor{
\textbf{Remark II:} A mixed displacement-pressure formulation that yields symmetric matrix system irrespective of the volumetric energy function can also be derived by using complementary potentials to volumetric energy functions, as discussed in Appendix A. However, such an approach requires {\it apriori} knowledge of complementary functions, which are uncommon in the constitutive modelling for hyperelastic materials. Alternative techniques based on numerical methods for the computation of complementary potentials and their derivatives pose additional difficulties for the cases with multiple roots, see Appendix A for a detailed discussion. The disadvantages of such a formulation based on complementary energy functions are also motivating factors for the proposed formulation.
}\\

\noindent
\mycolor{\textbf{Remark III:}} Note that the Lagrange multiplier ($\delta \lambda$) in the displacement-pressure (hybrid) formulation in ABAQUS \cite{abaqus-manual} is not the same as the independent approximation for the hydrostatic pressure and it is chosen intrusively to get the symmetric expression for the rate of virtual work (equation 3.2.3-11 in ABAQUS's theory manual \cite{abaqus-manual}). This makes ABAQUS's displacement-pressure formulation much more difficult to implement than the proposed formulation. Moreover, ABAQUS's formulation ignores changes in  the volume for the evaluation of the second variation ($d\delta \lambda$, in equation 3.2.3-13 in ABAQUS's theory manual) which has the potential to deteriorate rate of convergence of Newton-Raphson iterations, while the proposed formulation poses no such issues since it uses consistent linearisation. Furthermore, ABAQUS uses an artificial numerical constant, $\rho$ (refer to equation 3.2.3-4 in ABAQUS's theory manual), to overcome solver difficulties while the proposed formulation is free from such ad-hoc parameters.

\section{Comparison with the three-field formulation} \label{section-3field}
\mycolor{In this section, a comparison between the proposed two-field displacement-pressure mixed formulation and the well-established three-field displacement-pressure-Jacobian formulation \cite{SimoCMAME1985,book-fem-ZienkiewiczVol2} is presented. For this purpose, we consider the widely-used Q1/P0 finite element discretisation, in which it is possible to condense the pressure DOFs out. Other discretisations, for example, P2/P0 \cite{book-fem-BrezziFortin} or the recently proposed BT2/BT0 \cite{KadapaIJNME2019mixed}, are equally applicable.}

\subsection{Proposed two-field formulation : A condensed form}
For the Q1/P0 element, the coupled matrix of the proposed two-field formulation (\ref{matrixsystem-proposed}) can be condensed as
\begin{equation}  \label{eqn-matrix-2field-condensed}
\widehat{\mathbf{K}}_{\bm{u}\bm{u}} \, \Delta \mathbf{u} =
- \widehat{\mathbf{R}}_{\bm{u}},
\end{equation}
where
\begin{align}
\widehat{\mathbf{K}}_{\bm{u}\bm{u}} &= \mathbf{K}_{\bm{u}\bm{u}} - \mathbf{K}_{\bm{u} p} \, \mathbf{K}_{pp}^{-1} \, \mathbf{K}_{p\bm{u}}, \\
\widehat{\mathbf{R}}_{\bm{u}} &= \mathbf{R}_{\bm{u}} - \mathbf{K}_{\bm{u} p} \, \mathbf{K}_{pp}^{-1} \, \mathbf{R}_{p}.
\end{align}
Using $\mathbf{N}_{p}=1$, $\mathbf{K}_{pp}$ and $\mathbf{R}_{p}$ simplify to
\begin{align}
\mathbf{K}_{pp} &= - \, \IBi \widehat{\vartheta} \, \dV \label{eqn-Kpp-proposed-simplified} \\
\mathbf{R}_{p}  &= \IBi \left[ J - \widehat{J} - \widehat{\vartheta} \, p \right] \, \dV. \label{eqn-Rp-proposed-simplified}
\end{align}
Since the pressure is assumed to be discontinuous, the effective stiffness matrix, $\widehat{\mathbf{K}}_{\bm{u}\bm{u}}$ and the effective residual $\widehat{\mathbf{R}}_{\bm{u}}$ can be computed element-wise before assembling into the global stiffness matrix.

\subsection{Three-field mixed formulation}
For the three-field mixed displacement-pressure-Jacobian formulation, only those quantities that are relevant to the discussion in the present work are presented here. For comprehensive details on the three-field formulation, we refer the reader to the works of Zienkiewicz and Taylor \cite{book-fem-ZienkiewiczVol2} or Kadapa \cite{KadapaPhDThesis}.

\myred{
Considering,
\begin{equation}
 \bar{J} = 1 + \theta
\end{equation}
where, $\theta$ represents the change in the volume, and by taking approximations for variables $\bm{u}$, $p$ and $\theta$ as,
\begin{equation}
 \label{approx5}
 \bm{u}  = \mathbf{N}_u \, \mathbf{u}; \quad 
     p   = \mathbf{N}_p \, \mathbf{p}; \quad 
 \theta  = \mathbf{N}_{\theta} \, \bm{\theta}
\end{equation}
the resulting coupled matrix system for the three-field formulation with the energy functional given by Eq. (\ref{eqn-Pi-1}) can be written as,
}
\begin{align}  \label{eqn-matrix-3field}
 \begin{bmatrix}
  \mathbf{K}_{\bm{u}\bm{u}}   &  \mathbf{K}_{\bm{u}\theta}  &  \mathbf{K}_{\bm{u}p}    \\
  \mathbf{K}_{\theta\bm{u}}   &  \mathbf{K}_{\theta\theta}  &  \mathbf{K}_{\theta p}   \\
  \mathbf{K}_{p\bm{u}}        &  \mathbf{K}_{p\theta}    &  \bm{0}      \\  
\end{bmatrix}
 \begin{Bmatrix}  \Delta \mathbf{u} \\ \Delta \bm{\theta} \\ \Delta \mathbf{p} \end{Bmatrix}
=
- \begin{Bmatrix}  \mathbf{R}_{\bm{u}} \\ \mathbf{R}_{\theta} \\ \mathbf{R}_{p} \end{Bmatrix}
\end{align}
where,
\begin{align}
\mathbf{K}_{\bm{u}\bm{u}}   &= \IBi \mathbf{B}^{\T} \, \bar{\mathbf{D}}_{11} \, \mathbf{B} \; \bar{J} \, \dV \; + \; \mathbf{K}_G \\
\mathbf{K}_{\bm{u}\theta}   &= \IBi \mathbf{B}^{\T} \, \bar{\mathbf{D}}_{12} \, \mathbf{N}_{\theta} \, \dV = \mathbf{K}_{\theta \bm{u}}^{\T} \\
\mathbf{K}_{\bm{u}p}        &= \IBi \mathbf{B}^{\T} \, \mathbf{m} \, \mathbf{N}_{p} \; J \, \dV = \mathbf{K}_{p \bm{u}}^{\T} \\
\mathbf{K}_{\theta\theta}   &= \IBi \mathbf{N}_{\theta}^{\T} \, \bar{\mathbf{D}}_{22} \, \mathbf{N}_{\theta} \, \frac{1}{\bar{J}} \, \dV \\
\mathbf{K}_{\theta p}       &= - \IBi \mathbf{N}_{\theta}^{\T} \, \mathbf{N}_{p} \, \dV = \mathbf{K}_{p\theta}^{\T} \\
\mathbf{R}_{\bm{u}}         &=  \IBi \mathbf{B}^{\T} \, \breve{\bm{\sigma}} \, \bar{J} \, \dV - \mathbf{F}^{\mathrm{ext}}   \\
\mathbf{R}_{\theta}         &=  \IBi \mathbf{N}_{\theta}^{\T} \, \left[ \bar{p} - p \right] \, \dV \\
\mathbf{R}_{p}              &=  \IBi \mathbf{N}_p^{\T} \, \left[ J - \bar{J} \right] \, \dV  
\end{align}
with
\begin{align}
\bar{\mathbf{D}}_{11}
&= \dev{\bar{D}} -
\frac{2}{3} \, \left[ \mathbf{m} \, \dev{\bar{\bm{\sigma}}}^{\T} + \dev{\bar{\bm{\sigma}}} \, \mathbf{m}^{\T} \right] + 2 \, \left[ \bar{p} - \breve{p} \right] \, \mathbf{I}_{0} -
 \left[ \frac{2}{3} \, \bar{p} - \breve{p} \right] \, \mathbf{m} \, \mathbf{m}^{\T} \label{eqn-3field-D11} \\
\bar{\mathbf{D}}_{12} &= \frac{1}{3} \, \dev{I} \, \bar{\mathbf{D}} \, \mathbf{m} + \frac{2}{3} \, \dev{\bar{\bm{\sigma}}} \label{eqn-3field-D12} \\
\bar{\mathbf{D}}_{22} &= \frac{1}{9} \, \mathbf{m}^{\T} \, \bar{\mathbf{D}} \, \mathbf{m} - \frac{1}{3} \, \bar{p} \\
\bm{\breve{\sigma}}   &= \bm{\bar{\sigma}} + \mathbf{m} \, \left[ \breve{p}-\bar{p} \right] \\
\breve{p} &= \frac{J}{\bar{J}} \, p \\
\mathbf{I}_{0} &= \mathrm{diag} \left[ 1 \quad 1 \quad 1 \quad \frac{1}{2} \quad \frac{1}{2} \quad \frac{1}{2} \right] \\
\mathbf{m} &= \left[ 1 \quad 1 \quad 1 \quad 0 \quad 0 \quad 0 \right]^{\T}. 
\end{align}

For the strain energy functions decomposed into deviatoric and volumetric parts, it can be shown analytically that
\begin{align} \label{eqn-zeroing}
\bar{\mathbf{D}}_{12} = \bm{0}; \quad \text{and} \quad
 \frac{\bar{\mathbf{D}}_{22}}{\bar{J}} = \pderiv{^2\PsiVol(\bar{J})}{\bar{J}^2}.
\end{align}
See Chapter 4 in Kadapa \cite{KadapaPhDThesis} for the comprehensive details on the derivation of equations (\ref{eqn-zeroing}). Using the above equations and invoking the constant discontinuous approximation for the Jacobian variable and pressure, we obtain the element-wise stiffness matrices as,
\begin{align} \label{eqn-kut-zero}
\mathbf{K}_{\bm{u}\theta}^{e} &= \mathbf{K}_{\bm{u}\theta}^{e^{\T}} = \bm{0},
\end{align}
\begin{align}
\mathbf{K}_{p\theta}^{e}      &= \mathbf{K}_{\theta p}^{e} = - V^{e}_{0},
\end{align}
\begin{align}
\mathbf{K}_{\theta\theta}^{e} &= \int_{\mathcal{B}_{0}^{e}} \pderiv{^2\PsiVol(\bar{J})}{\bar{J}^2} \, \dV
\end{align}
where, $\mathcal{B}_{0}^{e}$ and $V^{e}_{0}$ are the domain and the volume of the element, respectively, in the reference configuration.

Now, the condensed form of the coupled matrix system (\ref{eqn-matrix-3field}), after simplification using the relation (\ref{eqn-kut-zero}), can be written as
\begin{equation}  \label{eqn-matrix-3field-condensed}
\widetilde{\mathbf{K}}_{\bm{u}\bm{u}} \, \Delta \mathbf{u} =
- \widetilde{\mathbf{R}}_{\bm{u}}
\end{equation}
where
\begin{align}
\widetilde{\mathbf{K}}_{\bm{u}\bm{u}} &= \mathbf{K}_{\bm{u}\bm{u}} - \mathbf{K}_{\bm{u} p} \, \widetilde{\mathbf{K}}_{pp}^{-1} \, \mathbf{K}_{p\bm{u}} \\
\widetilde{\mathbf{R}}_{\bm{u}} &= \mathbf{R}_{\bm{u}} - \mathbf{K}_{u p} \, \widetilde{\mathbf{K}}_{pp}^{-1} \, \widetilde{\mathbf{R}}_{p},
\end{align}
in which
\begin{align}
\widetilde{\mathbf{K}}_{pp} &= - \, \mathbf{K}_{p\theta} \, \mathbf{K}_{\theta\theta}^{-1} \, \mathbf{K}_{\theta p}, \\
\widetilde{\mathbf{R}}_{p} &=  \mathbf{R}_{p} - \mathbf{K}_{p\theta} \, \mathbf{K}_{\theta\theta}^{-1} \, \mathbf{R}_{\theta}
\end{align}

Using $\mathbf{N}_{p}=\mathbf{N}_{\theta}=1$, we obtain
\begin{align} \label{eqn-Kpp-tilde}
\widetilde{\mathbf{K}}_{pp}
%
= - \, \IBi \, \frac{1}{\pderiv{^2\PsiVol(\bar{J})}{\bar{J}^2}} \, \dV
= - \, \IBi \widetilde{\vartheta} \, \dV
\end{align}
\begin{align} \label{eqn-Rp-tilde}
\widetilde{\mathbf{R}}_{p}
&= \IBi \mathbf{N}_p^{\T} \left[ J - \bar{J} + \frac{1}{\pderiv{^2\PsiVol(\bar{J})}{\bar{J}^2}} \, \left[\pderiv{\PsiVol(\bar{J})}{\bar{J}} - p \right] \right] \dV  \nonumber  \\
&= \IBi \mathbf{N}_p^{\T} \left[ J - \widetilde{J} - \widetilde{\vartheta} \, p \right] \dV
\end{align}
where
\begin{align}
\widetilde{J} &= \bar{J} - \frac{\left.\pderiv{\PsiVol(J)}{J}\right\vert_{\bar{J}}}{\left.\pderiv{^2\PsiVol}{J^2}\right\vert_{\bar{J}}}; \qquad
\widetilde{\vartheta} = \frac{1}{\left.\pderiv{^2\PsiVol}{J^2}\right\vert_{\bar{J}}}.
\end{align}

Equations (\ref{eqn-Kpp-tilde}) and (\ref{eqn-Rp-tilde}) are exactly in the same forms as (\ref{eqn-Kpp-proposed-simplified}) and (\ref{eqn-Rp-proposed-simplified}), respectively, of the proposed two-field formulation. The only difference between the three-field formulation and the proposed two-field formulation is that, $\widehat{J}$ and $\widehat{\vartheta}$ in the proposed formulation are evaluated from the solution at the previously converged load step, while $\widetilde{J}$ and $\widetilde{\vartheta}$ in the three-field formulation are evaluated from the solution at the previous iteration of the current load step. This difference, as demonstrated with the numerical example in the following subsection, has a negligible effect on the accuracy of computed numerical results. \mycolor{It is worth pointing out at this stage that the proposed formulation does not require the computation of any complementary energy functions.}

\myred
{
\section{Numerical examples} \label{section-examples}
To demonstrate the accuracy of the proposed formulation, two numerical example are presented.
The first example consists of a cylindrical bar that undergoes a significant axial stretch, rendering it suitable for the demonstration of accuracy of different mixed formulations in this work. The second example is the widely-used benchmark example of a 3D block under a compressive load.

\subsection{Stretching of a cylindrical bar}
The  example consists of a circular cylindrical bar of radius 10 units and length 5 units that is constrained from moving axially on one end and with a prescribed axial displacement on the other end. Only a quarter portion of the bar is considered for the analysis due to the obvious symmetry in the geometry and boundary conditions. The geometry of the problem is as shown in Fig. \ref{fig-geom}. The boundary conditions are such that
\begin{align}
u_x &=  0 \quad \;\; \mathrm{at} \quad X=0, \\
u_y &=  0 \quad \;\; \mathrm{at} \quad Y=0, \\
u_z &=  0 \quad \;\; \mathrm{at} \quad Z=0, \\
u_z &= 20 \quad      \mathrm{at} \quad Z=5.
\end{align}
The displacement boundary condition is applied in ten increments in all the simulations.

The deviatoric part of the strain energy function is assumed to be the Gent model, as given by
\begin{align} \label{eqn-gentmodel}
\PsiDev = -\frac{\mu \, I_m}{2} \, \ln\left(1-\frac{I_{\overline{\bm{C}}} - 3}{I_m}\right)
\end{align}
where, $\mu$ is the shear modulus and $I_m$ is a material parameter that represents the upper limit of $\left[ I_{\overline{\bm{C}}} - 3 \right]$. The shear modulus is computed from the Young's modulus ($E$) and Poisson's ratio ($\nu$) using the relation $\mu=\frac{E}{2 \, (1+\nu)}$. For the volumetric part of the energy function, the function proposed by Hartmann and Neff \cite{HartmannIJSS2003}, $\PsiVol_{1}$ in Table. \ref{table-volmfunctions}, is considered in this work because this function results in significantly higher values of pressure for $J>1.5$ when compared with other volumetric energy functions, as shown in Fig. \ref{fig-volfuncs-plot2}; thus, enabling an effective demonstration of key differences in different mixed two-field formulations.

\subsubsection{Comparison with different two-field formulations}
The performance of the proposed two-field formulation against the other two two-field formulations is assessed by using the Ladyzhenskaya-Babuska-Brezzi (LBB)-stable BT2/BT1 element \cite{KadapaIJNME2019mixed} that employs a quadratic B\'ezier element for the displacement field and a linear B\'ezier element for the pressure field. We refer the reader to Kadapa \cite{KadapaIJNME2019bbar, KadapaIJNME2019mixed} for the comprehensive details on B\'ezier elements. The tetrahedral finite element mesh used for the analysis, as shown in Fig. \ref{fig-tet10-mesh}, consists of 6713 nodes and 4300 BT2/BT1 elements.

The Young's modulus is assumed to be $E=100$, and the parameter $I_m$ is taken as 30. Three different values of Poisson's ratio $\nu=\{0.3, 0.45, 0.4999\}$ are chosen to demonstrate the accuracy of the proposed scheme in the compressible as well as in the nearly incompressible regime. Note that, for the truly incompressible case, i.e., for $\nu=0.5$, all the two-field mixed formulations considered in this work become identical; hence, requiring no numerical validation.

The accuracy of numerical results is assessed by tracking the displacement and pressure values at node corresponding to point A (see Fig. \ref{fig-geom}). Graphs of Y-displacement and pressure at the node A against load step number for the three different values of Poisson's ratio are presented in Fig. \ref{fig-tet10-graphs}. The corresponding contour plots are shown in Figs. \ref{fig-contours-nu0p3}, \ref{fig-contours-nu0p45} and \ref{fig-contours-nu0p4999}. As shown, for all the values of Poisson's ratio, the results obtained with the proposed mixed formulation are in excellent agreement with those obtained with the Galerkin formulation. On the contrary, the perturbed Lagrangian approach fails to produce accurate results for $\nu=0.3$ as the applied displacement is increased beyond 40\% of the load, which corresponds to an axial stretch ratio of 2.6. For the nearly incompressible case ($\nu=0.4999$), the results obtained with different mixed formulations match well with each other, as expected. Thus, the proposed mixed displacement-pressure formulation yields numerical results of the same accuracy as that of the Galerkin approach while using a symmetric matrix solver.
\begin{figure}[H]
\centering
\subfloat[Geometry]{\includegraphics[clip, scale=0.3]{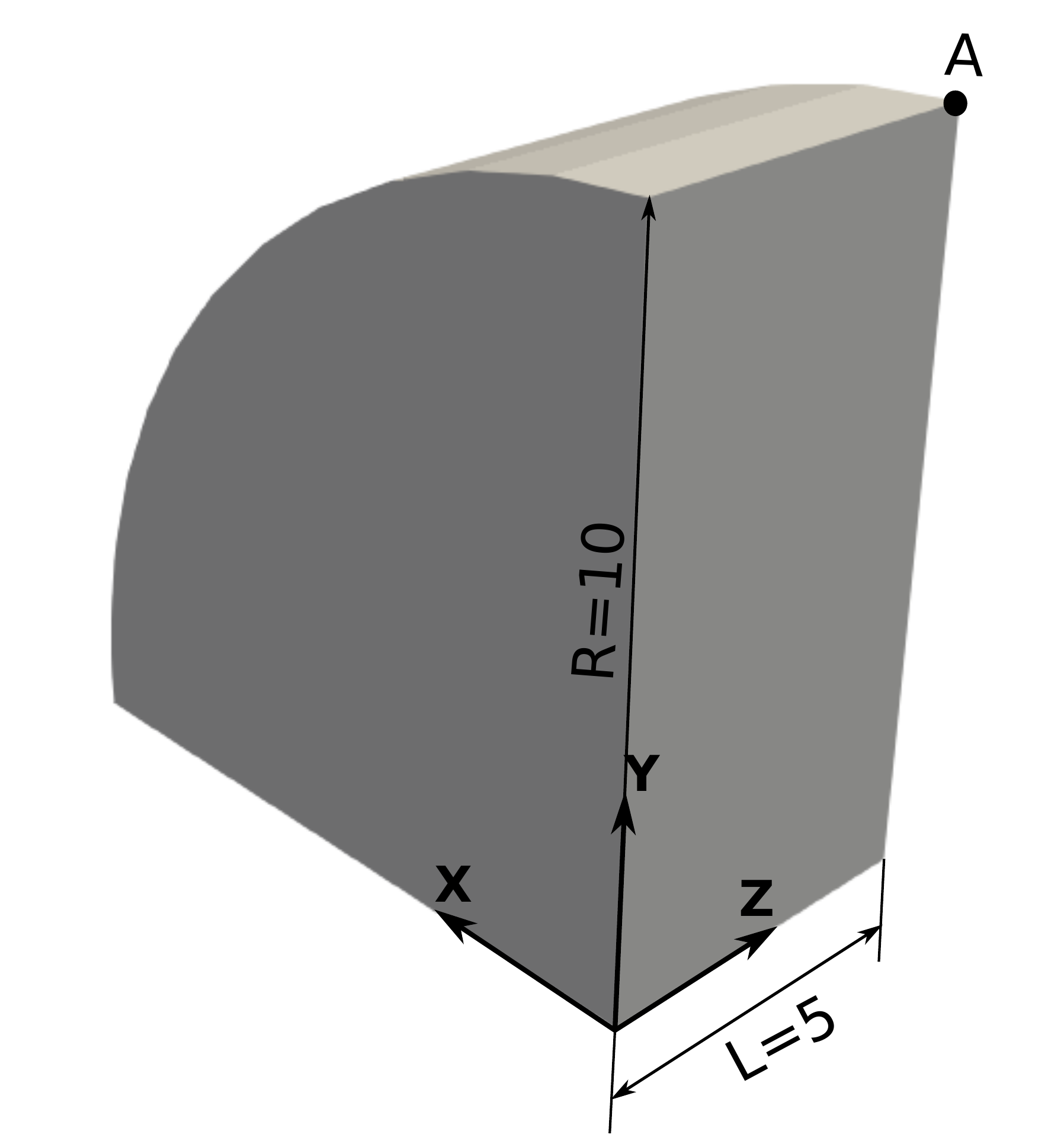} \label{fig-geom}}
\subfloat[Tetrahedral mesh]{\includegraphics[clip, scale=0.3]{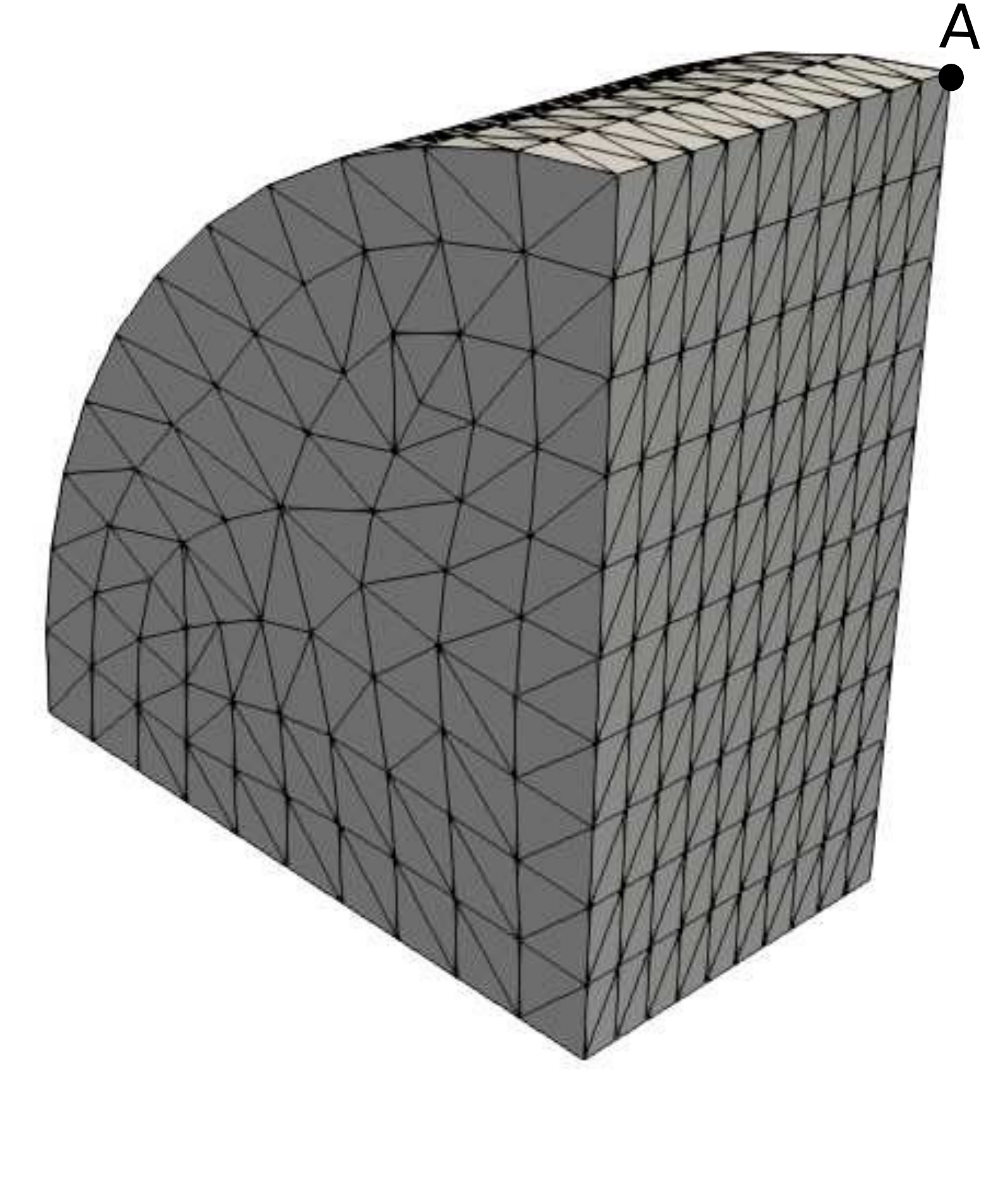} \label{fig-tet10-mesh}}
\subfloat[Hexhedral mesh]{\includegraphics[clip, scale=0.3]{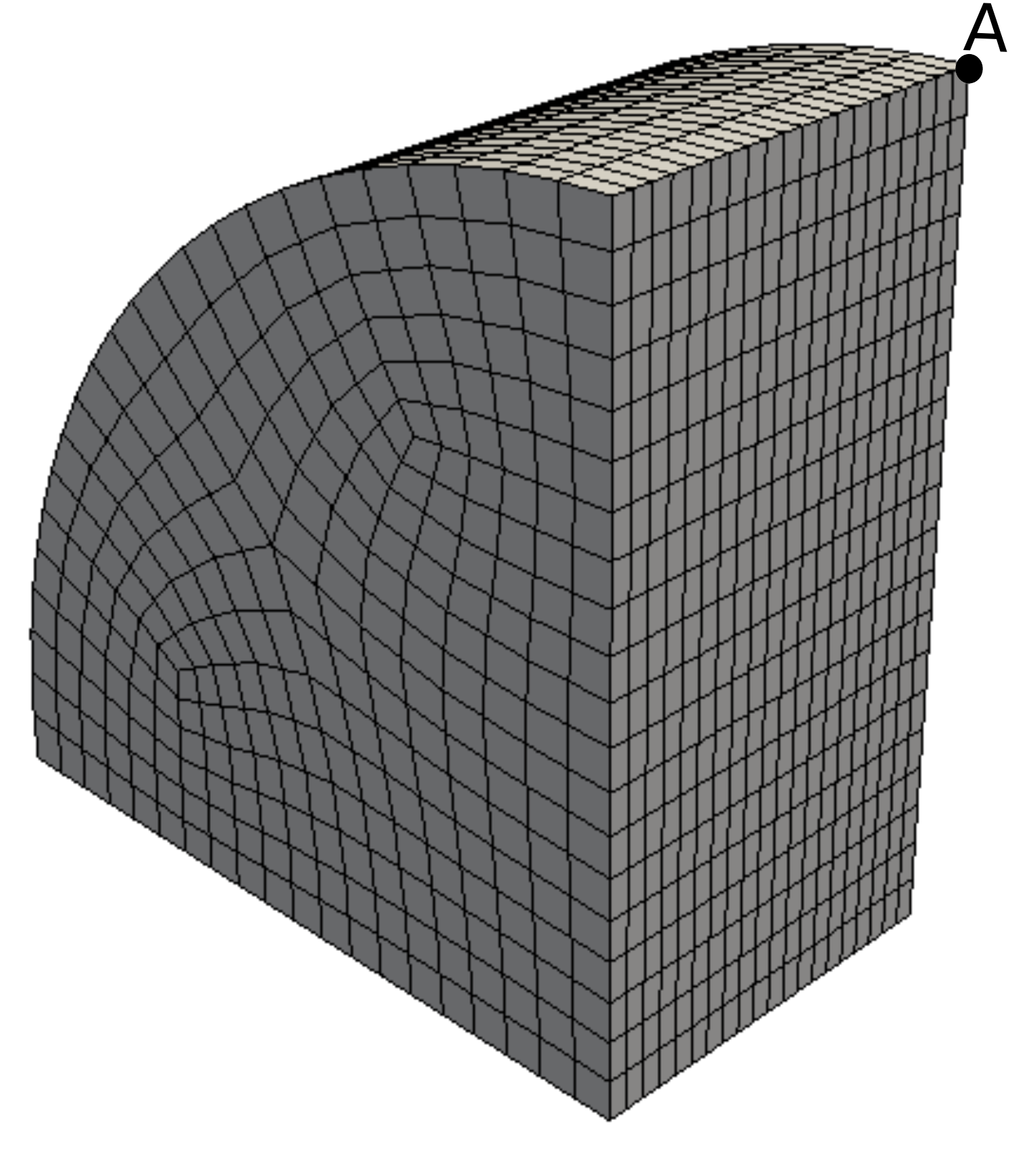} \label{fig-mesh-hex8} }
\caption{\mycolor{Cylindrical bar: (a) geometry of the problem, and (b) and (c) the finite element meshes.}}
\label{fig-geom-mesh}
\end{figure}

\begin{figure}[H]
\centering
\subfloat[$\nu=0.3$]{\includegraphics[clip, scale=0.5]{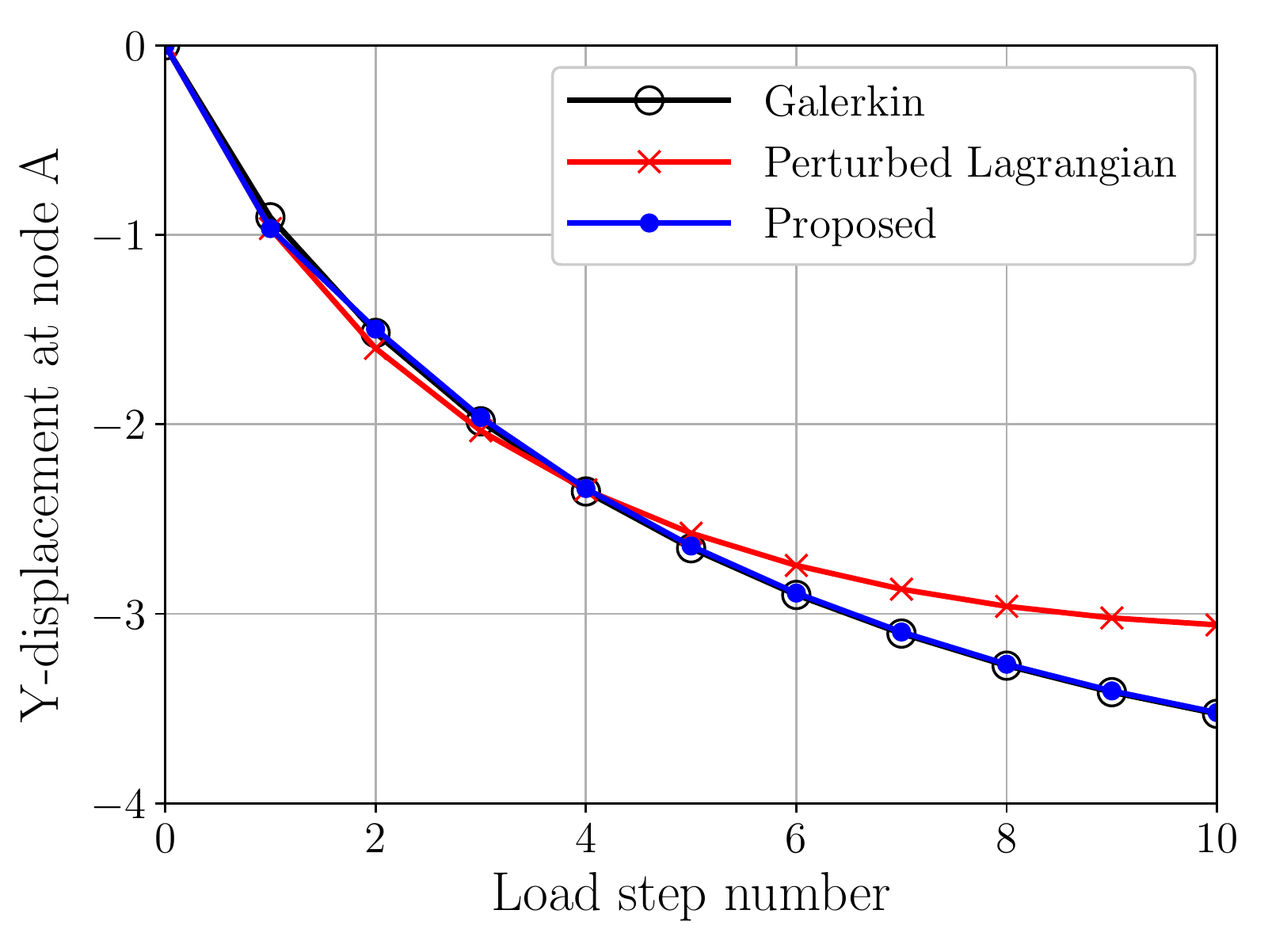}}
\subfloat[$\nu=0.3$]{\includegraphics[clip, scale=0.5]{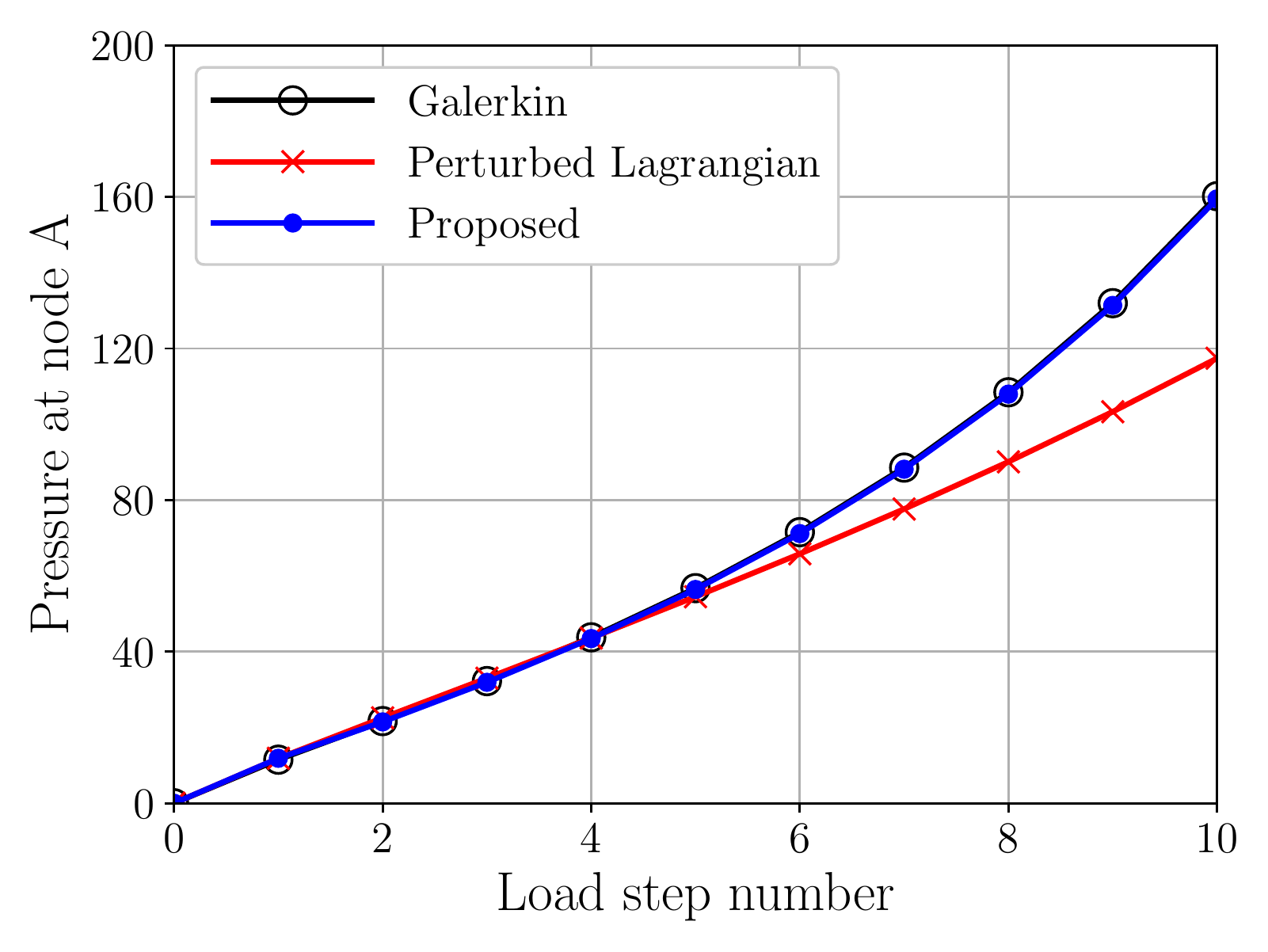}} \\
\subfloat[$\nu=0.45$]{\includegraphics[clip, scale=0.5]{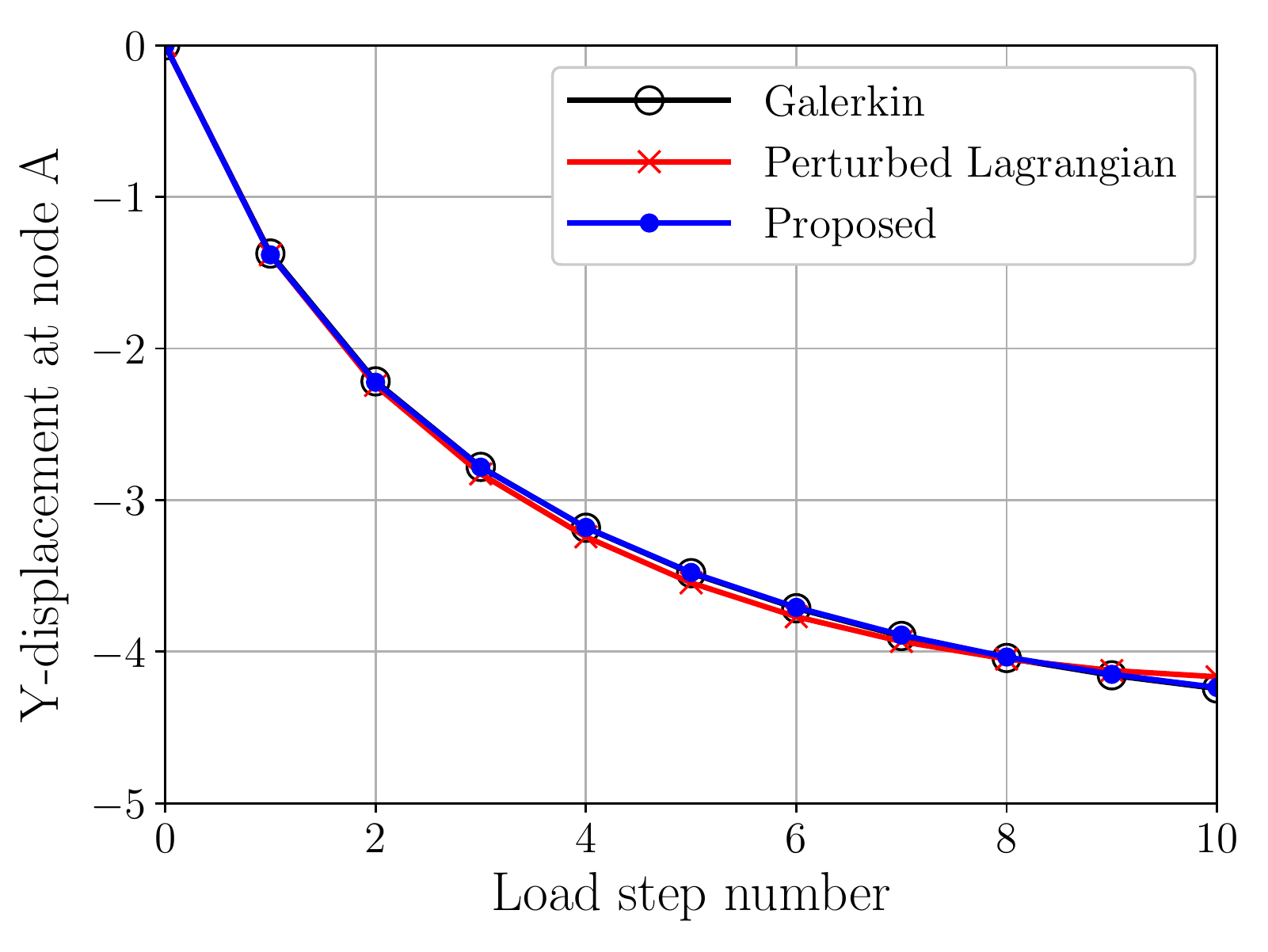}}
\subfloat[$\nu=0.45$]{\includegraphics[clip, scale=0.5]{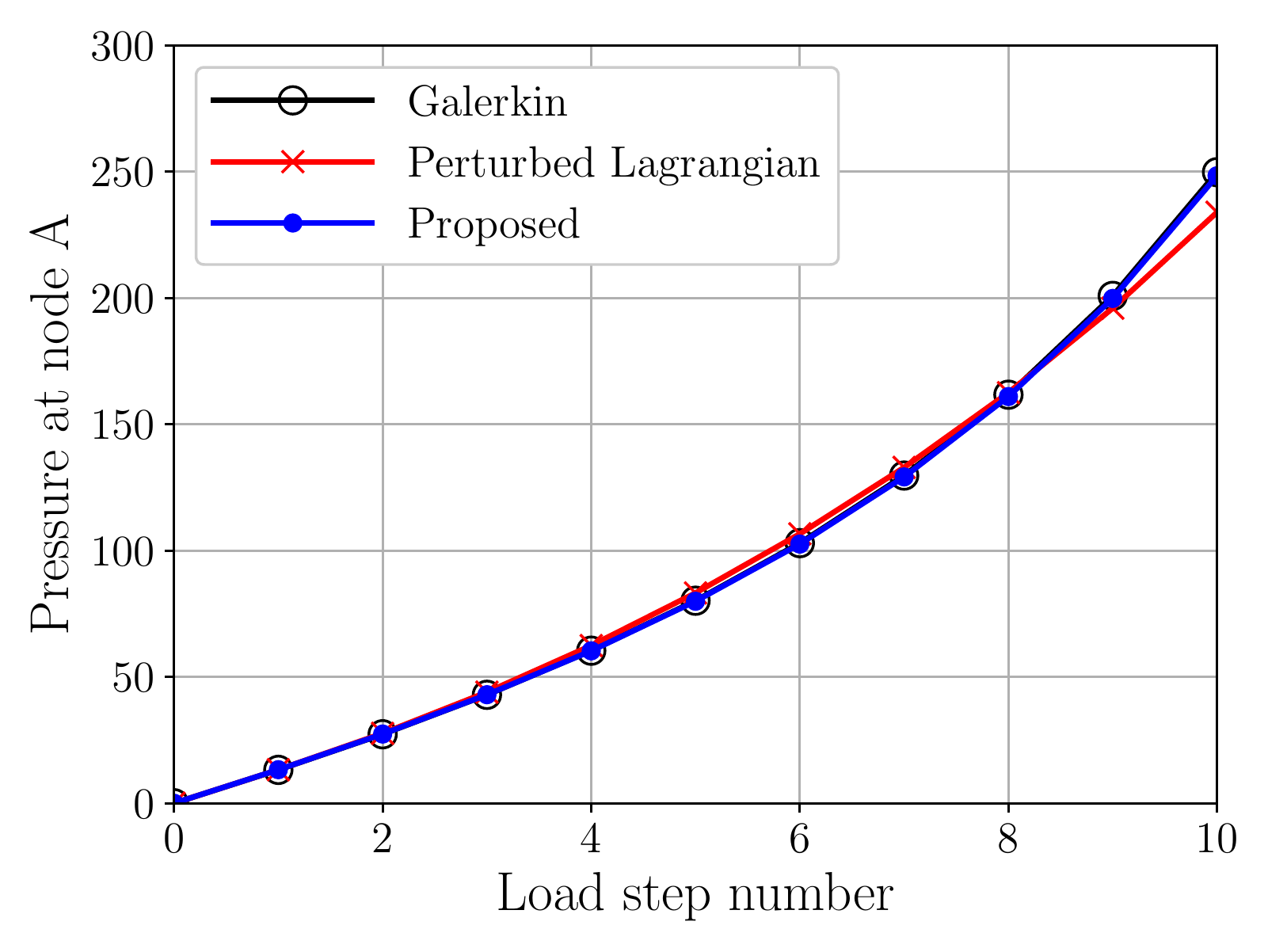}} \\
\subfloat[$\nu=0.4999$]{\includegraphics[clip, scale=0.5]{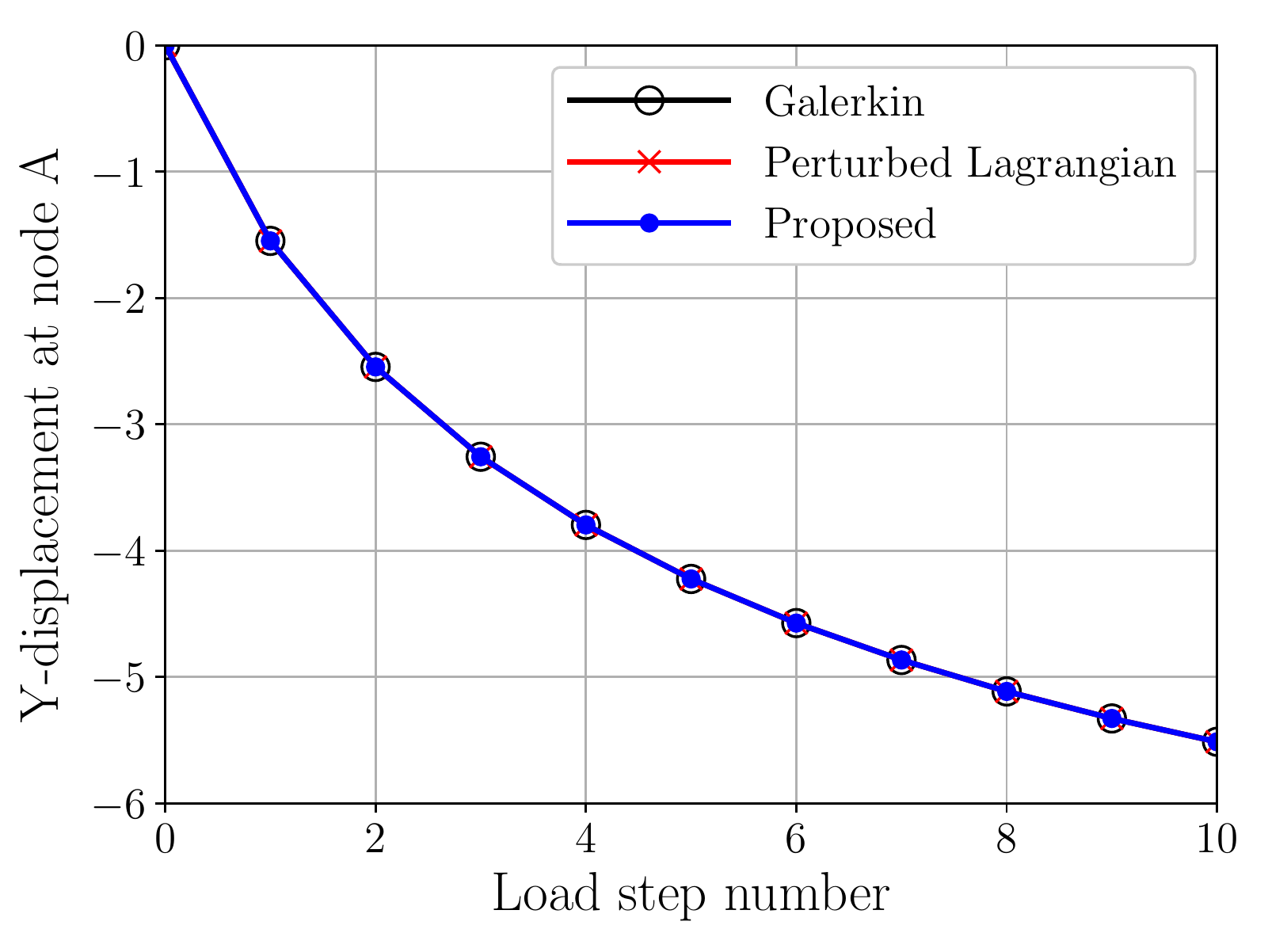}}
\subfloat[$\nu=0.4999$]{\includegraphics[clip, scale=0.5]{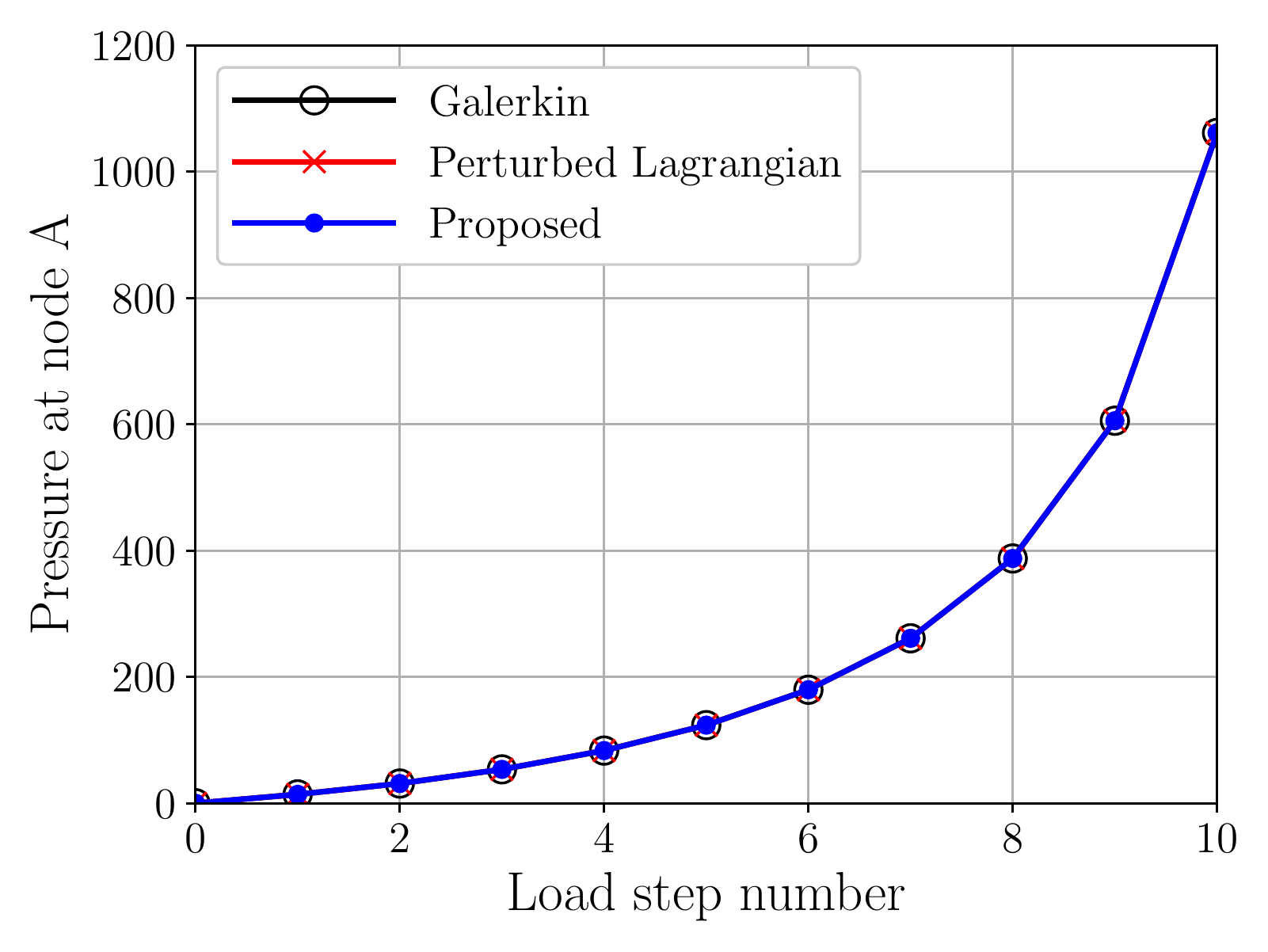}}
\caption{\mycolor{Cylindrical bar with BT2/BT1 element: evolution of Y-displacement and pressure at node A for $\nu=0.3$, $\nu=0.45$ and $\nu=0.4999$, obtained with different mixed formulations.}}
\label{fig-tet10-graphs}
\end{figure}
\begin{figure}[H]
\centering
\includegraphics[clip, scale=0.3]{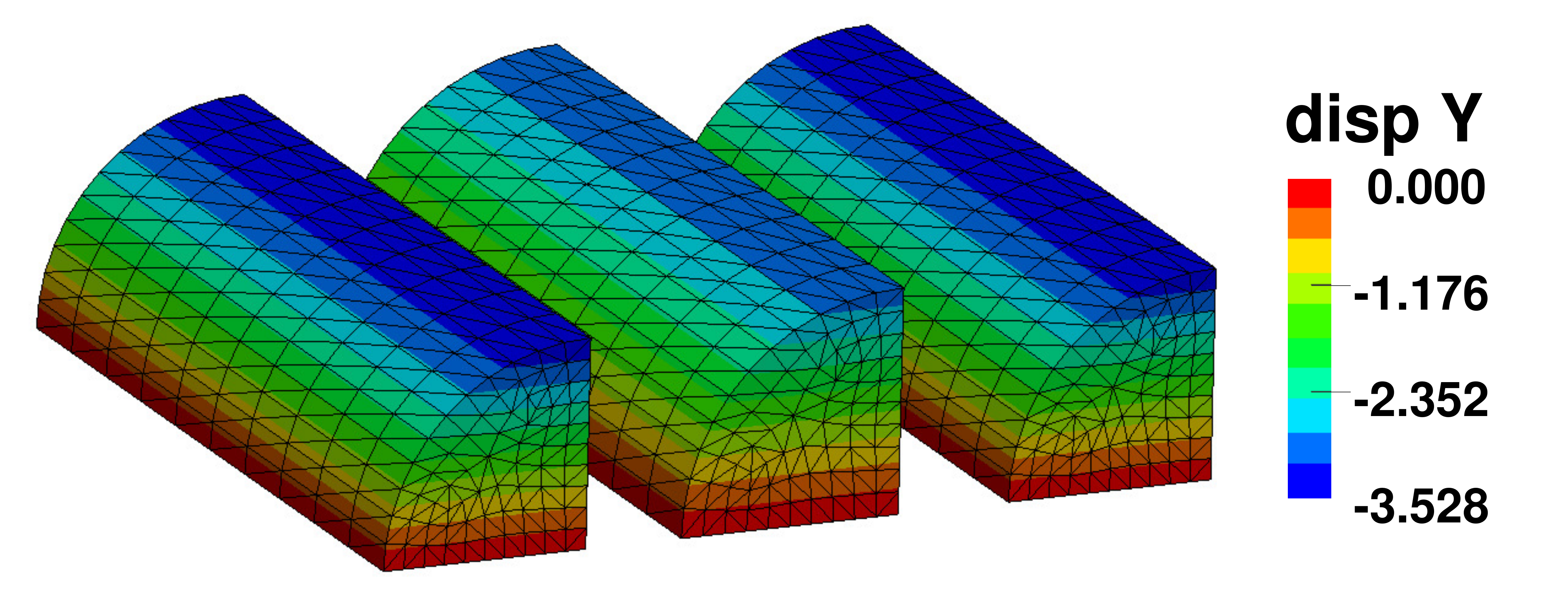} \\
\includegraphics[clip, scale=0.3]{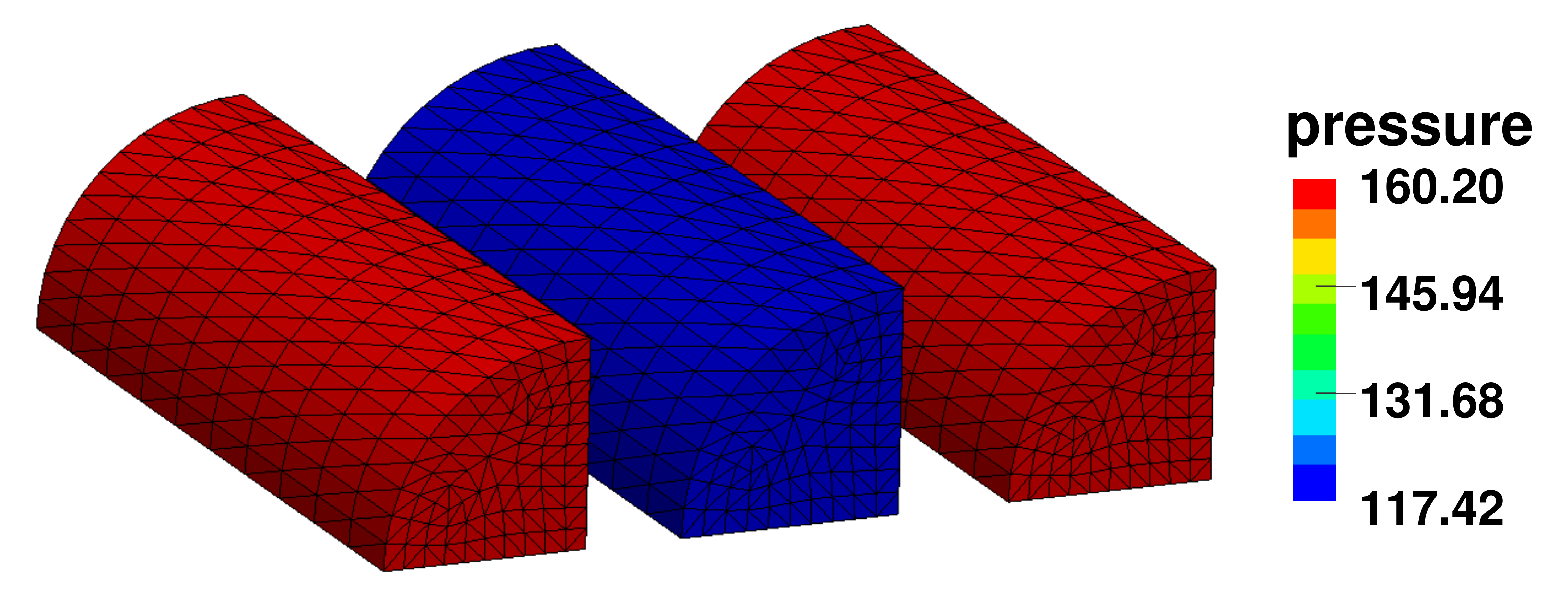}
\caption{\mycolor{Cylindrical bar with BT2/BT1 element: contour plots of Y-displacement and pressure for $\nu=0.3$ obtained with different mixed formulations. In each figure: (left) Galerkin formulation, (centre) perturbed Lagrangian formulation and (right) proposed formulation.}}
\label{fig-contours-nu0p3}
\end{figure}
\begin{figure}[H]
\centering
\includegraphics[clip, scale=0.3]{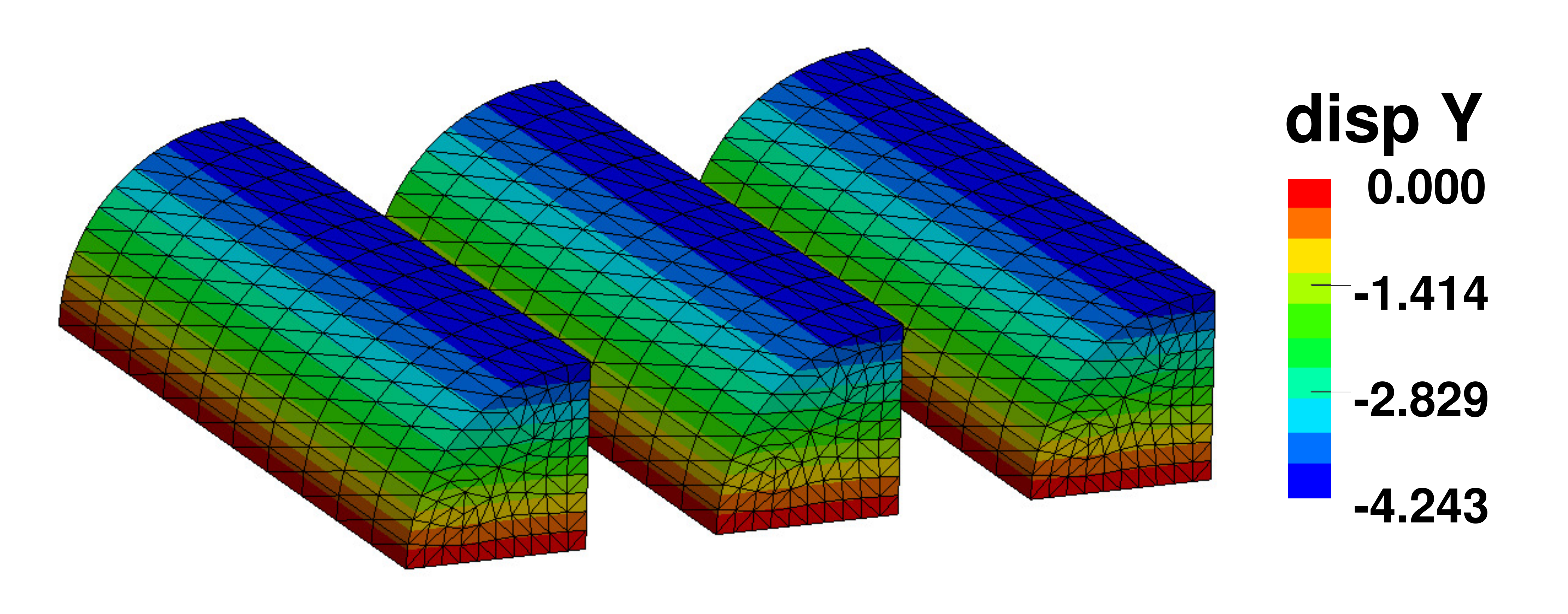} \\
\includegraphics[clip, scale=0.3]{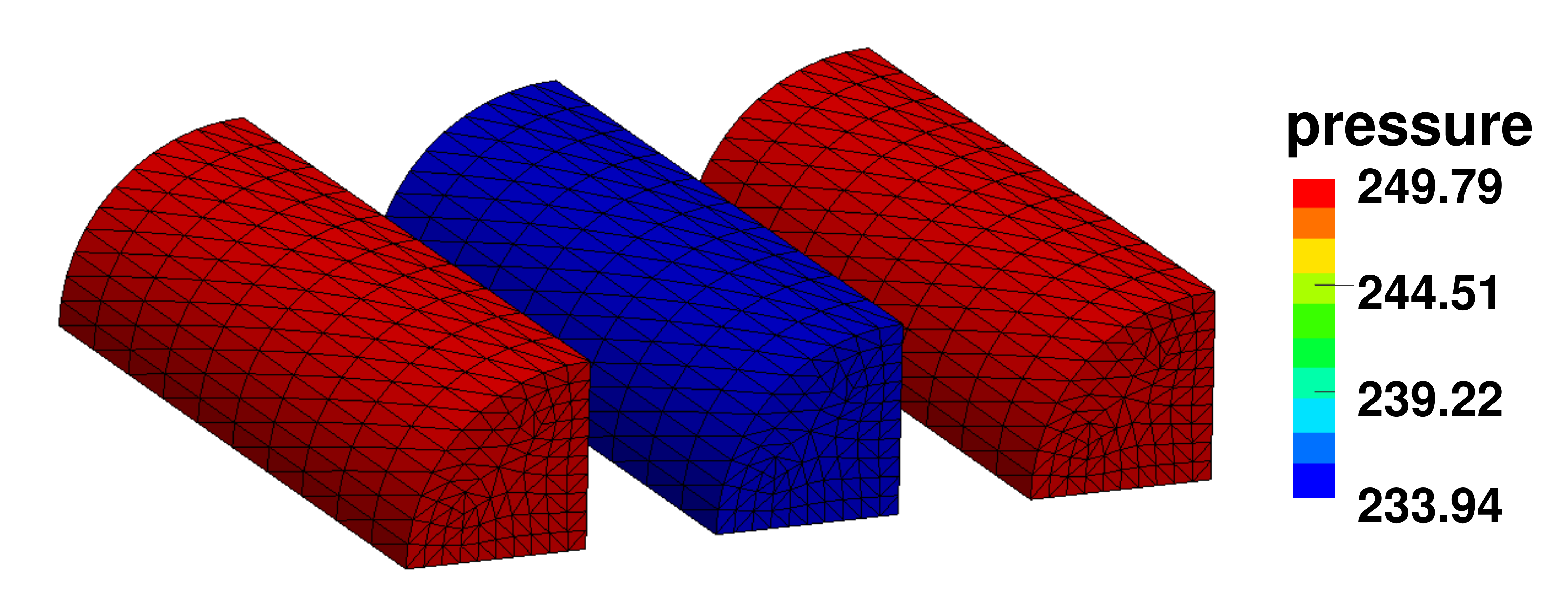}
\caption{\mycolor{Cylindrical bar with BT2/BT1 element: contour plots of Y-displacement and pressure for $\nu=0.45$ obtained with different mixed formulations. In each figure: (left) Galerkin formulation, (centre) perturbed Lagrangian formulation and (right) proposed formulation.}}
\label{fig-contours-nu0p45}
\end{figure}
\begin{figure}[H]
\centering
\includegraphics[clip, scale=0.3]{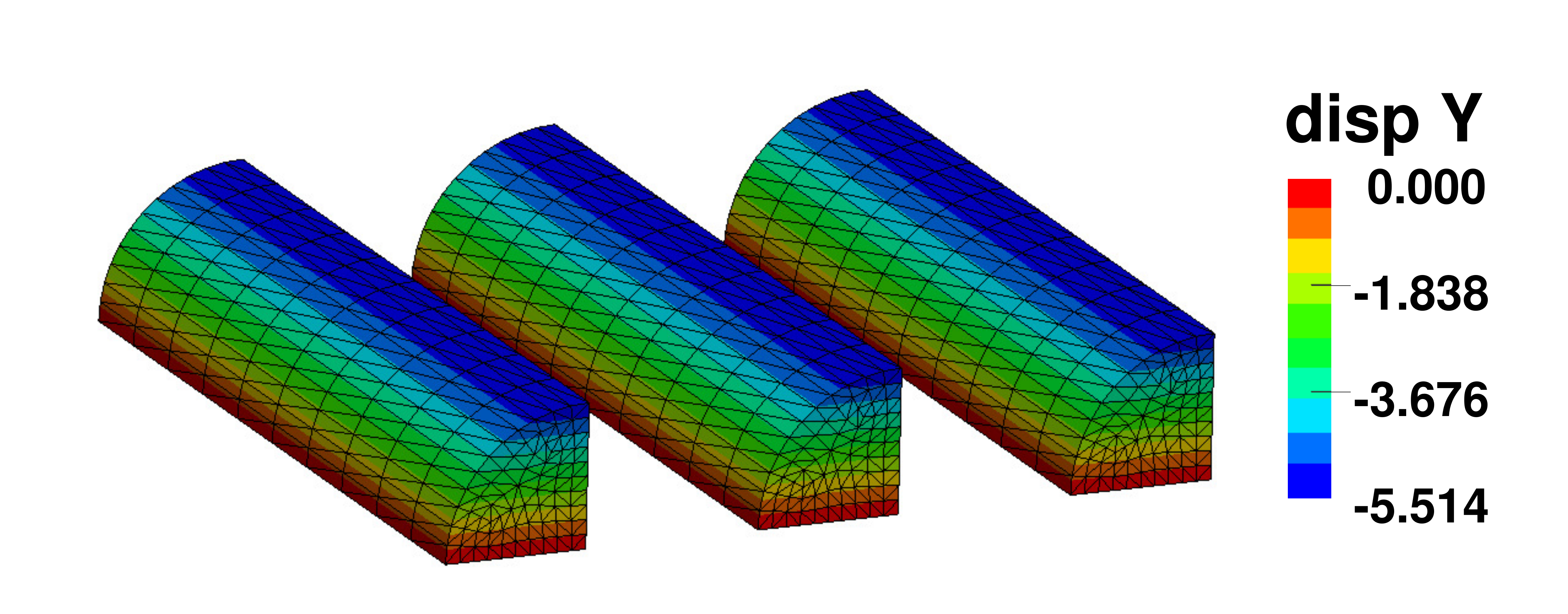} \\
\includegraphics[clip, scale=0.3]{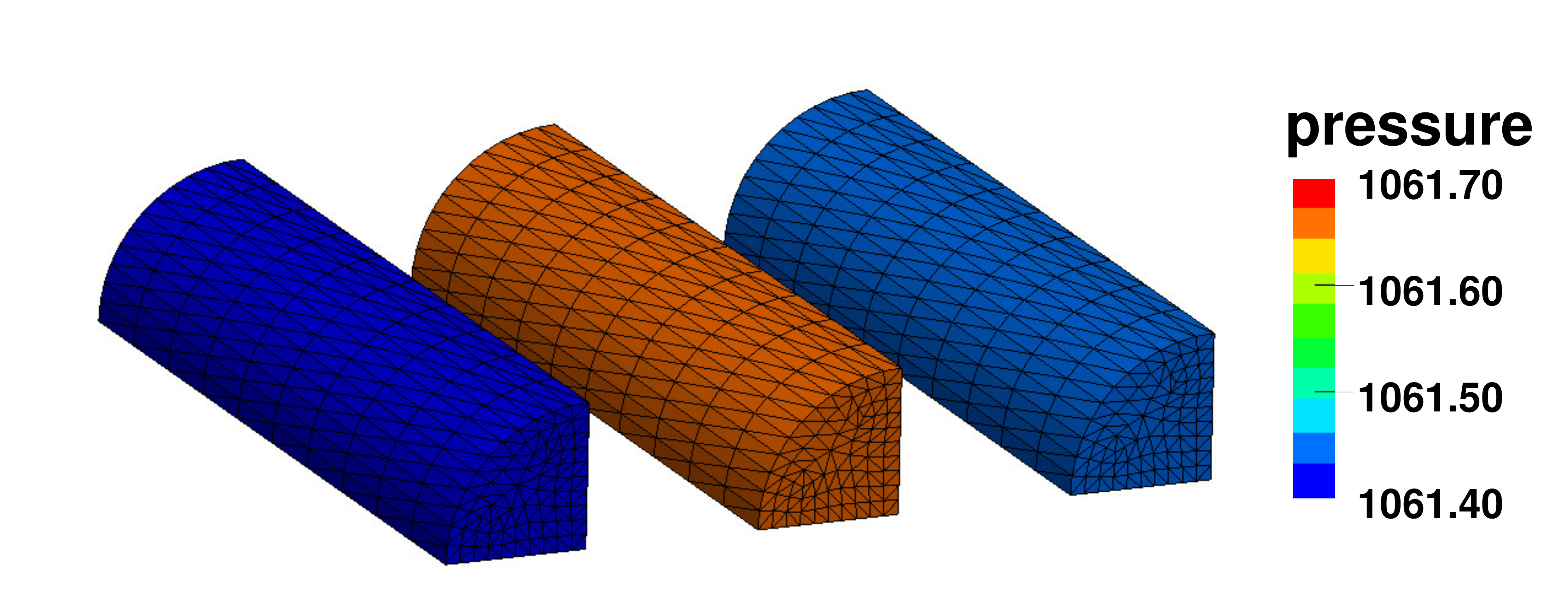}
\caption{\mycolor{Cylindrical bar with BT2/BT1 element: contour plots of Y-displacement and pressure for $\nu=0.4999$ obtained with different mixed formulations. In each figure: (left) Galerkin formulation, (centre) perturbed Lagrangian formulation and (right) proposed formulation.}}
\label{fig-contours-nu0p4999}
\end{figure}

\clearpage
\subsubsection{Comparison with the three-field formulation}
For comparison with the three-field formulation, the widely-used Q1/P0 element is considered in this work. For the Q1/P0 element in three dimensions, the displacement and pressure are approximated, respectively, using a trilinear hexahedron element and a constant discontinuous function. The finite element mesh shown in Fig. \ref{fig-mesh-hex8} consists of 6573 nodes and 5600 hexahedral elements. Similar to the BT2/BT1 element case, analysis is performed for three different values of Poisson's ratio under the same set of loading conditions and material properties. The graphs of Y-displacement and pressure at node A against load step, as illustrated in Fig. \ref{fig-hex8-graphs}, and the corresponding contour plots at the final deformed configuration as shown in Figs. \ref{fig-hex8-contours-nu0p3} and \ref{fig-hex8-contours-nu0p4999}, prove that the results obtained with the proposed two-field mixed formulation match very well with those obtained with the three-field formulation. Table. \ref{table-convergence}  illustrates that the convergence behaviour of Newton-Raphson iterations for the last load step using the proposed formulation follows the same trend as that of the three-field formulation.

\renewcommand{\arraystretch}{1.5}
\begin{table}[H]
\centering
\begin{tabular}{|c|c|c|c|c|}
\hline
\multirow{2}[2]{2cm}{Iteration number} &
   \multicolumn{2}{c|}{\multirow{1}[2]{*}{$\nu=0.3$}} & \multicolumn{2}{c|}{\multirow{1}[2]{*}{$\nu=0.4999$}} \\
\cline{2-5} & three-field & proposed & three-field & proposed \\
\hline
  1  &  6.4221 E$+$03 &  6.4637   E$+$03  &  9.0645   E$+$05 &  9.0645 E$+$05 \\
  2  &  8.7138 E$+$01 &  1.3193   E$+$02  &  1.6932   E$+$04 &  1.6992 E$+$04 \\
  3  &  6.1248 E$-$02 &  8.4520   E$-$02  &  2.4512   E$+$01 &  2.8163 E$+$01 \\
  4  &  6.4528 E$-$08 &  2.7272   E$-$07  &  4.2110   E$-$05 &  5.9347 E$-$05 \\
  5  &                &                   &  8.3370   E$-$09 &  8.5369 E$-$09 \\
\hline
\end{tabular}\vspace{1ex}
\caption{\mycolor{Cylindrical bar with Q1/P0 element: convergence of the norm of the residuals for the last load step with the three-field and the proposed two-field formulations.}}
\label{table-convergence}
\end{table}
\renewcommand{\arraystretch}{1.0}

\begin{figure}[H]
\centering
\subfloat[$\nu=0.3$]{\includegraphics[clip, scale=0.5]{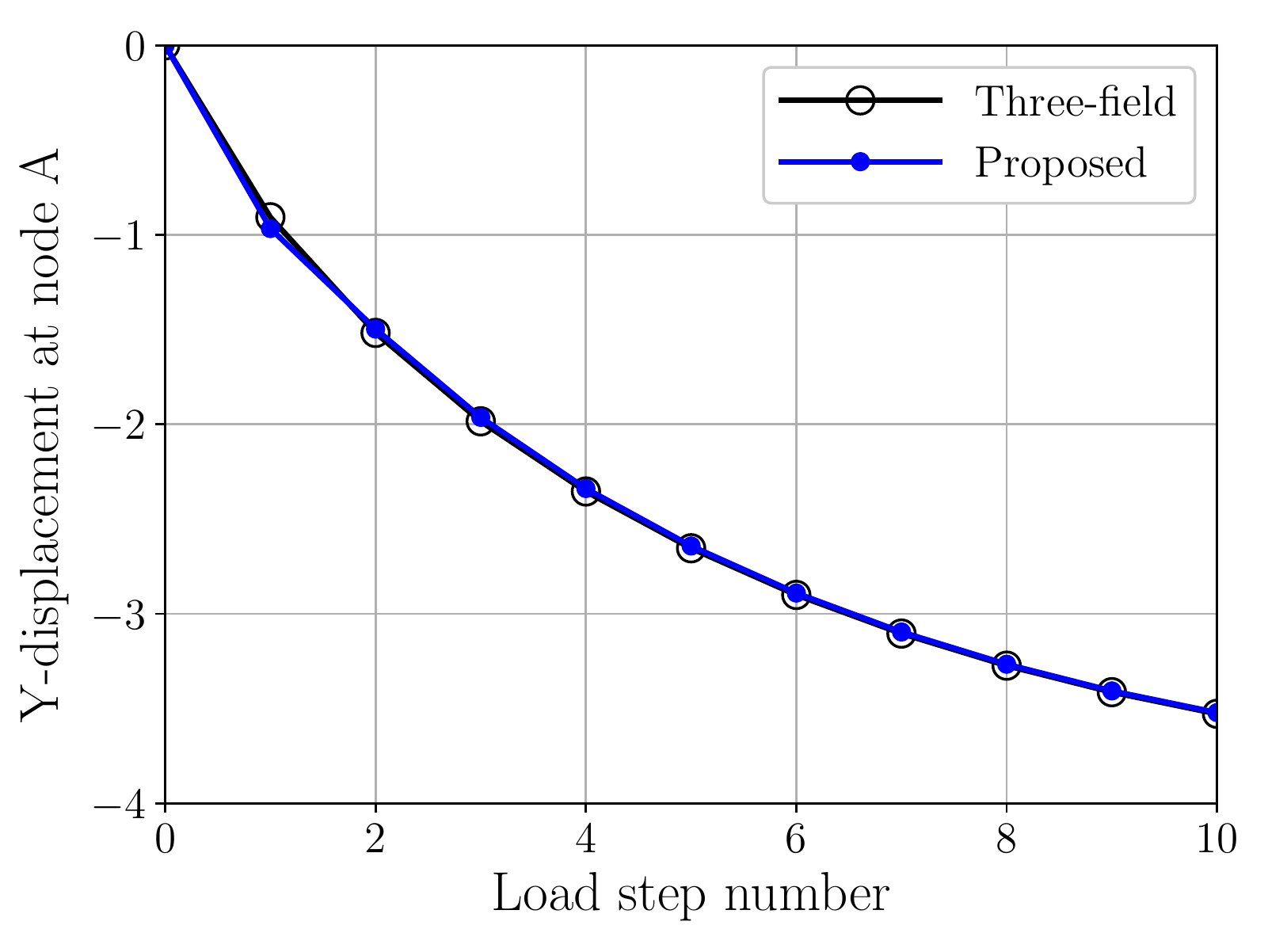}}
\subfloat[$\nu=0.3$]{\includegraphics[clip, scale=0.5]{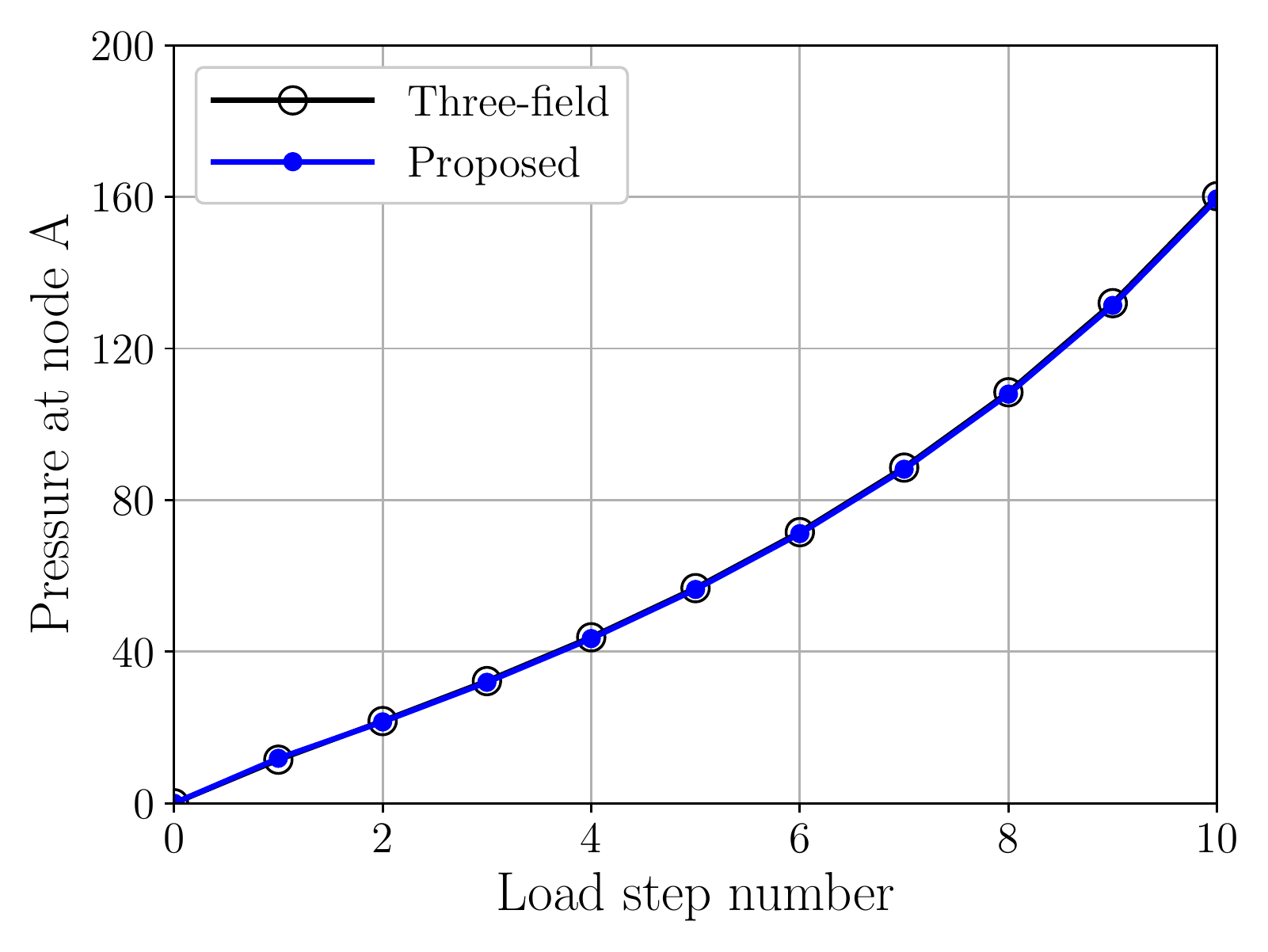}} \\
\subfloat[$\nu=0.45$]{\includegraphics[clip, scale=0.5]{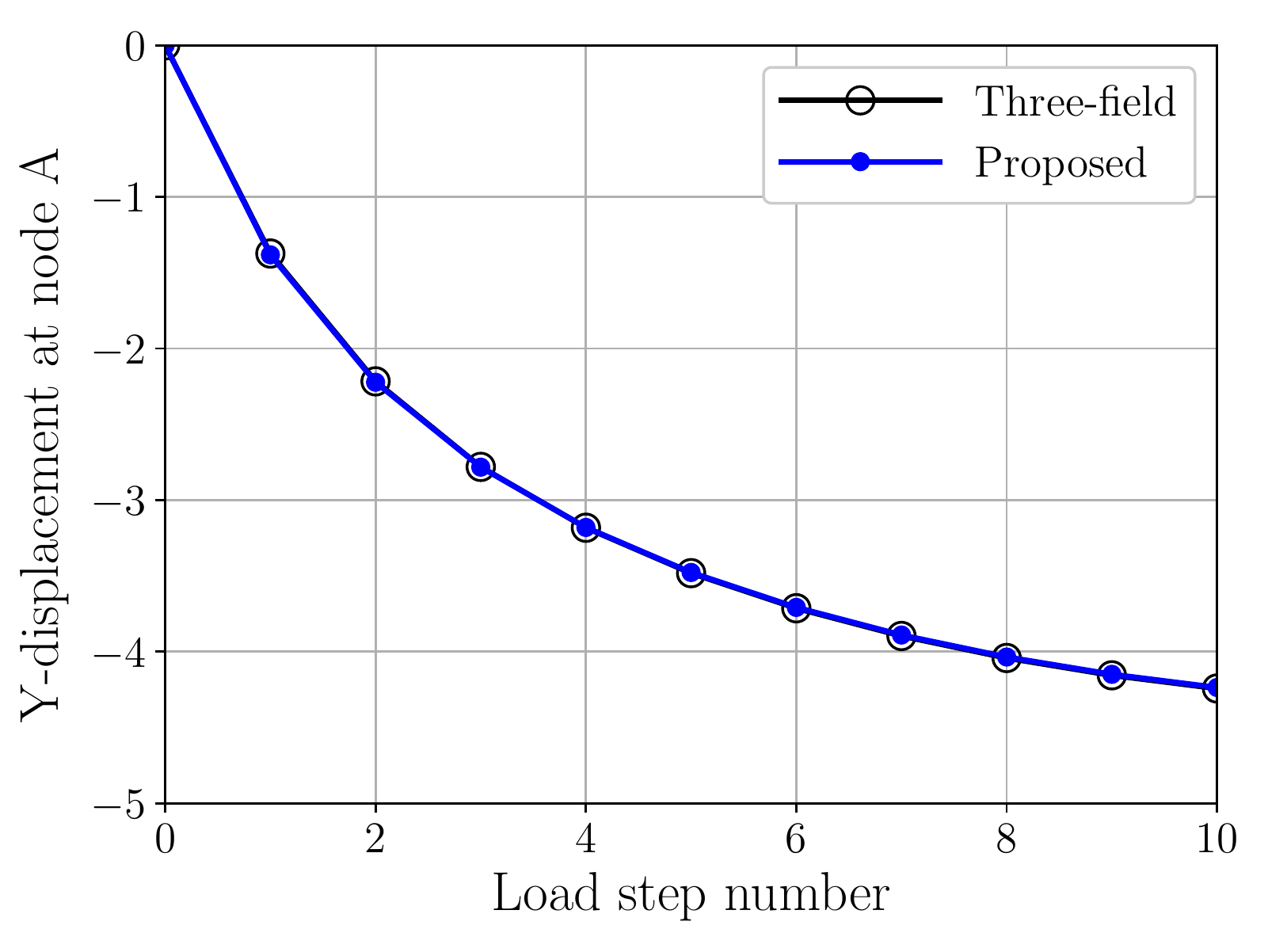}}
\subfloat[$\nu=0.45$]{\includegraphics[clip, scale=0.5]{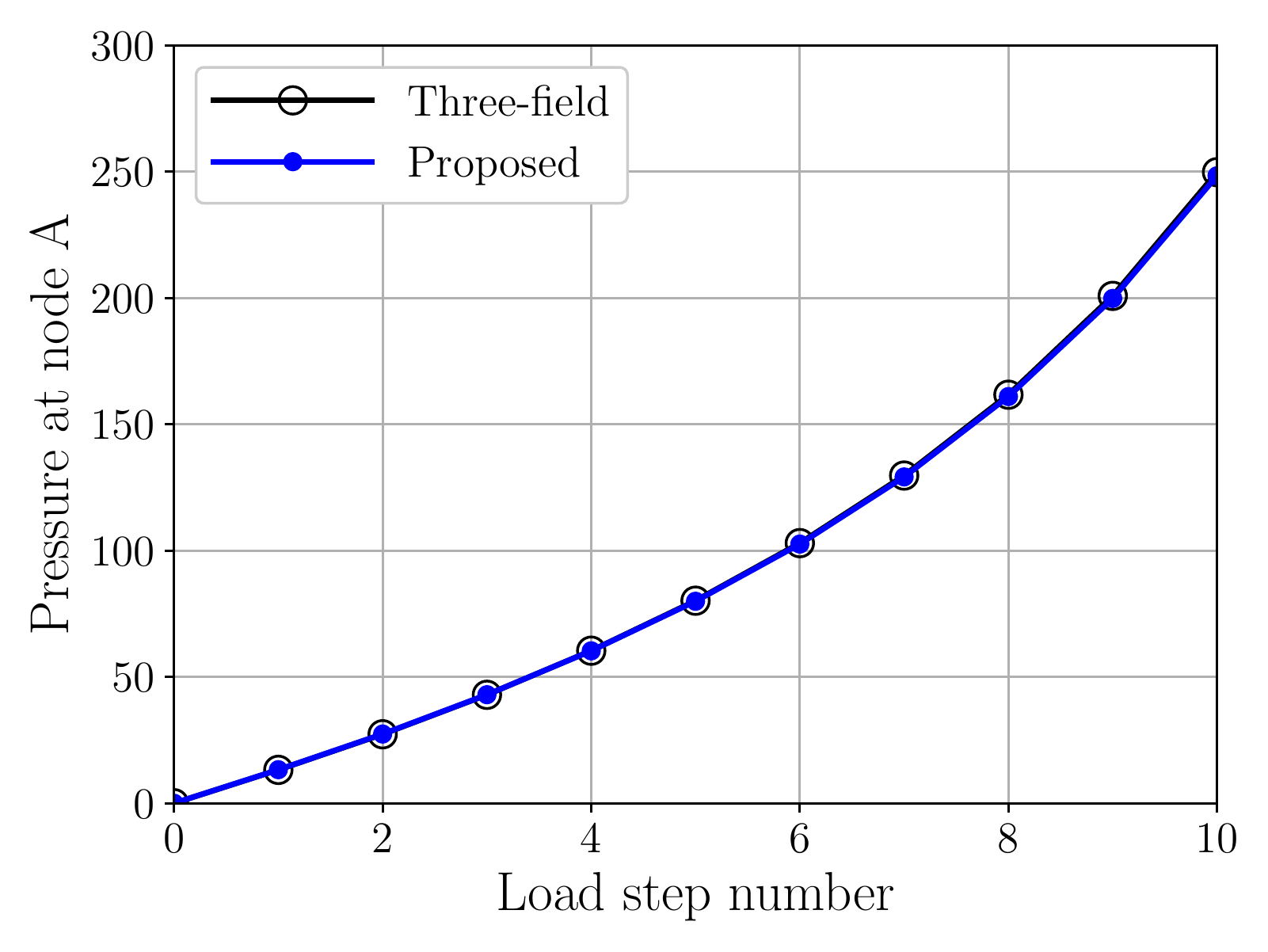}} \\
\subfloat[$\nu=0.4999$]{\includegraphics[clip, scale=0.5]{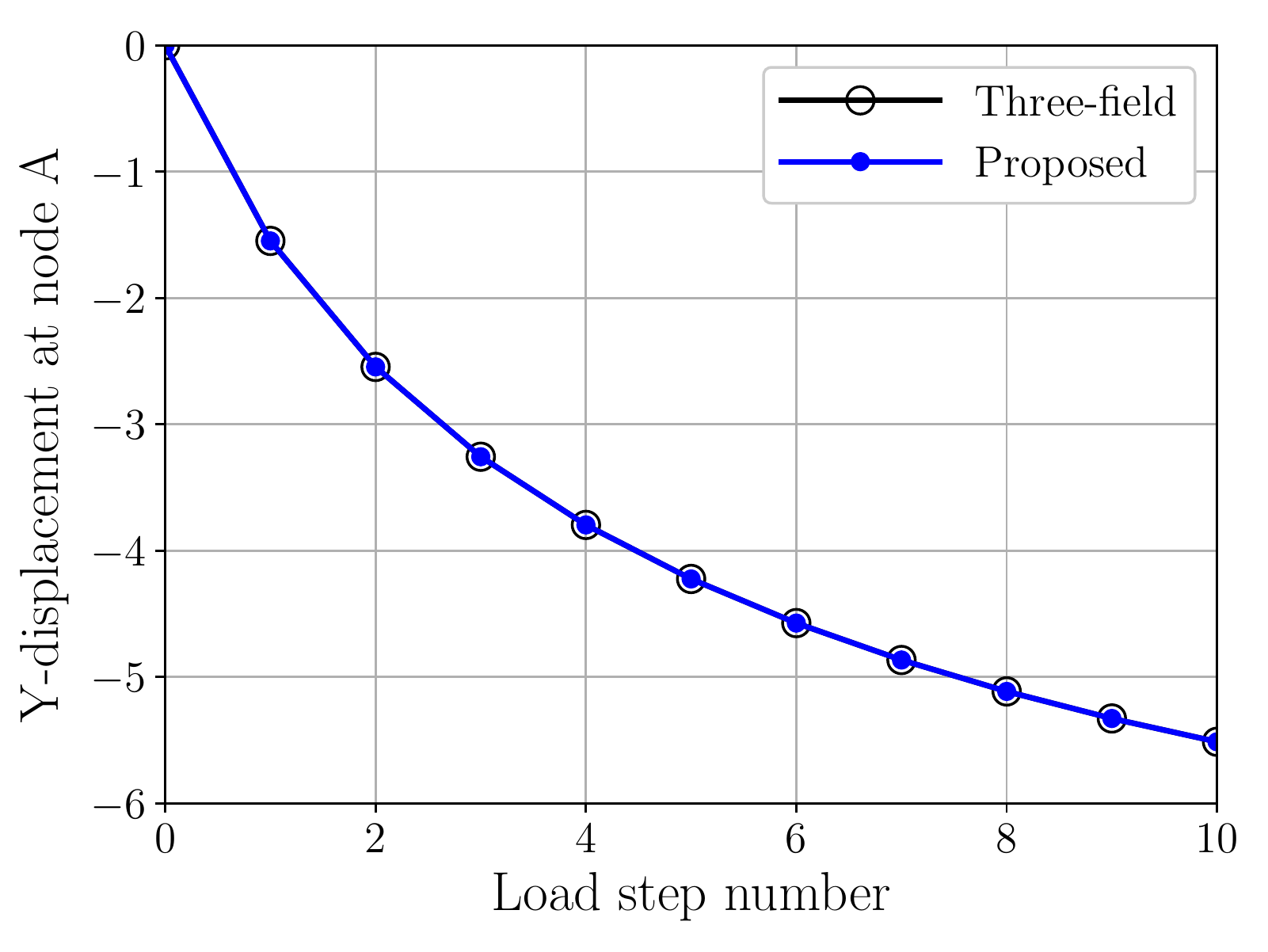}}
\subfloat[$\nu=0.4999$]{\includegraphics[clip, scale=0.5]{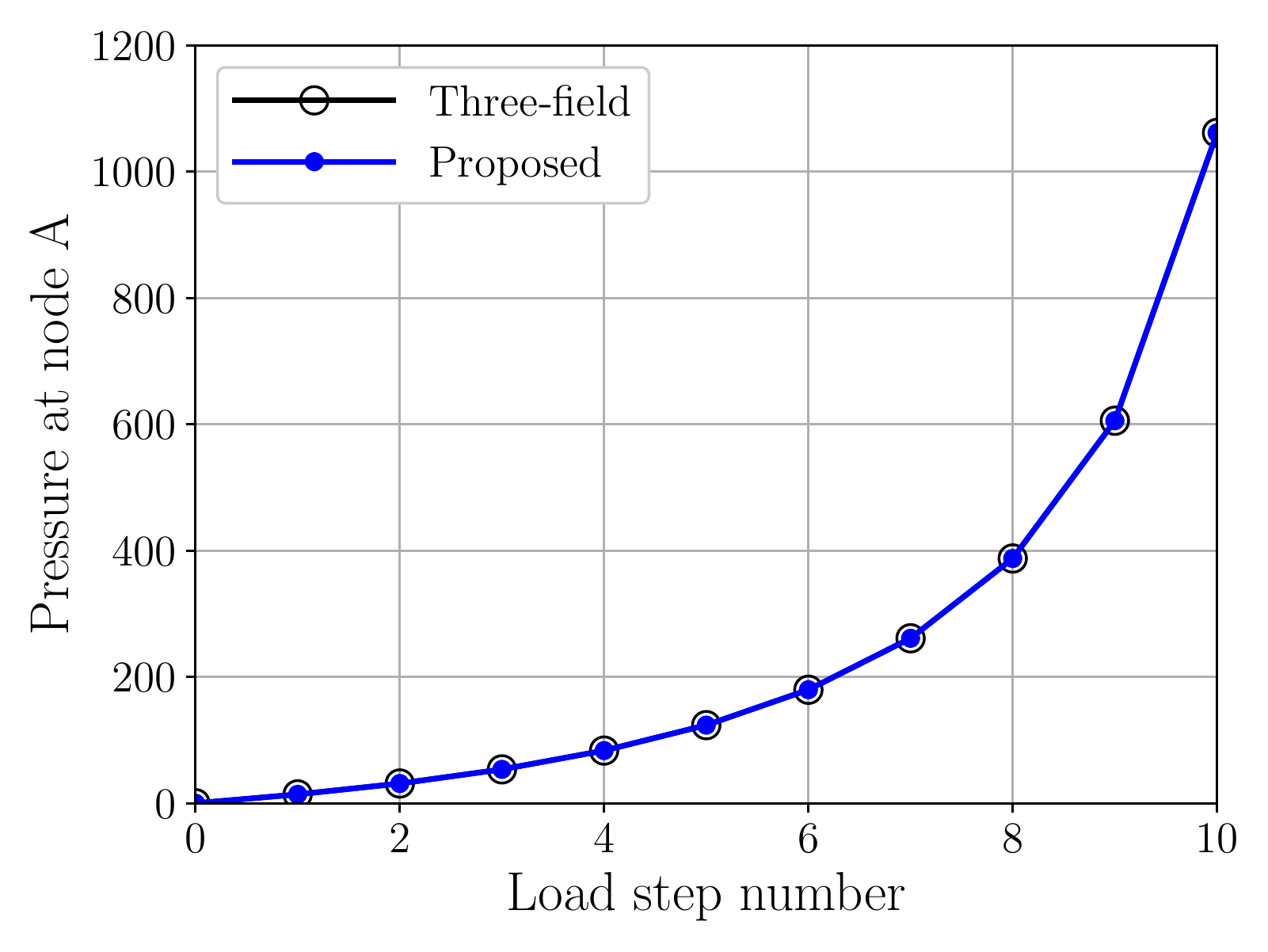}}
\caption{\mycolor{Cylindrical bar with Q1/P0 element: evolution of Y-displacement and pressure at node A for $\nu=0.3$, $\nu=0.45$ and $\nu=0.4999$, obtained with the three-field and the proposed mixed formulations.}}
\label{fig-hex8-graphs}
\end{figure}
\begin{figure}[H]
\centering
\includegraphics[clip, scale=0.35]{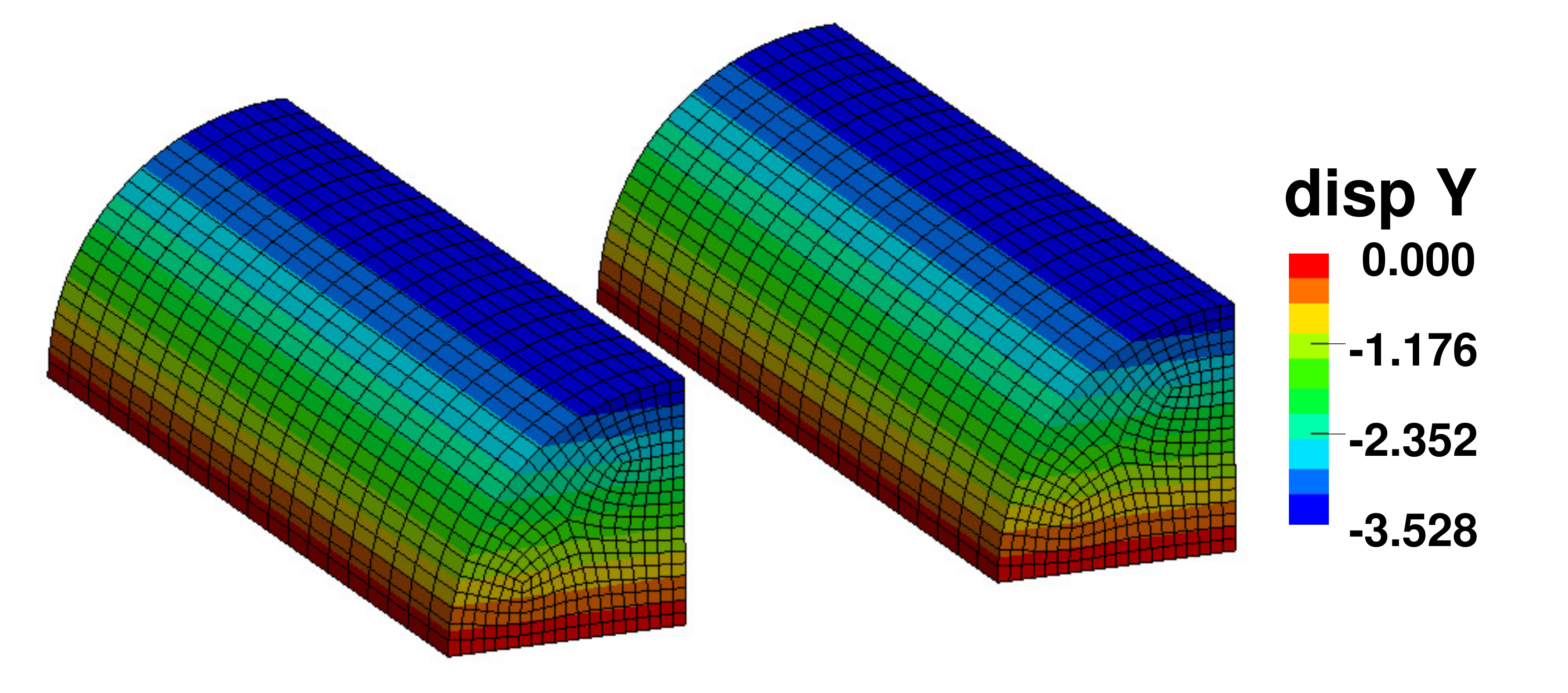} \\
\includegraphics[clip, scale=0.35]{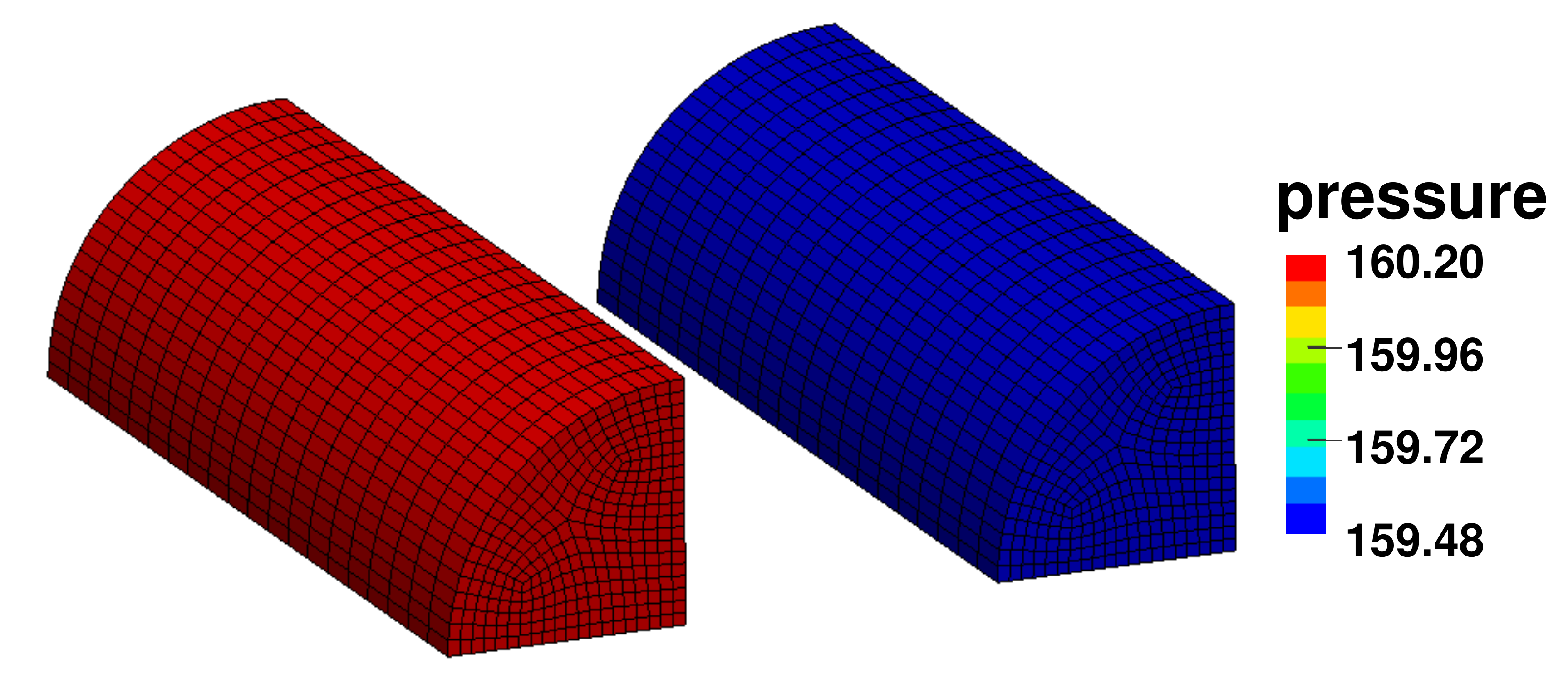}
\caption{\mycolor{Cylindrical bar with Q1/P0 element: contour plots of Y-displacement and pressure for $\nu=0.3$. In each figure: left) the standard three-field formulation and right) the proposed two-field formulation.}}
\label{fig-hex8-contours-nu0p3}
\end{figure}
\begin{figure}[H]
\centering
\includegraphics[clip, scale=0.35]{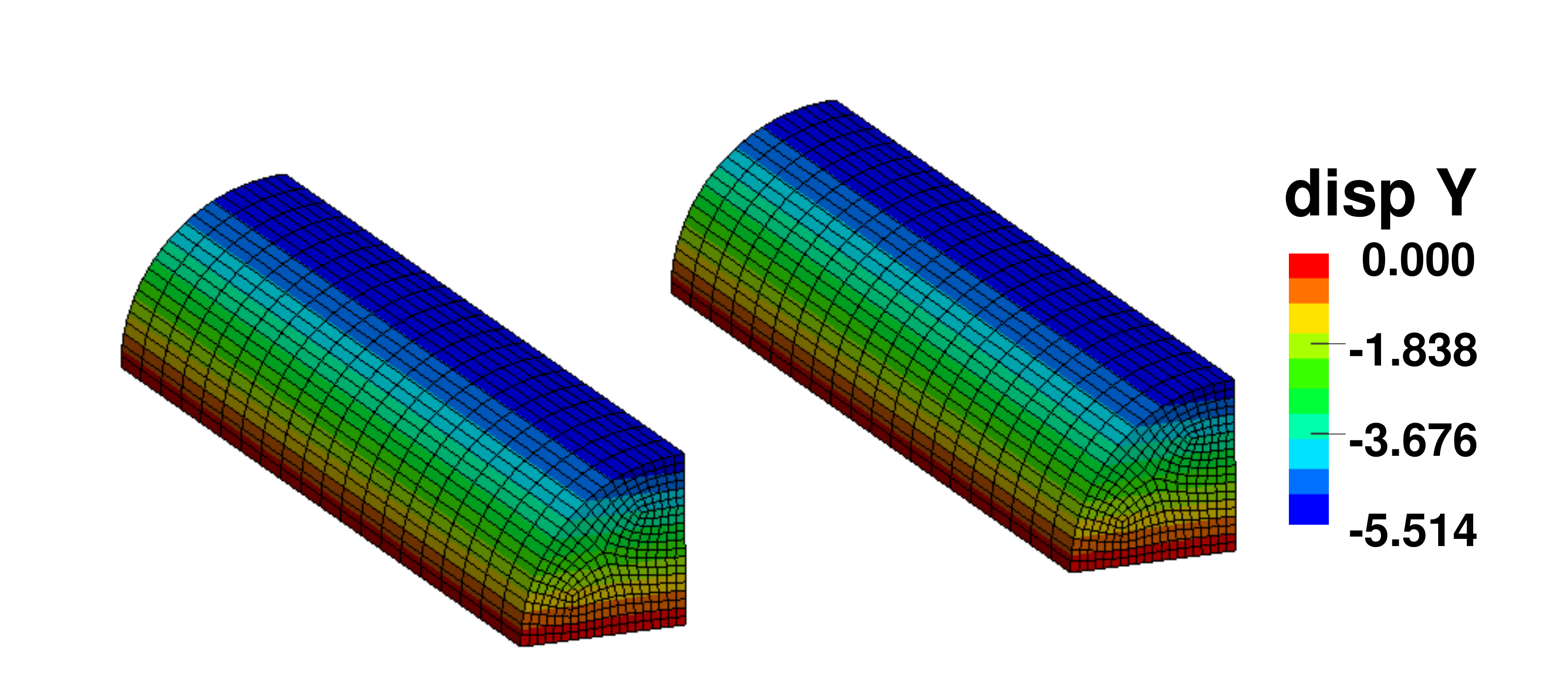} \\
\includegraphics[clip, scale=0.35]{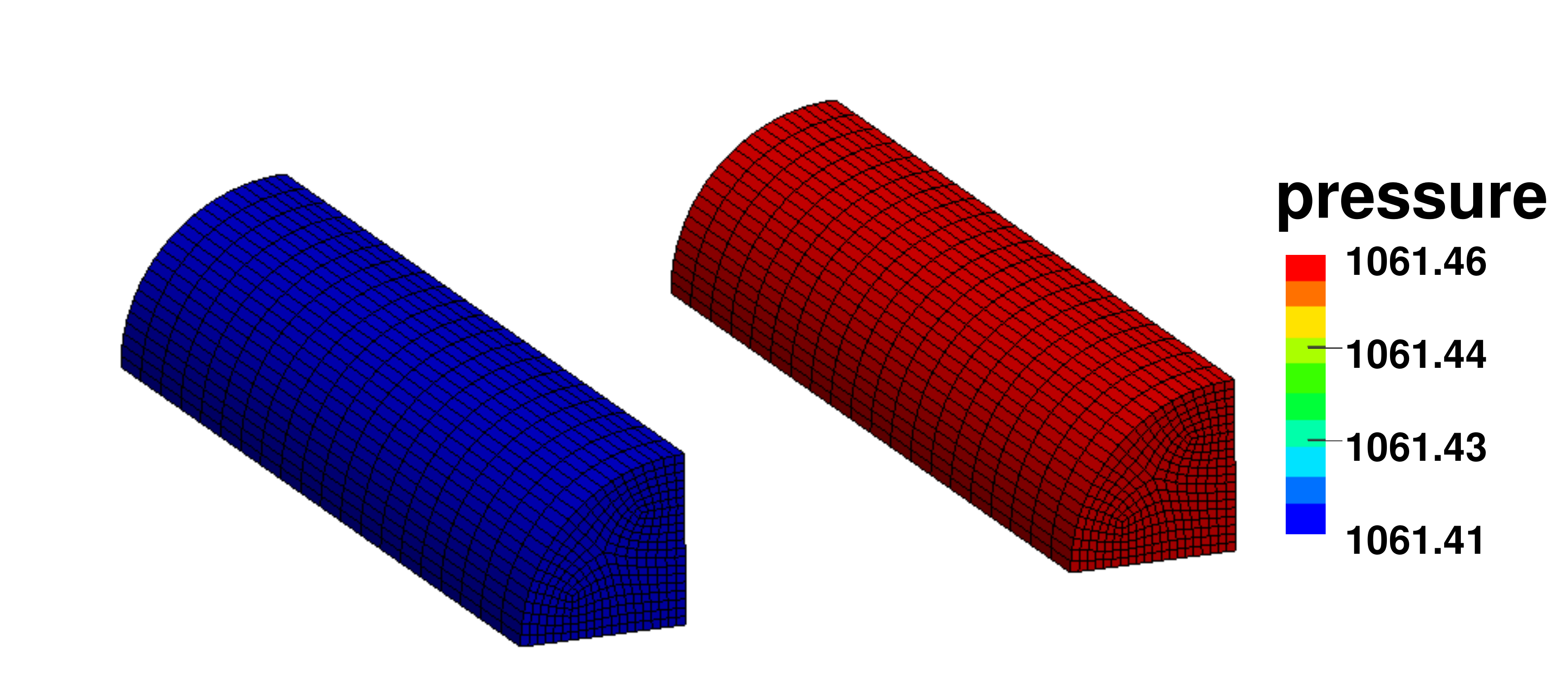}
\caption{\mycolor{Cylindrical bar with Q1/P0 element: contour plots of Y-displacement and pressure for $\nu=0.4999$. In each figure: left) the standard three-field formulation and right) the proposed two-field formulation.}}
\label{fig-hex8-contours-nu0p4999}
\end{figure}

\subsection{A block under a compressive load}
In this example, we demonstrate the accuracy of the proposed formulation against the three-field formulation using the widely-used benchmark example of a block under a compressible load \cite{KadapaPhDThesis, KadapaIJNME2019mixed, ReeseIJNME1999}. The geometry and boundary conditions of the problem are as shown in Fig. \ref{fig-block-geom}. Following Reese et al. \cite{ReeseIJNME1999}, the material model is assumed to be the Neo-Hookean model whose strain energy density function is given as
\begin{align}
\Psi = \frac{\mu}{2} \, \left[ I_{\overline{\bm{C}}} - 3 \right] + \frac{\kappa}{2} \, \left[ \ln(J) \right]^2.
\end{align}
The Young's modulus is, $E=240.565$ and the numerical study is carried out for three different values of Poisson's ratio, $\nu=\{ 0.3, 0.45, 0.4999 \}$ using four successively-refined finite element meshes as shown in Fig. \ref{fig-block-mesh}. The uniform pressure value applied on the highlighted area of the top surface (see Fig. \ref{fig-block-geom}) in the negative Z direction is $p/p_{0}=80$ with $p_0=20$ units. The load is assumed to be independent of the deformation and is applied in four uniform increments.

The Z-displacement of point A, referred in the literature as compression level, obtained with the three-field and the proposed two-field formulations at various load levels is presented in Fig. \ref{fig-block-graph-nu0p4999} together with those obtained with three other schemes, Q1SP \cite{ReeseIJNME1999}, H8MSGSO \cite{KryslIJNME2015} and $\bm{F}$-bar \cite{NetoIJSS1996}. As shown, the results obtained with the proposed two-field formulation are in excellent agreement with the solutions obtained with other schemes. The compression level obtained for different meshes, and at different load levels for each mesh, is presented in Figs. \ref{fig-block-graph-nu0p3} and \ref{fig-block-graph-nu0p45}, respectively, for $\nu=0.3$ and $\nu=0.45$. From Figs. \ref{fig-block-graph-nu0p4999}, \ref{fig-block-graph-nu0p3} and \ref{fig-block-graph-nu0p45}, it is evident that the results obtained with the proposed two-field formulation are indistinguishable from those of the three-field formulation. The final deformed shapes of the block along with the element-wise contour plots of the pressure are presented in Figs. \ref{fig-block-contours-nu0p3}, \ref{fig-block-contours-nu0p45} and \ref{fig-block-contours-nu0p4999}, respectively, for $\nu=0.3$, $\nu=0.45$ and $\nu=0.4999$. These contour plots illustrate that there is a negligible difference in the values of pressure obtained with the proposed two-field formulation and the conventional three-field formulation.
\begin{figure}[H]
\centering
\subfloat[]{\includegraphics[clip, scale=0.8]{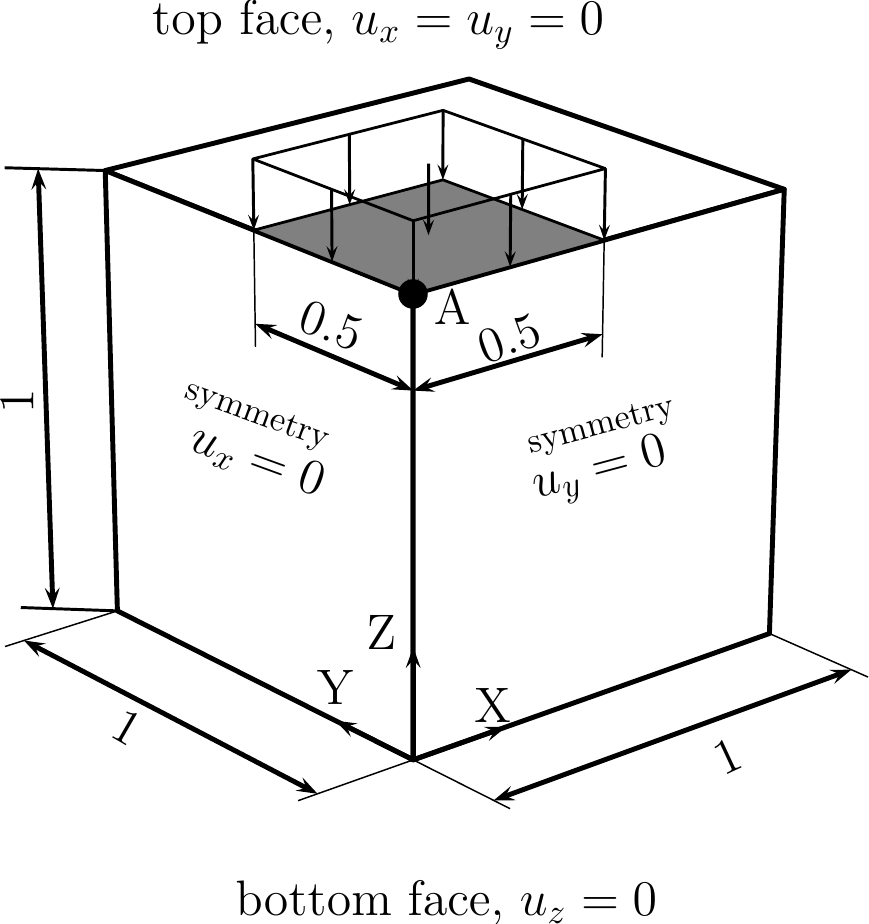} \label{fig-block-geom}
}
\subfloat[]{\includegraphics[clip, scale=0.6]{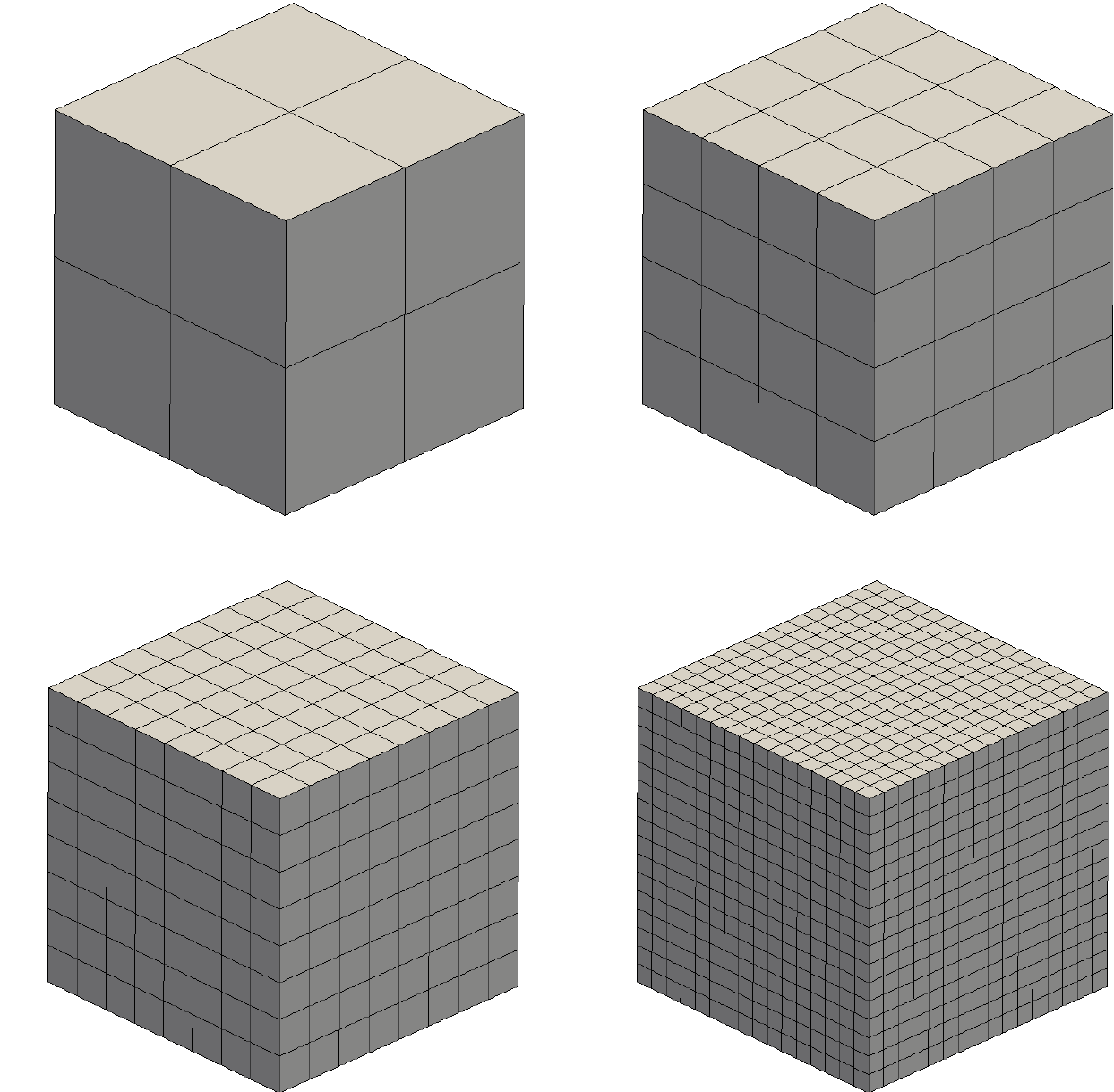} \label{fig-block-mesh}
}
\caption{\mycolor{3D block: (a) geometry and boundary conditions and, (b) finite element meshes used.}}
\end{figure}
\begin{figure}[H]
\centering
\includegraphics[trim=0mm 0mm 0mm 0mm, clip, scale=0.9]{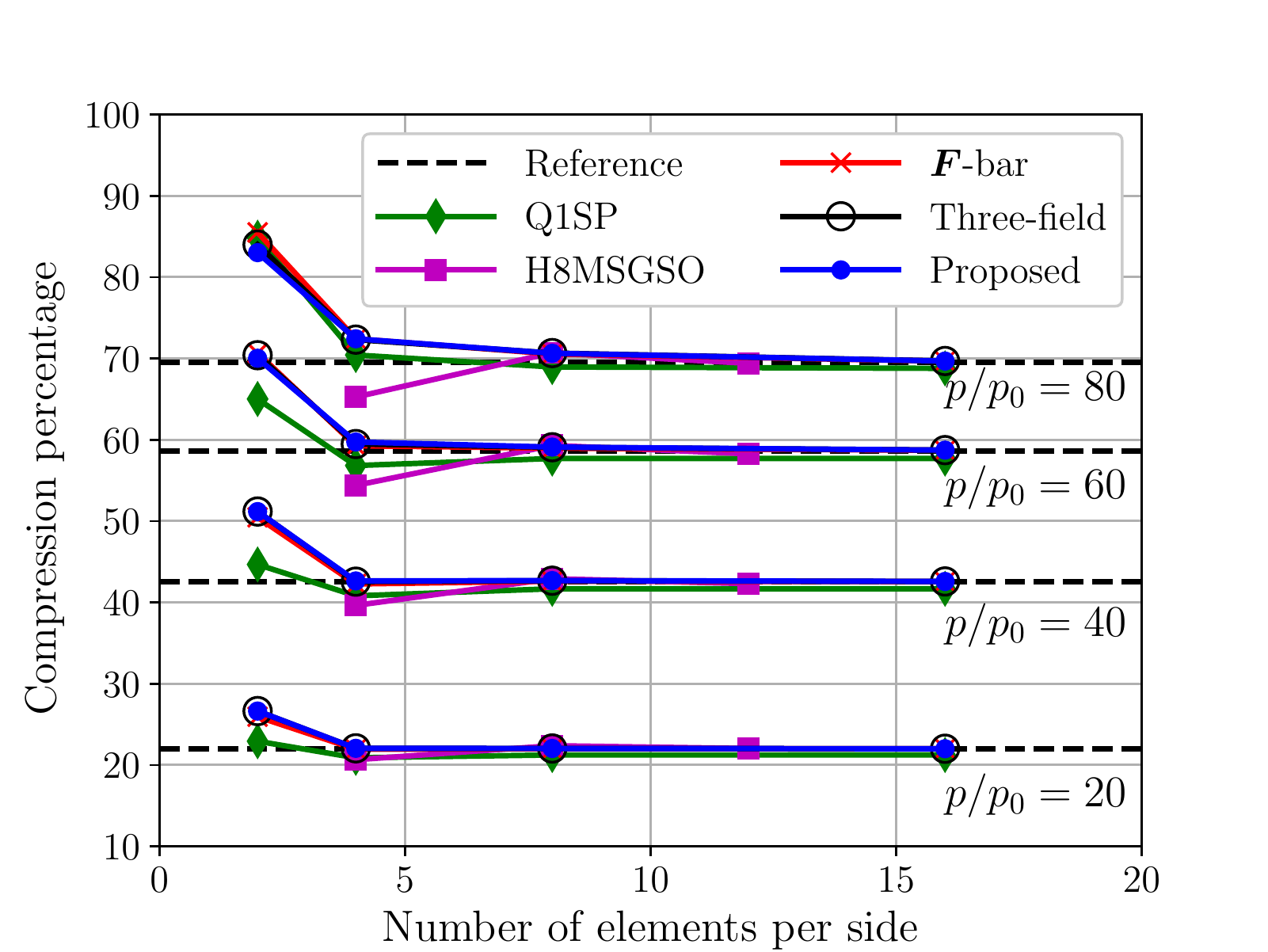}
\caption{\mycolor{3D block: comparison of compression level for different loading conditions for $\nu=0.4999$. The reference solution is taken from \cite{KadapaIJNME2019mixed}.}}
\label{fig-block-graph-nu0p4999}
\end{figure}
\begin{figure}[H]
\centering
\subfloat[$\nu=0.3$]{\includegraphics[trim=0mm 0mm 0mm 0mm, clip, scale=0.5]{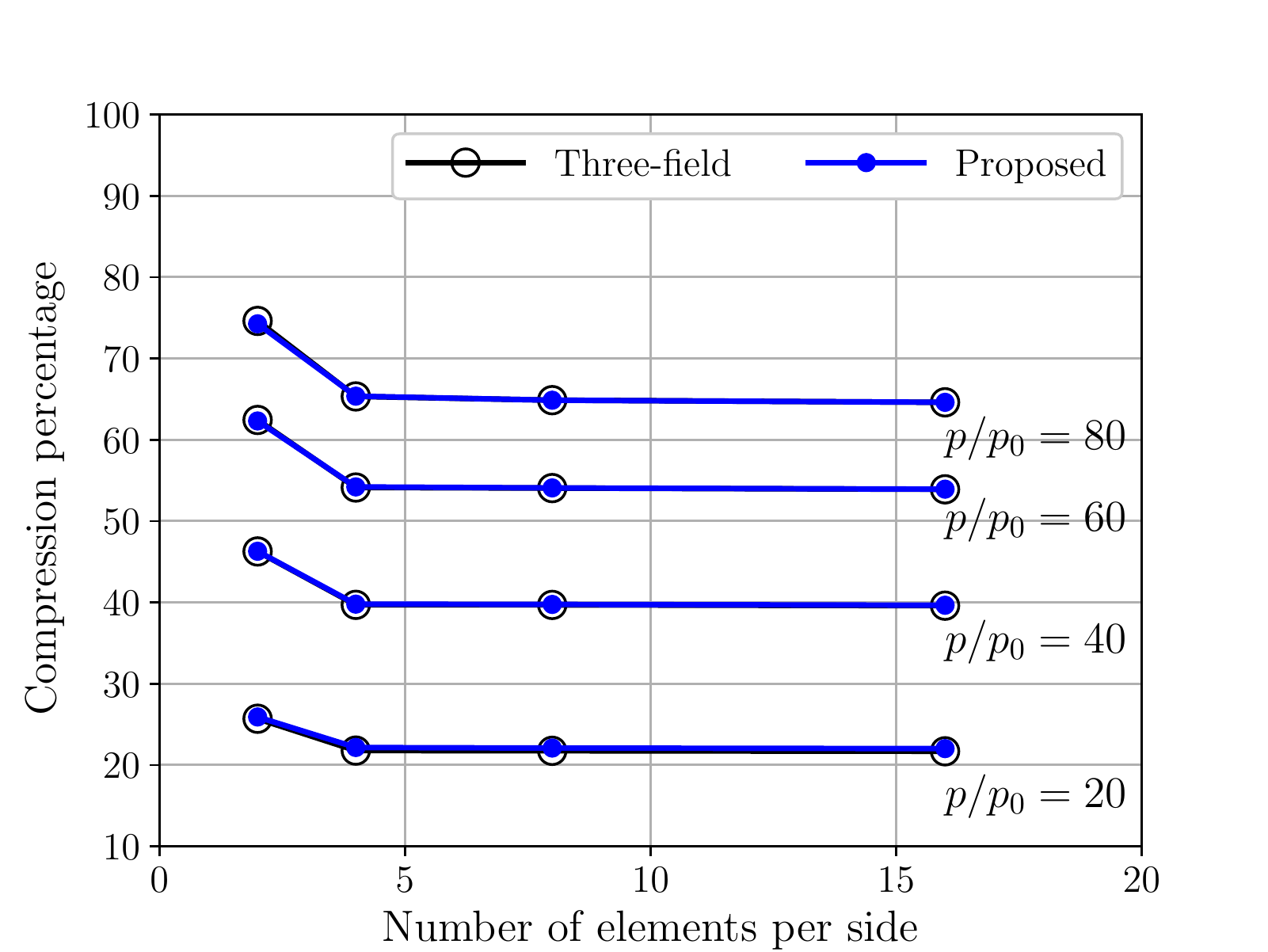} \label{fig-block-graph-nu0p3}}
\subfloat[$\nu=0.45$]{\includegraphics[trim=0mm 0mm 0mm 0mm, clip, scale=0.5]{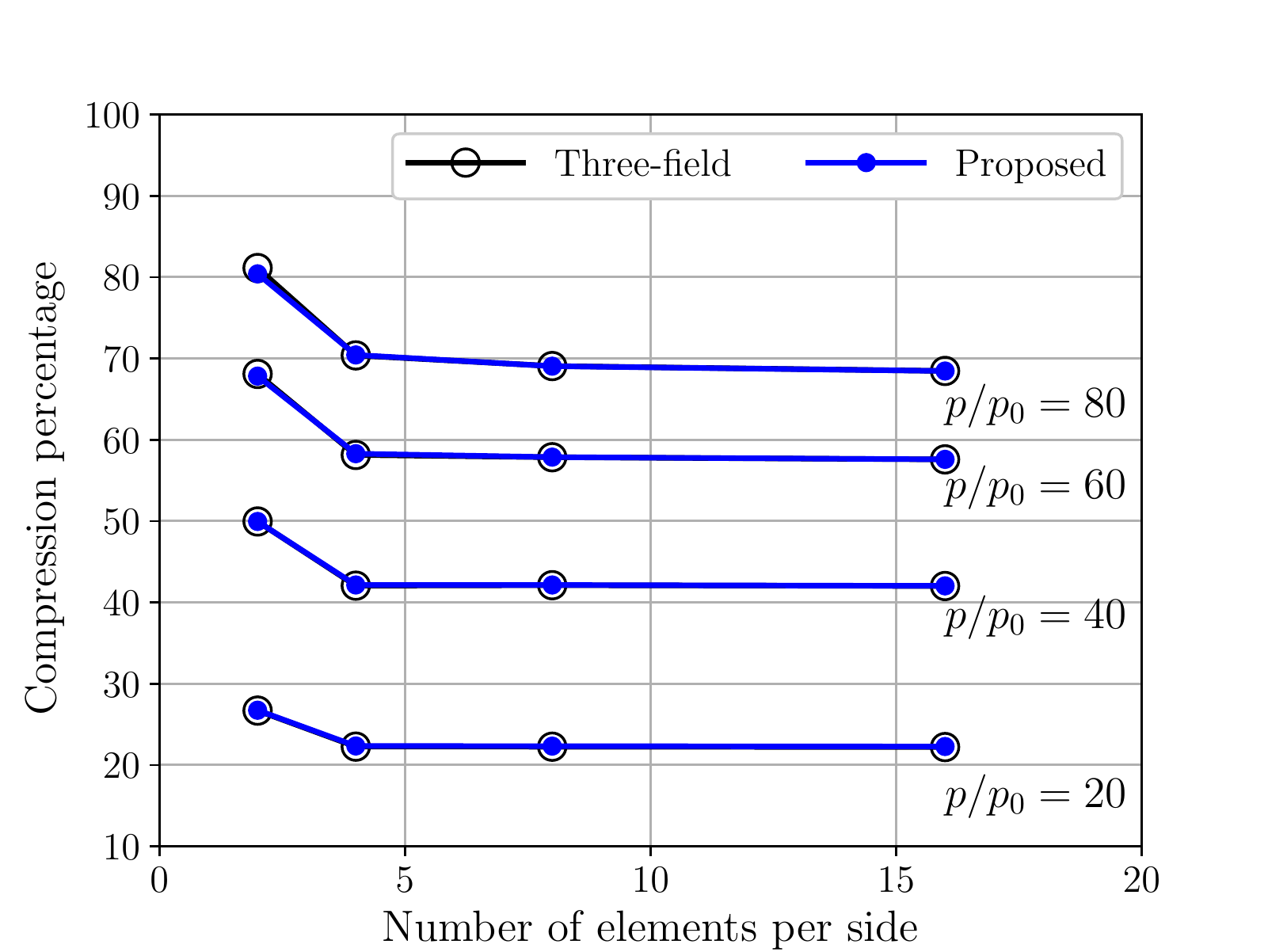} \label{fig-block-graph-nu0p45}}
\caption{\mycolor{3D block: comparison of compression level for different loading conditions for (a) $\nu=0.3$ and (b) $\nu=0.45$.}}
\label{fig-block-graph-nu0p3-nu0p45}
\end{figure}
\begin{figure}[H]
\centering
\includegraphics[trim=0mm 0mm 70mm 0mm, clip, scale=0.3]{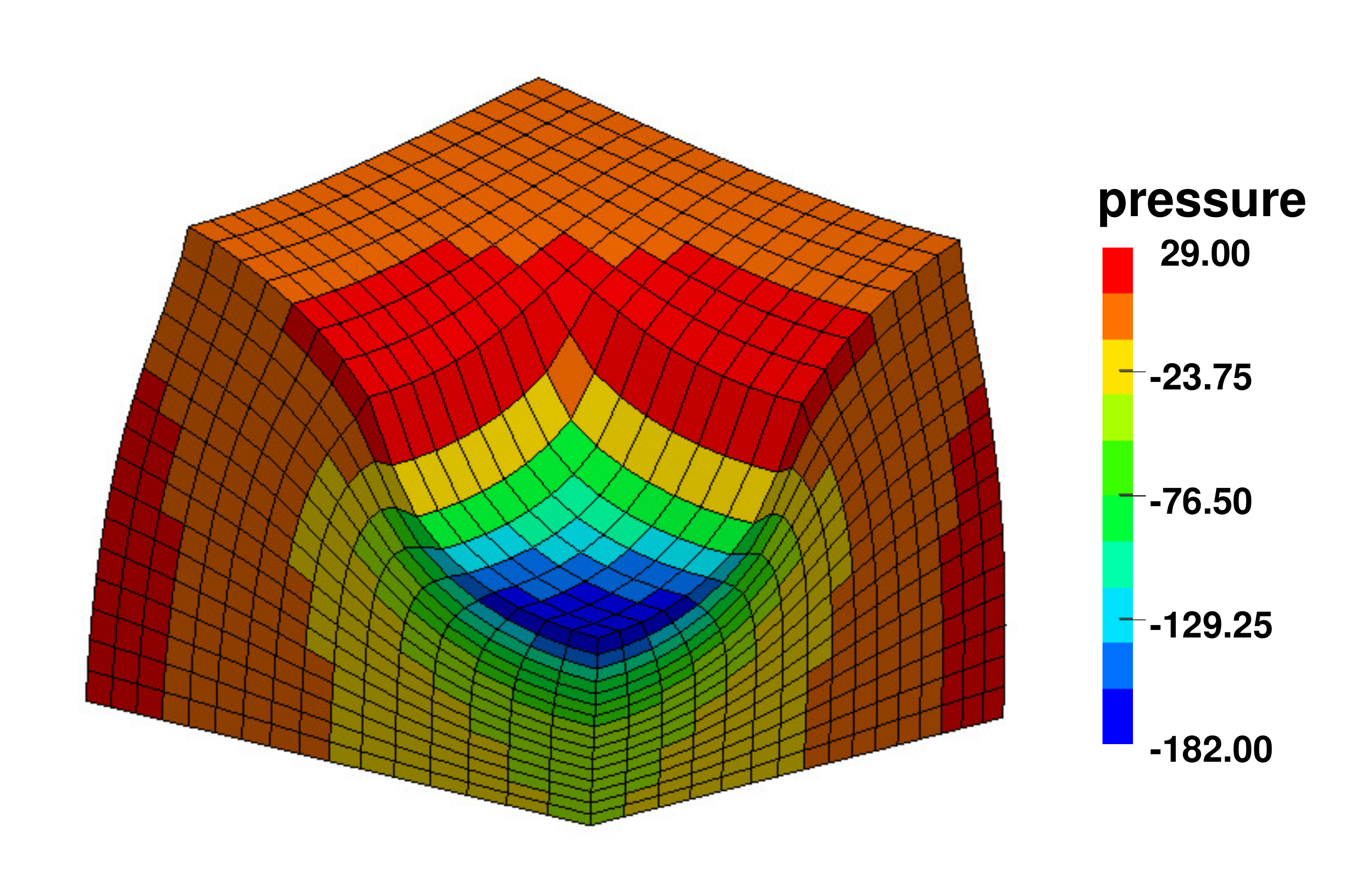}
\includegraphics[clip, scale=0.3]{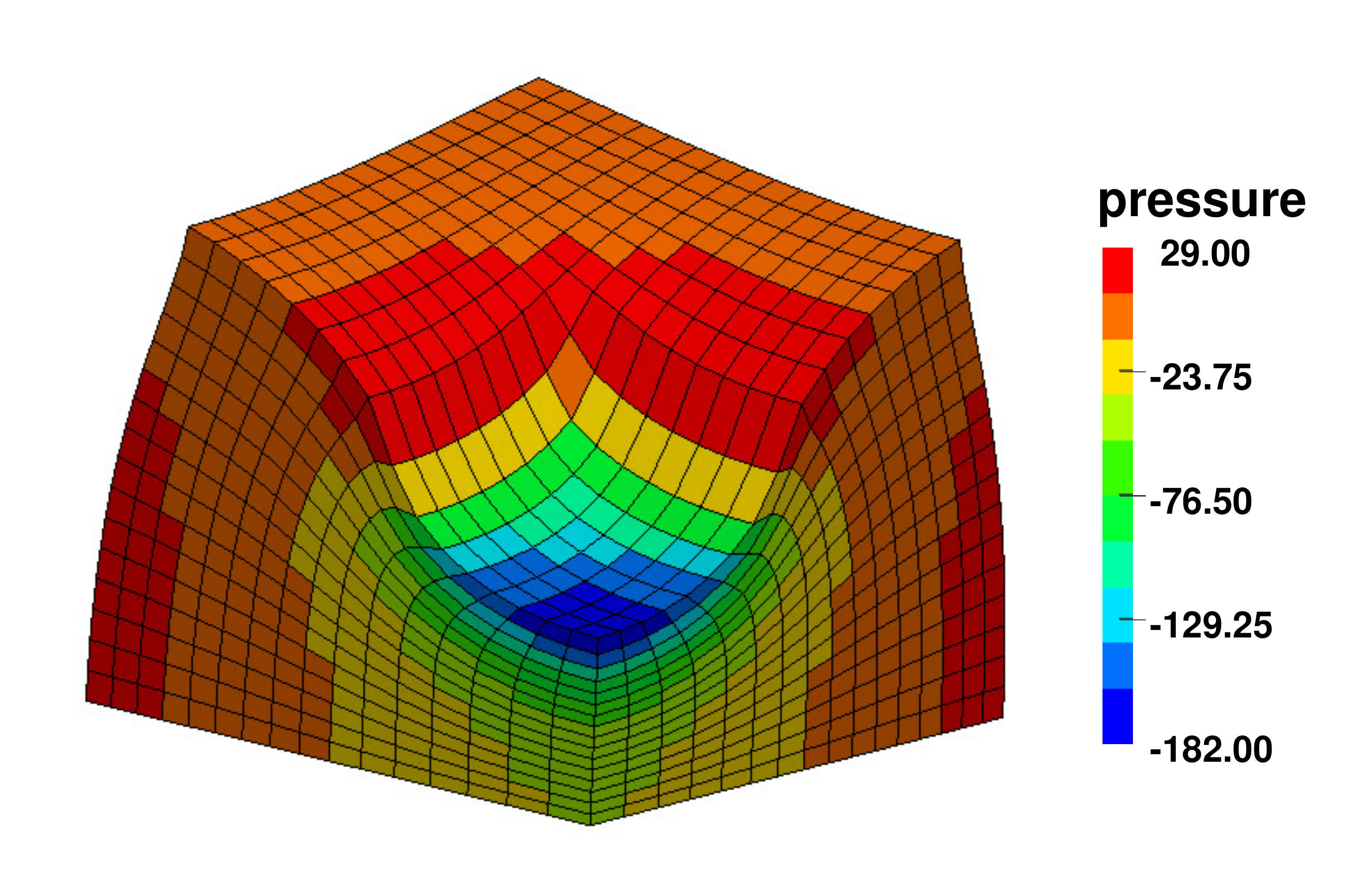}
\caption{\mycolor{3D block: contour plots of element-wise pressure for $\nu=0.3$. Left) three-field formulation and, right) proposed formulation.}}
\label{fig-block-contours-nu0p3}
\end{figure}
\begin{figure}[H]
\centering
\includegraphics[trim=0mm 0mm 70mm 0mm, clip, scale=0.3]{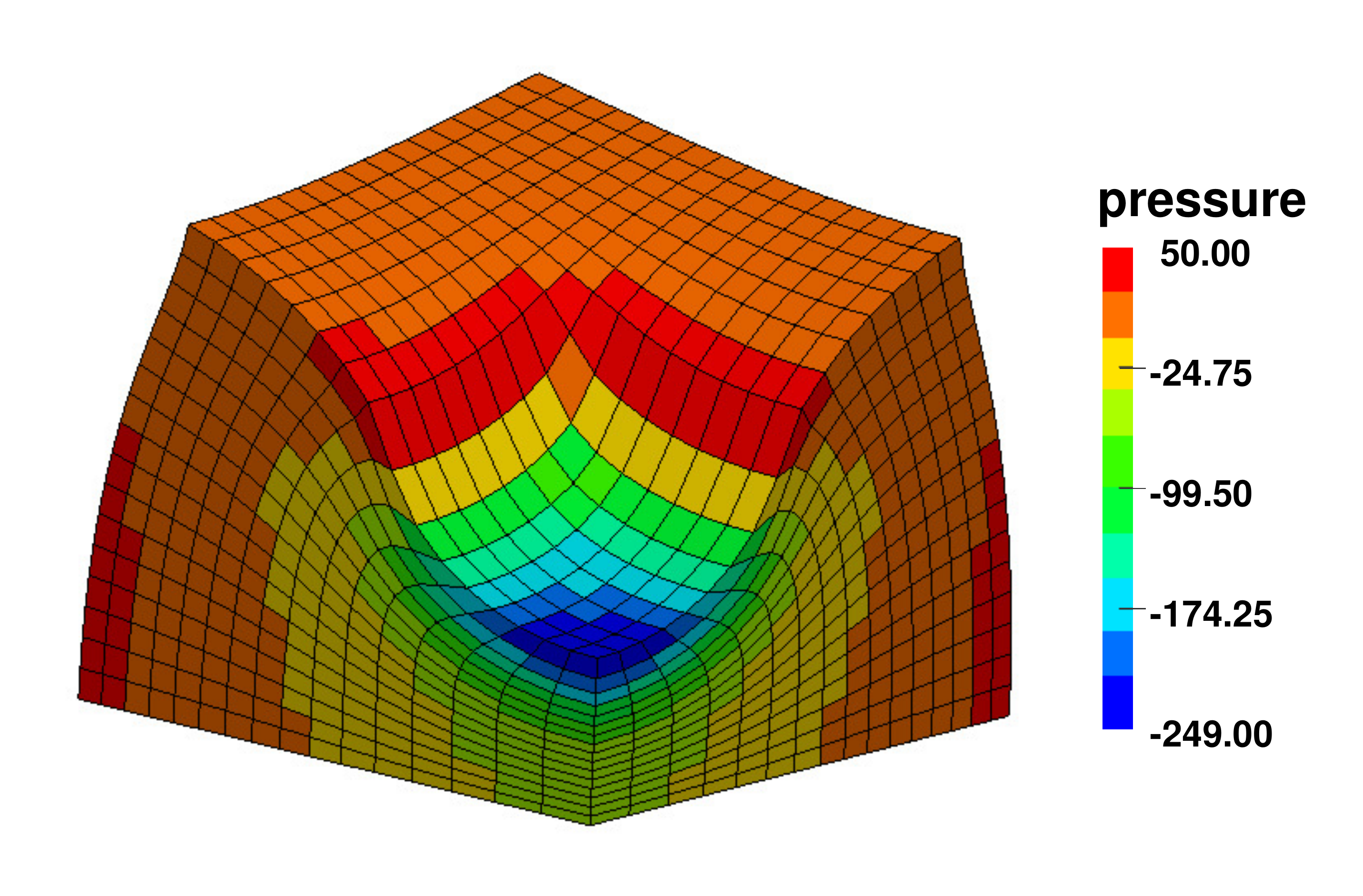}
\includegraphics[clip, scale=0.3]{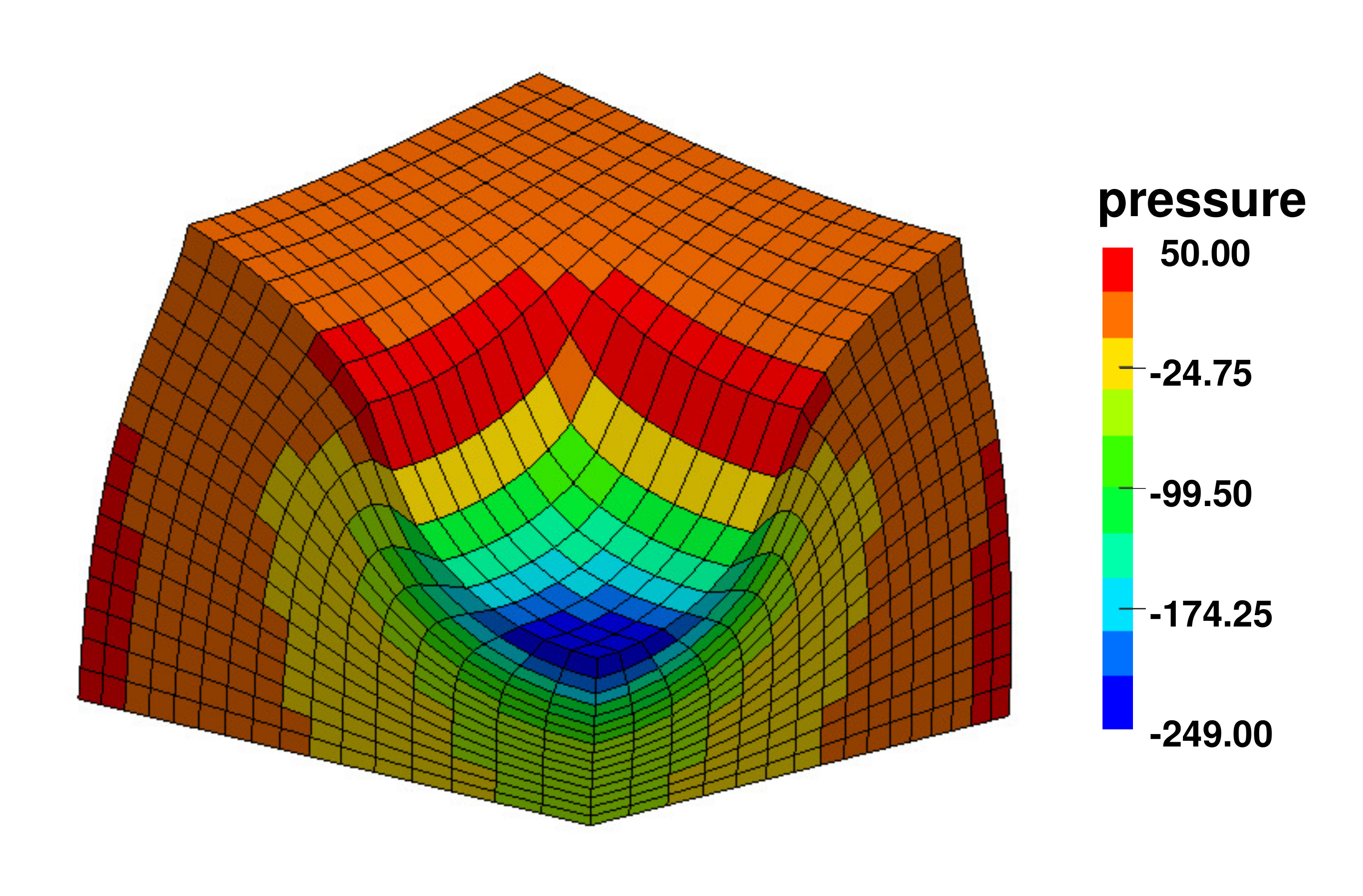}
\caption{\mycolor{3D block: contour plots of element-wise pressure for $\nu=0.45$. Left) three-field formulation and, right) proposed formulation.}}
\label{fig-block-contours-nu0p45}
\end{figure}
\begin{figure}[H]
\centering
\includegraphics[trim=0mm 0mm 70mm 0mm, clip, scale=0.3]{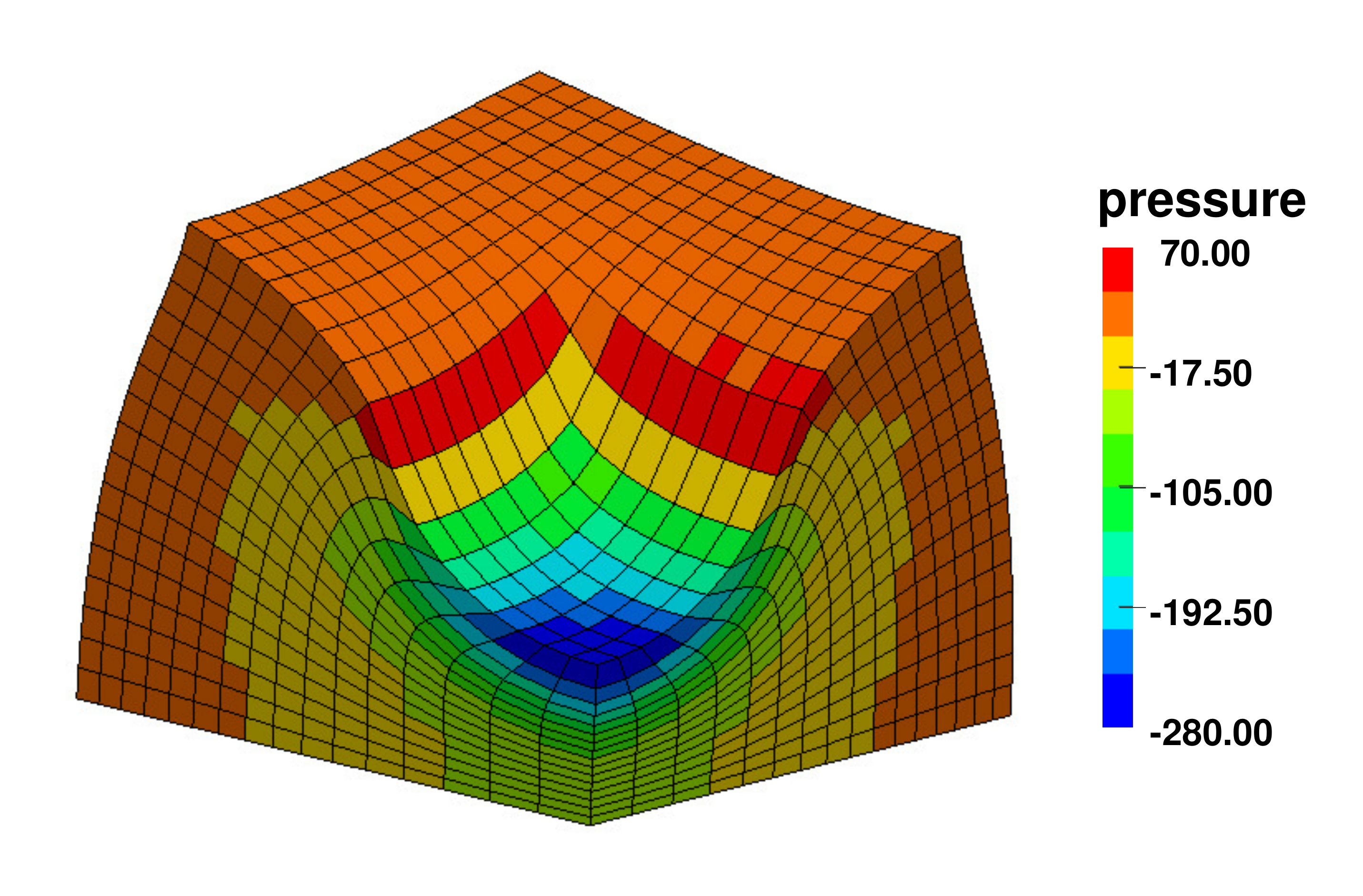}
\includegraphics[clip, scale=0.3]{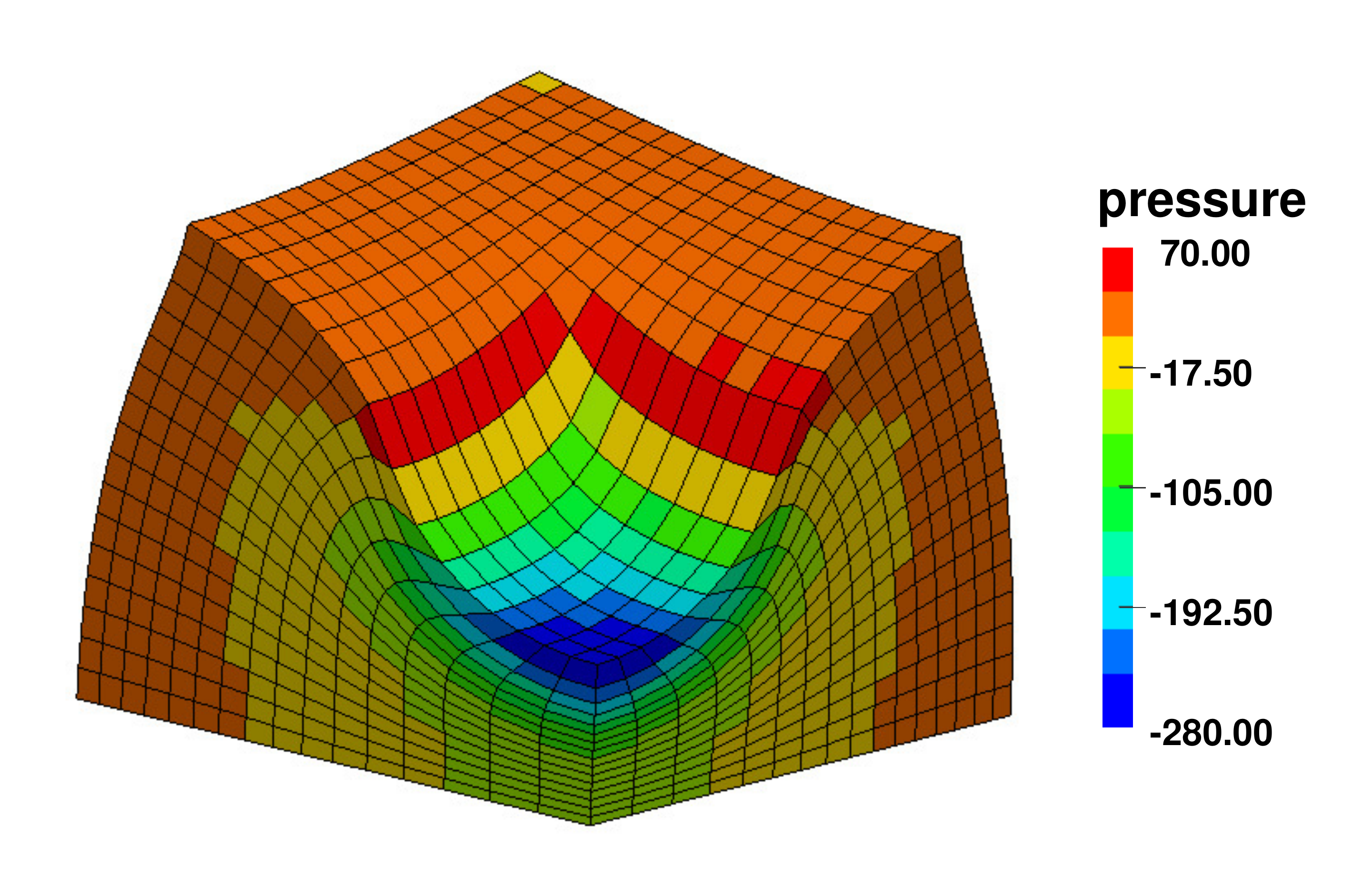}
\caption{\mycolor{3D block: contour plots of element-wise pressure for $\nu=0.4999$. Left) three-field formulation and, right) proposed formulation.}}
\label{fig-block-contours-nu0p4999}
\end{figure}

Therefore, numerical results that are in excellent agreement with the three-field formulation can be obtained with the proposed two-field formulation. This proves that there is no computational advantage in using the three-field formulation for hyperelastic materials whose strain energy function is decomposed into deviatoric and volumetric parts, because of the fact that the coupling terms $\mathbf{K}_{\bm{u}\theta}$ and $\mathbf{K}_{\theta \bm{u}}$ vanish for such material models. Moreover, the proposed formulation results in a simplified computer implementation when compared with the three-field formulation which involves evaluation of complicated expressions of material tensors in equations (\ref{eqn-3field-D11}) and (\ref{eqn-3field-D12}) at every integration of every element; such computations are completely avoided in the proposed formulation. Fewer DOFs and the reduced number of computations at every integration point make the proposed formulation computationally efficient relative to the three-field formulation.
}

\section{Summary and conclusions} \label{section-summary}
In this contribution, we have presented a generalised two-field mixed displacement-pressure formulation that not only is applicable for simulating the truly incompressible materials but also is consistent in the compressible regime. The proposed formulation yields symmetric matrix systems irrespective of the volumetric part of the energy function, and it does not require complementary functions.

\mycolor{
By recasting the three-field formulation as a two-field formulation, it is shown that the proposed two-field formulation is approximately equivalent to the three-field formulation. First, the accuracy of the novel mixed formulation is demonstrated using the LBB-stable BT2/BT1 element with the example of a cylindrical bar for three different values of Poisson's ratio, encompassing the compressible and the nearly incompressible regime. Later, with the widely-used Q1/P0 discretisation, the ability of the proposed two-field formulation in computing numerical results that are in excellent agreement with the three-field mixed displacement-pressure-Jacobian formulation is demonstrated.}

From the presented numerical results, it can be concluded that the proposed two-field mixed displacement-pressure formulation is a simplified and efficient alternative for the three-field formulation for hyperelastic material models whose strain energy functions are decomposed into deviatoric and volumetric parts. The fact that it is also applicable for the truly incompressible materials makes it a unified formulation for compressible as well as incompressible hyperelastic constitutive models and their extensions to thermoelasticity, viscoelasticity, electromechanics and magnetomechanics.

It is important to emphasize that the proposed formulation is not limited to the Q1/P0 discretisation. This novel formulation can be used with any combination of finite element spaces for the displacement and pressure fields, as demonstrated with the BT2/BT1 element in this paper. For shell problems that experience shear-locking issues, the proposed formulation can be used with higher-order discretisations, for example, BT2/BT0 and BT2/BT1 elements as presented in \cite{KadapaIJNME2019mixed}.

\section*{ACKNOWLEDGEMENTS}
The first author acknowledges the support of the Supercomputing Wales project, which is part-funded by the European Regional Development Fund (ERDF) via the Welsh Government. \mycolor{The authors are indebted to an anonymous reviewer for the useful insights into the idea of complementary energy functions presented in Appendix A.}

\renewcommand{\theequation}{A.\arabic{equation}}
\setcounter{equation}{0}
\myred{
\section*{Appendix A: Mixed displacement-pressure formulation using complementary energy functions}
With displacement, $\bm{u}$; Jacobian (or the determinant of the deformation gradient), $J$; and pressure, $p$, as independent variables, the three-field Hu-Washizu type energy functional (see \cite{book-fem-ZienkiewiczVol2}) can be written as,
\mycolor{
\begin{equation} \label{eqn-Pi-1}
\Pi_{\text{HW}}(\bm{u},p,\bar{J})
= \int_{\mathcal{B}_{0}} \left[ \PsiDev(\widehat{\bm{C}}) + p \, J + \underbrace{\PsiVol(\bar{J}) - p \, \bar{J}}_{\Gamma(p)} \right] \dV - \Pi_{\mathrm{ext}}
\end{equation}
}
where,
\begin{equation}
\widehat{\bm{C}} = \widehat{\bm{F}}^{\T} \, \widehat{\bm{F}} \quad \text{and} \quad
\bar{J} = \mathrm{det}(\widehat{\bm{F}}) \qquad \text{with} \quad
\widehat{\bm{F}} = \left( \frac{\bar{J}}{J}\right)^{1/3} \bm{F}.
\end{equation}

$\Gamma(p)$ in Eq. (\ref{eqn-Pi-1}) is called as the complementary energy functional (or the dual thermodynamical potential) to the volumetric potential $\PsiVol(\bar{J})$ \cite{OrtigosaCMAME2016}. For the given $\PsiVol(\bar{J})$, $\Gamma(p)$ is obtained by employing a Legendre transformation between the conjugates $\bar{J}$ and $p$, as given by
\begin{align} \label{eqn-argmin}
\Gamma(p) = \inf_{\bar{J}>0} \, \{ \PsiVol(\bar{J}) - p \, \bar{J} \}.
\end{align}

Using $\Gamma(p)$, the energy functional for the displacement-pressure formulation is given by
\begin{equation} \label{eqn-Pi-up-compl}
\Pi_{\text{CompPot}}(\bm{u},p)
= \int_{\mathcal{B}_{0}} \left[ \PsiDev(\overline{\bm{C}}) + p \, J + \Gamma(p) \right] \dV - \Pi_{\mathrm{ext}}.
\end{equation}

The following relations hold the for the variables $J$ and $p$, and energy potentials $\PsiVol$ and $\Gamma$.
\begin{align} \label{eqn-defs-pJ}
p = \pderiv{\PsiVol}{J}; \qquad \text{and} \qquad  J = - \, \pderiv{\Gamma(p)}{p}.
\end{align}

The procedure described up to this point illustrates that  it is possible, with limited scopes, to obtain a two field formulation in terms of displacements and pressure departing from  a three field mixed formulation. However, the process has several shortcomings. The crucial step involved in the derivation of mixed formulations using complementary energy functions is the evaluation of complementary energy functions themselves. This is because of the fact that the volumetric energy functions, but not their duals, are the de-facto energy functions developed in constitutive modelling for hyperelastic materials, see Hartmann and Neff \cite{HartmannIJSS2003} and Moermann et al. \cite{Moerman2019} for  details.

For the case of $\Psi^{vol}_3 = \frac{\kappa}{2} [J-1]^2$, we may obtain
\begin{subequations}
\begin{align}
p &= \kappa \, [J-1],  \label{eqn-relation-pJ} \\
J(p) &= 1 + p/\kappa,  \label{eqn-relation-Jp} \\
\Gamma_3(p) &= - p \left[ 1 + \frac{p}{2 \, \kappa} \right].  \label{eqn-relation-Gp}
\end{align}
\end{subequations}

While the closed-form expressions for the dual potentials might be  derived for a few simple volumetric functions as discussed above, such derivations are not possible for complicated volumetric energy functions encountered in practice, for example, $\PsiVol_1$, $\PsiVol_2$, $\PsiVol_5$ and the ones proposed recently in Moerman et al. \cite{Moerman2019}. For a complicated volumetric function, the dual potential and its derivatives with respect to pressure must be obtained numerically using an iterative technique, for example, a Newton-Raphson scheme. This, of course, is not straightforward due to the difficulties associated with Newton-Raphson scheme in finding the correct root for the cases in which the relations (\ref{eqn-defs-pJ}) may have multiple roots. Moreover, there are numerous volumetric energy functions available in the literature. Hence, the development of a computational framework by analysing energy functions case by case is obviously not elegant. Therefore, albeit  displacement-pressure mixed formulation based on the dual potentials yields symmetric matrix systems irrespective of the dual potential, the difficulties associated with the computation of the dual potentials and their derivatives further warrant the development of a generalised formulation that avoids complementary potentials altogether. The  formulation proposed herein is one such an approach.
}


\begin{thebibliography}{10}

\bibitem{book-fem-BrezziFortin}
F.~Brezzi and M.~Fortin.
\newblock {\em {Mixed and Hybrid Finite Element Methods}}.
\newblock Springer-Verlag, 1991.

\bibitem{book-fem-ZienkiewiczVol2}
O.~C. Zienkiewicz and R.~L. Taylor.
\newblock {\em {The Finite Element Method for Solid and Structural Mechanics}}.
\newblock Elsevier Butterworth and Heinemann, Oxford, England, {Sixth} edition,
  2005.

\bibitem{KadapaPhDThesis}
C.~Kadapa.
\newblock {\em {Mixed Galerkin and least-squares formulations for isogeometric
  analysis}}.
\newblock PhD thesis, College of Engineering, Swansea University, 2014.

\bibitem{KadapaCMAME2016elast}
C.~Kadapa, W.~G. Dettmer, and D.~Peri\'c.
\newblock {Subdivision based mixed methods for isogeometric analysis of linear
  and nonlinear nearly incompressible materials}.
\newblock {\em Computer Methods in Applied Mechanics and Engineering},
  305:241--270, 2016.

\bibitem{KadapaIJNME2019mixed}
C.~Kadapa.
\newblock {Novel quadratic B\'ezier triangular and tetrahedral elements using
  existing mesh generators: Extension to nearly incompressible implicit and
  explicit elastodynamics in finite strains}.
\newblock {\em International Journal for Numerical Methods in Engineering},
  119:75--104, 2019.

\bibitem{NetoIJSS1996}
E.~A. {de Souza Neto}, D.~Peri\'c, M.~Dutko, and D.~R.~J. Owen.
\newblock {Design of simple low order finite elements for large strain analysis
  of nearly incompressible solids}.
\newblock {\em International Journal of Solids Structures}, 33:3277--3296,
  1996.

\bibitem{NetoIJNME2005Fbarpatch}
E.~A. {de Souza Neto}, F.~M. Andrade~Pires, and D.~R.~J. Owen.
\newblock {F-bar-based linear triangles and tetrahedra for finite strain
  analysis of nearly incompressible solids. Part I: formulation and
  benchmarking}.
\newblock {\em International Journal for Numerical Methods in Engineering},
  62:353--383, 2005.

\bibitem{SimoJAM1986}
J.~C. Simo and T.~J.~R. Hughes.
\newblock {On the variational foundations of assumed strain methods}.
\newblock {\em Journal of Applied Mechanics}, 53:51--54, 1986.

\bibitem{BonetCNME1998}
J.~Bonet and A.~J. Burton.
\newblock {A simple average nodal pressure tetrahedral element for
  incompressible and nearly incompressible dynamic explicit applications}.
\newblock {\em Communications in Numerical Methods in Engineering},
  14:437--449, 1998.

\bibitem{PiresCNME2004}
F.~M.~A. Pires, E.~A. de~Souza~Neto, and J.~L. de~la Cuesta~Padilla.
\newblock {An assessment of the average nodal volume formulation for the
  analysis of nearly incompressible solids under finite strains}.
\newblock {\em Communications in Numerical Methods in Engineering},
  20:569--583, 2004.

\bibitem{FlanaganIJNME1981}
D.~P. Flanagan and T.~Belytschko.
\newblock {A uniform strain hexahedron and quadrilateral with orthogonal
  hourglass control}.
\newblock {\em International Journal for Numerical Methods in Engineering},
  17:679--706, 1981.

\bibitem{BelytschkoCMAME1984hglass}
T.~Belytschko, J.~S. Ong, W.~K. Liu, and J.~M. Kennedy.
\newblock {Hourglass control in linear and nonlinear problems}.
\newblock {\em Computer Methods in Applied Mechanics and Engineering},
  43:251--276, 1984.

\bibitem{WriggersCM1996}
P.~Wriggers and J.~Korelc.
\newblock {On enhanced strain methods for small and finite deformations}.
\newblock {\em Computational Mechanics}, 18(6):413--428, 1996.

\bibitem{KorelcCM2010}
J.~Korelc, U.~$\check{S}$olinc, and P.~Wriggers.
\newblock {An improved EAS brick element for finite deformation}.
\newblock {\em Computational Mechanics}, 46:641--659, 2010.

\bibitem{ChenCMAME1997}
J.~S. Chen, W.~Han, C.~T. Wu, and W.~Duan.
\newblock {On the perturbed Lagrangian formulation for nearly incompressible
  and incompressible hyperelasticity}.
\newblock {\em Computer Methods in Applied Mechanics and Engineering},
  142:335--351, 1997.

\bibitem{SimoCMAME1985}
J.~C. Simo, R.~L. Taylor, and K.~S. Pister.
\newblock {Variational and projection methods for the volume constraint in
  finite deformation elasto-plasticity}.
\newblock {\em Computer Methods in Applied Mechanics and Engineering},
  51:177--208, 1985.

\bibitem{ChiumentiCMAME2002}
M.~Chiumenti, Q.~Valverde, C.~A. de~Saracibar, and M.~Cervera.
\newblock {A stabilized formulation for incompressible elasticity using linear
  displacement and pressure interpolations}.
\newblock {\em Computer Methods in Applied Mechanics and Engineering},
  191:5253--5264, 2002.

\bibitem{CerveraCMAME2010a}
M.~Cervera, M.~Chiumenti, and R.~Codina.
\newblock {Mixed stabilized finite element methods in nonlinear solid
  mechanics. Part I: formulation}.
\newblock {\em Computer Methods in Applied Mechanics and Engineering},
  199:2559--2570, 2010.

\bibitem{CerveraCMAME2010b}
M.~Cervera, M.~Chiumenti, and R.~Codina.
\newblock {Mixed stabilized finite element methods in nonlinear solid
  mechanics. Part II: strain localization}.
\newblock {\em Computer Methods in Applied Mechanics and Engineering},
  199:2571--2589, 2010.

\bibitem{ScovazziIJNME2016}
G.~Scovazzi, B.~Carnes, X.~Zeng, and S.~Rossi.
\newblock {A simple, stable, and accurate linear tetrahedral finite element for
  transient, nearly, and fully incompressible solid dynamics: a dynamic
  variational multiscale approach}.
\newblock {\em International Journal for Numerical Methods in Engineering},
  106:799--839, 2016.

\bibitem{ScovazziCMAME2017velocity}
G.~Scovazzi, T.~Song, and X.~Zeng.
\newblock {A velocity/stress mixed stabilized nodal finite element for
  elastodynamics: Analysis and computations with strongly and weakly enforced
  boundary conditions}.
\newblock {\em Computer Methods in Applied Mechanics and Engineering},
  325:532--576, 2017.

\bibitem{AbboudIJNME2018}
N.~Abboud and G.~Scovazzi.
\newblock {Elastoplasticity with linear tetrahedral elements: A variational
  multiscale method}.
\newblock {\em International Journal for Numerical Methods in Engineering},
  115:913--955, 2018.

\bibitem{FrancaNM1988}
L.~P. Franca, T.~J.~R. Hughes, A.~F.~D. Loula, and I.~Miranda.
\newblock {A new family of stable elements for nearly incompressible elasticity
  based on a mixed Petrov-Galerkin finite element formulation}.
\newblock {\em Numerische Mathematik}, 53:123--141, 1988.

\bibitem{KlassCMAME1999}
O.~Klaas, A.~Maniatty, and M.~S. Shephard.
\newblock {A stabilized mixed finite element method for finite elasticity.
  Formulation for linear displacement and pressure interpolation}.
\newblock {\em Computer Methods in Applied Mechanics and Engineering},
  180:65--79, 1999.

\bibitem{MasudJAM2005}
A.~Masud and K.~Xia.
\newblock {A stabilized mixed finite element method for nearly incompressible
  elasticity}.
\newblock {\em Journal of Applied Mechanics}, 72:711--720, 2005.

\bibitem{PakravanIJNME2017b}
A.~Pakravan and P.~Krysl.
\newblock {Mean-strain 10-node tetrahedron with energy-sampling stabilization
  for nonlinear deformation}.
\newblock {\em International Journal for Numerical Methods in Engineering},
  111:603--623, 2017.

\bibitem{BijalwanIJMSI2019}
A.~Bijalwan and B.~P. Patel.
\newblock {A new 3D finite element for the finite deformation of nearly
  incompressible hyperelastic solids}.
\newblock {\em International Journal of Materials and Structural Integrity},
  13(1/2/3):67--80, 2019.

\bibitem{KadapaIJNME2015}
C.~Kadapa, W.~G. Dettmer, and D.~Peri\'c.
\newblock {NURBS based least-squares finite element methods for fluid and solid
  mechanics}.
\newblock {\em International Journal for Numerical Methods in Engineering},
  101:521--539, 2015.

\bibitem{ManteuffelSIAMJNA2006}
T.~A. Manteuffel, S.~F. McCormick, J.G. Schmidt, and C.R. Westphal.
\newblock {First-order system least squares for geometrically nonlinear
  elasticity}.
\newblock {\em SIAM Journal on Numerical Analysis}, 44:2057--2081, 2006.

\bibitem{MajidiSIAMJNA2001}
M.~Majidi and G.~Starke.
\newblock {Least-squares Galerkin methods for parabolic problems I:
  semi-discretization in time}.
\newblock {\em SIAM Journal on Numerical Analysis}, 39:1302--1323, 2001.

\bibitem{MajidiSIAMJNA2002}
M.~Majidi and G.~Starke.
\newblock {Least-squares Galerkin methods for parabolic problems II: the fully
  discrete case and adaptive algorithms}.
\newblock {\em SIAM Journal on Numerical Analysis}, 39:1648--1666, 2002.

\bibitem{ElguedjCMAME2008}
T.~Elguedj, Y.~Bazilevs, V.M. Calo, and T.~J.~R. Hughes.
\newblock {\={B} and \={F} projection methods for nearly incompressible linear
  and non-linear elasticity and plasticity using higher-order NURBS elements}.
\newblock {\em Computer Methods in Applied Mechanics and Engineering},
  197:2732--2762, 2008.

\bibitem{LeiEC2016}
Z.~Lei, E.~Rougier, E.~E. Knight, L.~Frash, J.~W. Carey, and H.~Viswanathan.
\newblock {A non-locking composite tetrahedron element for the combined finite
  discrete element method}.
\newblock {\em Engineering Computations}, 33(7):1929--1956, 2016.

\bibitem{MehnertMMS2017}
M.~Mehnert, J.~P. Pelteret, and P.~Steinmann.
\newblock {Numerical modelling of nonlinear thermo-electro-elasticity}.
\newblock {\em Mathematics and Mechanics of Solids}, 22(11):2196--2213, 2017.

\bibitem{WulfinghoffCMAME2017}
S.~Wulfinghoff, H.~R. Bayat, A.~Alipour, and S.~Reese.
\newblock {A low-order locking-free hybrid discontinuous Galerkin element
  formulation for large deformations}.
\newblock {\em Computer Methods in Applied Mechanics and Engineering},
  323:353--372, 2017.

\bibitem{WriggersCM2017}
P.~Wriggers, B.~D. Reddy, w.~Rust, and B.~Hudobivnik.
\newblock {Efficient virtual element formulations for compressible and
  incompressible finite deformations}.
\newblock {\em Computational Mechanics}, 60:253--268, 2017.

\bibitem{JiangIJCM2018}
C.~Jiang, X.~Han, Z-Q. Zhang, G.~R Liu, and G-J. Gao.
\newblock {A Locking-Free Face-Based S-FEM via Averaging Nodal Pressure using
  4-Nodes Tetrahedrons for 3D Explicit Dynamics and Quasi-statics}.
\newblock {\em International Journal for Computational Methods}, 15(6):1850043,
  2018.

\bibitem{BayatAMSES2018}
H.~R. Bayat, S.~Wulfinghoff, S.~Kastian, and S.~Reese.
\newblock {On the use of reduced integration in combination with discontinuous
  Galerkin discretization: application to volumetric and shear locking
  problems}.
\newblock {\em Advanced Modelling and Simulation in Engineering Sciences},
  5:10, 2018.

\bibitem{BayatCM2018}
H.~R. Bayat, J.~Kr\"amer, L.~Wunderlick, S.~Wulfinghoff, S.~Reese, B.~Wohlmuth,
  and C.~Wieners.
\newblock {Numerical evaluation of discontinuous and nonconforming finite
  element methods in nonlinear solid mechanics}.
\newblock {\em Computational Mechanics}, 65:1413--1427, 2018.

\bibitem{CoombsCMAME2018}
W.~M. Coombs, T.~J. Charlton, M.~Cortis, and C.~E. Augarde.
\newblock {Overcoming volumetric locking in material point methods}.
\newblock {\em Computer Methods in Applied Mechanics and Engineering},
  333:1--21, 2018.

\bibitem{SevillaIJNME2018}
R.~Sevilla, M.~Giacomini, A.~Karkoulias, and A.~Huerta.
\newblock {A superconvergent hybridisable discontinuous Galerkin method for linear elasticity}.
\newblock {\em International Journal for Numerical Methods in Engineering},
  116:91--116, 2018.

\bibitem{ToghipourCMA2018}
A.~Toghipour, J.~Parvizian, S.~Heinze, and A.~D\"uster.
\newblock {The finite cell method for nearly incompressible finite strain
  plasticity problems with complex geometries}.
\newblock {\em Computers \& Mathematics with Applications}, 75(9):3298--3316,
  2018.

\bibitem{MoutsanidisCPM2019}
G.~Moutsanidis, J.~J. Koester, M.~R. Tupek, J-S. Chen, and Y.~Bazilevs.
\newblock {Treatment of near-incompressibility in meshfree and
  immersed-particle methods}.
\newblock {\em Computational Particle Mechanics}, pages 1--19, 2019.

\bibitem{DalIJNME2019}
H.~Dal.
\newblock {A quasi incompressible and quasi inextensible element formulation
  for transversely isotropic materials}.
\newblock {\em International Journal for Numerical Methods in Engineering},
  117(1):118--140, 2019.

\bibitem{OnishiIJNME2017}
Y.~Onishi, R.~Iida, and K.~Amaya.
\newblock {F-bar aided edge-based smoothed finite element method using
  tetrahedral elements for finite deformation analysis of nearly incompressible
  solids}.
\newblock {\em International Journal for Numerical Methods in Engineering},
  109:1582--1606, 2017.

\bibitem{OnishiIJCM2019}
Y.~Onishi.
\newblock {F-Bar Aided Edge-Based Smoothed Finite Element Method with 4-Node
  Tetrahedral Elements for Static Large Deformation Elastoplastic Problems}.
\newblock {\em International Journal for Computational Methods}, 16(5):1840010,
  2019.

\bibitem{SevillaCS2019}
R.~Sevilla, M.~Giacomini, and A.~Huerta.
\newblock {A locking-free face-centred finite volume (FCFV) method for linear elastostatics}.
\newblock {\em Computers and Structures}, 212:43-57, 2019.

\bibitem{connolly2018}
S.~J. Connolly, D.~Mackenzie, and Y.~Gorash.
\newblock {Isotropic hyperelasticity in principal stretches: explicit
  elasticity tensors and numerical implementation}.
\newblock {\em Computational Mechanics}, 64(5):1273--1288, 2019.

\bibitem{connolly2019}
S.~J. Connolly, D.~Mackenzie, and Y.~Gorash.
\newblock {Higher-order and higher floating‐point precision numerical
  approximations of finite strain elasticity moduli}.
\newblock {\em International Journal for Numerical Methods in Engineering},
  120(10):1184--1201, 2019.

\bibitem{ViebahnAMSES2019}
N.~Viebahn, J.~Schr\"oder, and P.~Wriggers.
\newblock {An extension of assumed stress finite elements to a general
  hyperelastic framework}.
\newblock {\em Advanced Modeling and Simulation in Engineering Sciences}, 6:9,
  2019.

\bibitem{book-fem-BonetWood}
J.~Bonet and R.~D. Wood.
\newblock {\em {Nonlinear continuum mechanics for finite element analysis}}.
\newblock Cambridge University Press, 1997.

\bibitem{SchroderCM2017}
J.~Schr\"oder, N.~Viebahn, P.~Wriggers, F.~Auricchio, and K.~Steeger.
\newblock {On the stability analysis of hyperelastic boundary value problems
  using three- and two-field mixed finite element formulations}.
\newblock {\em Computational Mechanics}, 60:479--492, 2017.

\bibitem{BisharaMMS2018}
D.~Bishara and M.~Jabareen.
\newblock {A reduced mixed finite-element formulation for modeling the
  viscoelastic response of electro-active polymers at finite deformation}.
\newblock {\em Mathematics and Mechanics of Solids}, 24(5):1578--1610, 2018.

\bibitem{ParkIJSS2012}
H.~S. Park, Z.~Suo, J.~Zhou, and P.~A. Klein.
\newblock {A dynamic finite element method for inhomogeneous deformation and
  electromechanical instability of dielectric elastomer transducers}.
\newblock {\em International Journal of Solids and Structures},
  49(15-16):2187--2194, 2012.

\bibitem{ParkSS2013}
H.~S. Park and T.~D. Nguyen.
\newblock {Viscoelastic effects on electromechanical instabilities in
  dielectric elastomers}.
\newblock {\em Soft Matter}, 9:1031--1042, 2013.

\bibitem{SeifiCMAME2018}
S.~Seifi, K.~C. Park, and H.~S. Park.
\newblock {A staggered explicit-implicit finite element formulation for
  electroactive polymers}.
\newblock {\em Computer Methods in Applied Mechanics and Engineering},
  337:150--164, 2018.

\bibitem{AskIJNLM2012}
A.~Ask, A.~Menzel and M.~Ristinmaa.
\newblock {Phenomenological modeling of viscous electrostrictive polymers}.
\newblock {\em International Journal of Non-Linear Mechanics}, 47(2):156--165,
  2012.

\bibitem{PelteretIJNME2016}
J.~P. Pelteret, D.~Davydov, A.~McBride, D.~K. Vu, and P.~Steinmann.
\newblock {Computational electro- and magneto-elasticity for
  quasi-incompressible media immersed in free space}.
\newblock {\em International Journal for Numerical Methods in Engineering},
  108(11):1307--1342, 2016.

\bibitem{JabareenPIUTAM2015}
M.~Jabareen.
\newblock {On the modeling of electromechanical coupling in electro-active
  polymers using the mixed finite element formulation}.
\newblock {\em Procedia IUTAM}, 12:105--115, 2015.

\bibitem{mehnert2018}
M.~Mehnert, M.~Hossain, and P.~Steinmann.
\newblock {Numerical modeling of thermo-electro-viscoelasticity with
  field-dependent material parameters}.
\newblock {\em International Journal of Non-Linear Mechanics}, 106:13--24,
  2018.

\bibitem{mehnert2019}
M.~Mehnert, M.~Hossain, and P.~Steinmann.
\newblock {Experimental and numerical investigations of the
  electro-viscoelastic behavior of VHB 4905}.
\newblock {\em European Journal of Mechanics-A/Solids}, 77:103797, 2019.

\bibitem{SteinmannAAM2012}
P.~Steinmann, M.~Hossain, and G.~Possart.
\newblock {Hyperelastic models for rubber-like materials: Consistent tangent
  operators and suitability of Treloar's data}.
\newblock {\em Archive of Applied Mechanics}, 82(9):1183--1217, 2012.

\bibitem{HossainJMBM2013}
M.~Hossain and P.~Steinmann.
\newblock {More hyperelastic models for rubber-like materials: Consistent
  tangent operator and comparative study}.
\newblock {\em Journal of the Mechanical Behaviour of Materials},
  22(1-2):27--50, 2013.

\bibitem{HossainJMBM2015}
M.~Hossain, N.~Kabir, and A.~F. M.~S. Amin.
\newblock {Eight-chain and full-network models and their modified versions for
  rubber hyperelasticity: A comparative study}.
\newblock {\em Journal of the Mechanical Behaviour of Materials},
  24(1-2):11--24, 2015.

\bibitem{MarckmannRCT2006}
G.~Marckmann and E.~Verron.
\newblock {Comparison of hyperelastic models for rubber-like materials}.
\newblock {\em Rubber Chemistry and Technology, American Chemical Society},
  79(5):835--858, 2006.

\bibitem{DollJAM2000}
S.~Doll and K.~Schweizerhof.
\newblock {On the Development of Volumetric Strain Energy Functions}.
\newblock {\em Journal of Applied Mechanics}, 67(1):17--21, 2000.

\bibitem{Moerman2019}
K.~M. Moerman, B.~Fereidoonnezhad, and P.~McGarry.
\newblock {Novel Hyperelastic Models for Large Volumetric Deformations}.
\newblock {\em \url{https://engrxiv.org/cfxdr}}, 2019.

\bibitem{HartmannIJSS2003}
S.~Hartmann and P.~Neff.
\newblock {Polyconvexity of generalized polynomial type hyperelastic strain
  energy functions for near incompressibility}.
\newblock {\em International Journal of Solids and Structures}, 40:2767--2791,
  2003.

\bibitem{Ansys2000}
Ansys Inc.
\newblock {\em ANSYS Theory Manual}, 2000.

\bibitem{SimoCMAME1991}
J.~C Simo and R.~L. Taylor.
\newblock {Quasi-incompressible finite elasticity in principal stretches.
  continuum basis and numerical algorithms}.
\newblock {\em Computer Methods in Applied Mechanics and Engineering},
  85:273--310, 1991.

\bibitem{SimoCMAME1982}
J.C. Simo and R.~L. Taylor.
\newblock {Penalty function formulations for incompressible nonlinear elastostatics}.
\newblock {\em Computer Methods in Applied Mechanics and Engineering},
  35:107--118, 1982.

\bibitem{LiuEC1994}
C.~H. Liu, G.~Hofstetter, and H.~A. Mang.
\newblock {3D finite element analysis of rubber-like materials at finite strains}.
\newblock {\em Engineering Computations}, 11:111--128, 1994.

\bibitem{MeiheIJNME1994}
C.~Meihe.
\newblock {Aspects of the formulation and finite element implementation of large strain isotropic elasticity}.
\newblock {\em International Journal for Numerical Methods in Engineering},
  37:1981--2004, 1994.

\bibitem{KadapaIJNME2019bbar}
C.~Kadapa.
\newblock {Novel quadratic B\'ezier triangular and tetrahedral elements using existing mesh generators: Applications to linear nearly incompressible elastostatics and implicit and explicit elastodynamics}.
\newblock {\em International Journal for Numerical Methods in Engineering},
  117:543--573, 2019.

\bibitem{book-Matrices-Duff}
I.~S. Duff, A.~M. Erisman, and J.~K. Reid.
\newblock {\em {Direct Methods for Sparse Matrices}}.
\newblock Oxford Science Publications, second edition, 2017.

\bibitem{GouldACMTOMS2007}
N.~I. Gould, J.~A. Scott, and Y.~Hu.
\newblock {A numerical evaluation of sparse direct solvers for the solution of large sparse symmetric linear systems of equations}.
\newblock {\em ACM Transactions on Mathematical Software}, 33(2):10, 2007.

\bibitem{book-linalgebra-Dongarra}
J.~Dongarra, I.~S. Duff, D.~C. Sorensen, and H.~A. van~der Vorst.
\newblock {\em {Numerical Linear Algebra for High-performance Computers}}.
\newblock SIAM, 1998.

\bibitem{book-Matrices-Pissanetzky}
S.~Pissanetzky.
\newblock {\em {Sparse Matrix Technology}}.
\newblock Academic Press, 1984.

\bibitem{book-Matrices-Saad}
Y.~Saad.
\newblock {\em {Iterative methods for sparse linear systems}. Vol. 82. SIAM, 2003.}
\newblock SIAM, second edition, 2003.

\bibitem{book-Matrices-Barrett}
R.~Barrett, M.~W. Berry, T.~F. Chan, J.~Demmel, J.~Donato, J.~Dongarra, V.~Eijkhout, R.~Pozo, C.~Romine, and H.~Van~der Vorst.
\newblock {\em {Templates for the solution of linear systems: building blocks for iterative methods}}.
\newblock SIAM, 1994.

\bibitem{NachtigalSIAMJMA1992}
N.~M. Nachtigal, S.~C. Reddy, and L.~N. Trefethen.
\newblock {How fast are nonsymmetric matrix iterations?}
\newblock {\em SIAM Journal on Matrix Analysis and Applications},
  13(3):778--795, 1992.

\bibitem{book-Matrices-Watkins}
D.~S. Watkins.
\newblock {\em {Fundamentals of Matrix Computations}}.
\newblock John Wiley \& Sons, New York, second edition, 2002.

\bibitem{ReeseIJNME1999}
S.~Reese, M.~Kussner, and B.~D. Reddy.
\newblock {A new stabilization technique for finite elements in non-linear elasticity}.
\newblock {\em International Journal for Numerical Methods in Engineering},
  44:1617--1652, 1999.

\bibitem{KryslIJNME2015}
P.~Krysl.
\newblock {Mean-strain eight-node hexahedron with optimized energy-sampling stabilization for large-strain deformation}.
\newblock {\em International Journal for Numerical Methods in Engineering},
  103:650--670, 2015.
  
\bibitem{abaqus-manual}
Abaqus theory manual. \url{https://classes.engineering.wustl.edu/2009/spring/mase5513/abaqus/docs/v6.6/books/stm/default.htm?startat=ch03s02ath61.html}

\bibitem{RajagopalMAMS2018}
A.~Rajagopal, M.~Kraus, and P.~Steinmann.
\newblock {Hyperelastic analysis based on a polygonal finite element method}
\newblock {\em Mechanics of Advanced Materials and Structures},
25:930--942, 2018.

\myred{
\bibitem{OrtigosaCMAME2016}
R.~Ortigosa, A.~J.~Gil, and C.~H.~Lee.
\newblock {A computational framework for large strain nearly and truly incompressible electromechanics based on convex multi-variable strain energies},
\newblock {\em Computer Methods in Applied Mechanics and Engineering}, 310:297--334, 2016.

\bibitem{Bercovier}
M. Bercovier. Perturbation of mixed variational problems. Application to mixed finite element methods, RAIRO. Analyse numérique, 12, 211-236, 1978.

\bibitem{Wriggers}
P. Wriggers, J. Schr\"oder, F. Auricchio. Finite element formulations for large strain anisotropic material with inextensible fibers, Advanced Modeling and Simulation in Engineering Sciences, 3:25, 2016.

\bibitem{Zienkiewicz}
O. C. Zienkiewicz, R. L. Taylor, J. Z. Zhu. The Finite Element Method: Its Basis and Fundamentals, Sixth Edition, Elsevier Butterworth and Heinemann, Oxford, United Kingdom, 2005.

\bibitem{Simo}
J. C. Simo, P. Wriggers, R. L. Taylor. A perturbed Lagrangian formulation for the finite element solution of contact problems, Computer Methods in Applied Mechanics and Engineering, 50:163-180, 1985.

\bibitem{Tur}
M. Tur, J. Albelda. J. M. Navarro-Jimenez, J, J, Rodenas. A modified perturbed Lagrangian formulation for contact problems, Computational Mechanics, 55:737-754, 2015.
}

\end{thebibliography}

\end{document}